\documentclass[10pt,twoside,twocolumn]{article}
\usepackage[english]{babel}
\usepackage{times,subeqnarray}
\usepackage{url}
\newif\myifpdf
\ifx\pdfoutput\undefined
   \usepackage[dvips]{graphicx}
\else
   \pdfoutput=1        
   \usepackage[pdftex]{graphicx}
\fi

\usepackage{apatitlepages}
\usepackage{psydraft}
\usepackage{apa}       

\newcommand{\wij}{w_{ij}}

\newcommand{\oneo}[1]{\frac{1}{#1}}

\myifpdf
  \DeclareGraphicsExtensions{.pdf,.eps,.png,.jpg,.mps,.tif}
\fi

\def\myheading{ Deep Predictive Learning }

\begin{document}
\bibliographystyle{apa}

\sloppy
\raggedbottom

\def\mytitle{ Deep Predictive Learning: A Comprehensive Model of Three Visual Streams }

\def\myauthor{Randall C. O'Reilly, Dean R. Wyatte, and John Rohrlich\\
  Department of Psychology and Neuroscience \\
  University of Colorado Boulder, 345 UCB, Boulder, CO 80309\\
  {\small randy.oreilly@colorado.edu}\\}

\def\mynote{
  Supported by: ONR grants N00014-14-1-0670 / N00014-16-1-2128, N00014-15-1-2832, N00014-13-1-0067, D00014-12-C-0638.  We thank Tom Hazy, Seth Herd, Kai Krueger, Tim Curran, David Sheinberg, Lew Harvey, Jessica Mollick, Will Chapman, Helene Devillez, and the rest of the CCN Lab for many helpful comments and suggestions.   R. C. O'Reilly is Chief Scientist at eCortex, Inc., which may derive indirect benefit from the work presented here. }

\def\myabstract{
  How does the neocortex learn and develop the foundations of all our high-level cognitive abilities?  We present a comprehensive framework spanning biological, computational, and cognitive levels, with a clear theoretical continuity between levels, providing a coherent answer directly supported by extensive data at each level.  Learning is based on making predictions about what the senses will report at 100 msec (alpha frequency) intervals, and adapting synaptic weights to improve prediction accuracy.  The pulvinar nucleus of the thalamus serves as a projection screen upon which predictions are generated, through deep-layer 6 corticothalamic inputs from multiple brain areas and levels of abstraction.  The sparse driving inputs from layer 5 intrinsic bursting neurons provide the target signal, and the temporal difference between it and the prediction reverberates throughout the cortex, driving synaptic changes that approximate error backpropagation, using only local activation signals in equations derived directly from a detailed biophysical model.  In vision, predictive learning requires a carefully-organized developmental progression and anatomical organization of three pathways (What, Where, and What * Where), according to two central principles: top-down input from compact, high-level, abstract representations is essential for accurate prediction of low-level sensory inputs; and the collective, low-level prediction error must be progressively and opportunistically partitioned to enable extraction of separable factors that drive the learning of further high-level abstractions.  Our model self-organized systematic invariant object representations of 100 different objects from simple movies, accounts for a wide range of data, and makes many testable predictions.}


\twocolumn

\titlesamepage{\mytitle}{\myauthor}{\mynote}{\myabstract}







\section{Introduction}

What is the nature of the remarkable neocortical learning and maturational mechanisms that result in the development of our considerable perceptual and cognitive abilities?  In other words, where does our knowledge come from?  Phenomenologically, it appears to magically emerge after several months of slobber-filled gaping at the world passing by --- what is the magic recipe for extracting high-level knowledge from an ongoing stream of perceptual experience?  Answering this central question has been the ultimate goal of many lines of research, at many levels of analysis from synapses up to machine learning algorithms and psychological theories.  Despite many advances at each of these levels of analysis, we still lack an overall framework with the key elements of a comprehensive answer to this question: integration across these different levels in a mutually compatible way, with the account at each level having direct empirical support, and directly connecting to the adjacent levels, leaving no obvious theoretical roadblocks.  In this paper, such a framework is proposed, providing a broad and deep integration of many different sources of data and theoretical ideas coming from many different researchers.  This framework is implemented in a computer model that demonstrates both its computational function and its ability to account for a wide range of data.  Many important issues remain to be addressed and our simple, first-pass model has many limitations, but we suggest that it represents a coherent skeleton upon which future work can build.

Our model encompasses most of the posterior visual neocortex, including both the dorsal {\em Where} (and {\em How}) and ventral {\em What} pathways, along with a proposed {\em third} visual stream, that serves to integrate information from these other two streams (i.e., a {\em What * Where} stream).  The model watches a simple movie of brief scenes where one out of 100 different possible objects moves in a random direction (or remains still), and makes random saccades every 200 msec.  This captures the most basic aspects of the visual world: objects are generally stable over time, and Newton's first law of motion, while also incorporating the main reliable form of motor control that a baby has (moving their eyes).  We are primarily interested in two questions: can the model learn to accurately predict what it will see next, and can it develop higher-level abstract representations of this environment across these three visual pathways.  Specifically, can it learn to separate the {\em What} from the {\em Where} by developing invariant representations of the objects that can be used to recognize the specific object being viewed, regardless of where it appears and moves?  Such representations are widely recognized as having great adaptive value to an organism, and form the foundation of much of our semantic understanding of the world.  However, they typically require training with explicit, invariant category labels due to the strong anti-correlation between the similarity structure at the retinal inputs (where different objects in the same location are more similar than the same object at different locations) and the desired invariant object representations that discriminate between different objects.

\pagestyle{myheadings}

A critical idea, advanced by many different researchers, for how substantial amounts of learning can emerge from the largely passive sensory experience of babies, is that each moment can be turned into a {\em predictive learning} problem: learning to predict what visual input will arise next, given what has come before.  Achieving accurate such predictions requires extensive (implicit) knowledge of physics and optics (and many more things for richer environments, including other biological organisms, conspecifics, family, machines, etc), and thus it seems plausible that learning from prediction errors should be capable of shaping an effective internal model of the external world.  However, actually getting this idea to work in practice is quite challenging for a variety of reasons explained below.  One of the main contributions of the present work is showing that a specific set of principles and mechanisms for achieving effective predictive learning aligns remarkably well with detailed properties of the visual system, at multiple levels and across developmental maturation.

Anatomically, we hypothesize that the pulvinar nucleus of the thalamus plays the role of a {\em projection screen} where the predictions are represented (similar to Mumford's (1991) {\em blackboard} conception).  These predictions are generated every 100 msec (10 hz, alpha rhythm), collaboratively by the entire visual neocortex, conveyed to the pulvinar via extensive corticothalamic projections from cortical deep layers.  These predictions are followed immediately by driving inputs (via layer 5IB intrinsic bursting neurons) from V1 and other cortical areas that reflect the bottom-up {\em ground truth} training signal (i.e., what is actually seen).  The temporal difference between the activity state representing the prediction versus the state with the ground-truth training signal (i.e., the {\em prediction error}) propagates throughout the network, driving synaptic changes according to a biologically-based learning mechanism, which shapes representations across the network to reduce the prediction error.  The shared, collaborative nature of this pulvinar projection screen is critical for coordinating and specializing representations across different visual areas.

Computationally, our framework is a form of a {\em hierarchical generative model}, which have been widely explored as models of brain / cognitive function (and we restrict our discussion to that subset, broadly defined, as opposed to the broader machine learning field).  These models are typically trained progressively from the bottom-up (i.e., layer-by-layer), and according to a relatively strict hierarchy where each layer learns to predict the behavior of the layer below it.  We found this approach to have significant limitations, and instead discovered two critical principles that were necessary for the development of systematic, high-level, abstract knowledge representations in our model: 1. Compact, high-level abstract representations are essential for accurate prediction generation at the lowest levels, and thus there must be extensive top-down short-cut projections from the highest levels of the hierarchy down to the lowest levels; and 2. The overall prediction error (broadcast by the pulvinar as a temporal difference) must be progressively and opportunistically partitioned by differentially-specialized such high-level pathways, breaking it down into separable factors that can then drive learning to reflect these factors.  In the case of vision, the spatial ({\em Where}) aspect of prediction can be learned first, independent of the {\em What} aspect, and having systematic and accurate high-level spatial predictions projecting to the common pulvinar area then partitions away that aspect of the prediction error, leaving a residual that is more about object identity ({\em What}).  Furthermore, the development of systematic, invariant object-identity representations required a third visual pathway that absorbed the prediction error associated with integrating {\em What} and {\em Where} information together.

Thus, by incorporating key developmental and anatomical constraints on top of a core predictive learning framework based on properties of the pulvinar and deep neocortical layers, the interacting visual pathways learn to represent separable factors ({\em Where, What, What * Where}) that jointly yield highly accurate and generalizable predictions of subsequent visual inputs.  Critically, the {\em What} pathway of the model develops abstract, invariant object representations without any explicit object category inputs (i.e., in a purely self-organizing manner).  Thus, our model shows how abstract knowledge can emerge from passive observation of a sensory stream, potentially explaining the apparently magical emergence of knowledge in infants over the first few months of life.

Although many generative models are discussed in terms of generating predictions, many of them do not actually include an explicit temporal divide, and instead end up learning by reconstructing the {\em current} sensory input (e.g., an {\em auto-encoder} in neural network terms).  These kinds of auto-encoders require various constraints to avoid degenerate solutions, and it remains unclear whether such models can produce systematic abstract internal representations in a purely self-organizing manner (typically they are subsequently trained with standard explicit object category labels, for example).  By contrast, the task of predicting the {\em future} sensory input avoids many of these problems because, as the saying goes, prediction is difficult, {\em especially about the future}.  We reserve the term {\em predictive} here exclusively for the {\em about the future} sense, and discuss the relationship to existing models in detail in the General Discussion section.

A signature example of predictive behavior at the neural level in the brain is the {\em predictive remapping} of visual space in anticipation of a saccadic eye movements \cite{DuhamelColbyGoldberg92,ColbyDuhamelGoldberg97,GottliebKusunokiGoldberg98,NakamuraColby02,MarinoMazer16}.  Here, parietal neurons start to fire at the {\em future} receptive field location where a currently-visible stimulus will appear after a planned saccade is actually executed.  We argue that this is just one example of a far more pervasive predictive process operating throughout the neocortex to predict what will be experienced next.  A major consequence of this predictive process is the perception of a stable, coherent visual world despite constant saccades and other sources of visual change (to appreciate the importance of these predictive mechanisms, try gently nudging your eyeballs to experience what an unpredictable sensory experience feels like).  Our overall framework is consistent with the account of predictive remapping given by \incite{Wurtz08} and \incite{CavanaghHuntAfrazEtAl10}, who argue that the key remapping takes place at the high levels of the dorsal stream, which then drive top-down activation of the predicted location in lower areas, instead of the alternative where lower-levels remap themselves based on saccade-related signals.  The lower-level visual layers are simply too large and distributed to be able to remap across the relevant degrees of visual angle.

This same lesson applies broadly for generating predictions about all aspects of the world, and is why we believe that top-down activation from compact, high-level, abstract representations is essential for the success of predictive learning.  However, it also represents a notoriously challenging catch-22 problem: how can high-level abstract representations develop prior to the lower-level representations that they build upon?  How can we develop the abstract generalization of ``cat'' when we don't yet know anything about fur, paws, teeth, etc?  Our model successfully addresses this challenge using a variety of different pragmatic solutions, as we detail below.

\subsection{Core Mechanisms of Predictive Learning}

Predictive learning is an old and widely-explored idea \cite{Elman90,Elman91,Jordan89,SchusterPaliwal97,HawkinsBlakeslee04,GeorgeHawkins09}, which is also gaining renewed interest in some recent deep neural network models \cite{LotterKreimanCox16}.  In motor control, the notion of a predictive {\em forward model} that anticipates the outcomes of actions is well-established \cite{KawatoFurukawaSuzuki87,JordanRumelhart92,MiallWolpert96}, and the current framework advances the notion that the entire neocortex is a forward model for sensory and motor outcomes.  An important contribution of our model is to provide a detailed biological mapping of this predictive learning idea, that provides a clear continuity in going from low-level mechanisms of synaptic plasticity up to brain-area structure.  Specifically, our model provides biologically-sound answers to all of the following essential questions:

\begin{itemize}
\item {\em How do local synaptic signals drive plasticity in a way that produces highly-functional learning in the context of a large complex network of interacting brain areas?}  Although the biological data, and locality constraints, appear to favor some variant of a Hebbian learning mechanism, computational models consistently show that this form of learning is incapable of solving real-world problems, and that instead some form of error-driven learning is required.  The recent resurgence of interest in backpropagation learning \cite{RumelhartHintonWilliams86} reinforces the idea that this is the most powerful form of neural learning, and we have long argued that the relevant synaptic mechanisms are readily available to support this form of learning \cite{OReilly96,OReillyHazyHerd15,OReillyMunakataFrankEtAl12,OReillyMunakata00}.  Specifically, we argue that known biological mechanisms can readily support learning that is sensitive to a temporal difference in the state of both sending and receiving neurons across the synapse, where this temporal difference reflects an error signal as explained below.  We were able to derive this learning rule directly from a detailed and well-validated model of spike-timing-dependent-plasticity (STDP) \cite{UrakuboHondaFroemkeEtAl08}.  However, all of our previous models have relied upon implausible sources (and timing) of error signals (e.g., explicit category label inputs for object recognition, as in most current deep neural network models) --- a critical gap that is filled in the current framework.

\item {\em What is the source of error-driven learning signals?}  One of the most appealing features of predictive learning is that the relevant error signals are ubiquitous and ``free'': these systems learn by comparing what actually happens next versus a prediction generated just prior.  In this sense they are effectively {\em unsupervised} or {\em self-organizing} learning systems, because they do not require any additional source of learning signals.  However, unlike Hebbian-learning based self-organizing models, predictive learning can leverage the power of error backpropagation to drive learning in a deep hierarchy of areas, in a coordinated fashion, to produce much more powerful results.  Also, as noted above, we argue that predictive learning is better than the related, but simpler, goal of {\em auto-encoding} or reconstructing the current inputs (i.e., by learning a generative model that is capable of regenerating these input patterns).  Current deep-neural-network auto-encoder models typically adopt a de-noising framework in order to avoid the network learning a degenerate ``mindless copying'' solution to the problem: the inputs are presented with noise added, and the network is trained to produce the de-noised version \cite{BengioYaoAlainEtAl13,Valpola14,RasmusBerglundHonkalaEtAl15}.  By contrast, prediction is sufficiently challenging already, and adding the dynamic, temporal aspect to the problem adds many important dimensions of relevance to the real-world survival of organisms, so we think it is overall a much more likely goal for biological learning.  Nevertheless, predictive learning can be viewed as a form of auto-encoding (i.e., a {\em predictive auto-encoder}) in the sense that it is generating low-level visual representations to match actual inputs, and many lessons from auto-encoder networks should be applicable here as well.

\item {\em How are the prediction and actual outcome separately represented, and how is the timing of the prediction and outcome coordinated \& organized?}  Predictive learning models immediately raise these important and challenging questions, which fortunately admit to direct experimental testing and falsification.  The space of possibilities here is large, but we were able to find a particular solution that fits well with some otherwise rather peculiar features of the biology.  Specifically, we hypothesize that the higher-order thalamus (i.e., the {\em pulvinar}) provides the neural substrate for both the predicted and actual outcome, with alternating phases of prediction and outcome organized within the 100 msec / 10 Hz {\em alpha} cycle that is characteristic of both thalamic and deep neocortical layer firing (this is an evolution of our earlier proposal; \nopcite{KachergisWyatteOReillyEtAl14,OReillyWyatteRohrlich14}).  Thus, instead of having distinct neural substrates dedicated to representing either predictions or outcomes, we hypothesize a particular economy of shared functionality for this common substrate, which is particularly important in supporting the form of biologically-plausible synaptic-level error-driven learning that we had previously developed \cite{OReilly96}.  Specifically, this form of error-driven learning compares two states of network activation over time: an earlier {\em minus phase} state representing the network's best guess or expectation, versus a subsequent {\em plus phase} state reflecting the actual outcome (these terms, and the overall temporal-difference framework, were developed originally in the Boltzmann machine; \nopcite{AckleyHintonSejnowski85}).  Whereas we previously had only general speculations about how these phases were organized over time, and what constituted the actual plus phase signal, the predictive-learning-over-the-pulvinar hypothesis, organized in alternating phases within the alpha cycle, provides concrete, testable predictions that we evaluate below.

Although the prediction and outcome are encoded over the same pulvinar substrate in our model, we do hypothesize that the superficial (4,2,3) and deep (5,6) layers of the neocortex play distinct roles, with the superficial layers representing the {\em current state} of the environment and the ongoing internal ``mental'' state of the organism, while the deep layers are specifically responsible for {\em generating the prediction} about what will happen next (via direct projections into the pulvinar).  Well-established patterns of neocortical connectivity combine with phasic burst firing properties of a subset of deep-layer neurons to effectively shield the deep layers from direct knowledge of the current state, creating the opportunity to generate a prediction.  Metaphorically, the deep layers of the model are briefly closing their ``eyes'' so that they can have the challenge of predicting what they will ``see'' next.  This phasic disconnection from the current state is essential for predictive learning (even though it willfully hides information from a large population of neurons, which may seem counter-intuitive), and the remarkable convergence of biological properties supporting this phasic disconnection property in the deep neocortical layers provides strong support for our overall hypothesis.

\item {\em How are error signals represented, and transmitted to drive learning?}  Many attempts to map error-driven learning and generative models into the brain hypothesize the presence of neurons that explicitly represent the error signal in their firing.  However, as reviewed in detail below, we find the available evidence in support of such neurons in the neocortex or thalamus to be weak and subject to compelling alternative explanations.  Thus, we favor the {\em implicit} temporal-difference version of the error signal in our model as described above.  Mathematically, these temporal differences reflect the same error gradient as computed by the explicit error backpropagation algorithm \cite{OReilly96}, and we have shown that these error gradients propagated as activation signals through multiple interconnected areas are sufficient to train powerful deep object recognition networks \cite{OReillyWyatteHerdEtAl13,WyatteHerdMingusEtAl12,WyatteCurranOReilly12}.  Biologically, within the first 75 msec period of the overall 100 msec alpha cycle, the entire network interactively {\em settles} or converges on an integrated representation of the current state throughout the superficial layers, while the deep layers generate their best prediction of what will happen next, and project this to the pulvinar.  The full network of brain areas can thus work together to collaboratively produce the best possible representation, with individual pyramidal neurons sending standard excitatory signals to other pyramidal neurons, amid a background of dynamic {\em surround} inhibition.  Then, when the plus-phase outcome state is experienced over the last 25 msec of the alpha cycle (driven by burst firing of deep layer 5IB intrinsic bursting neurons that send strong feed-forward driving inputs to pulvinar thalamic relay cells (TRC's), as elaborated below), any differences between this outcome state and the prior prediction state are experienced as ripples of propagating activation-state differences emanating from the pulvinar and penetrating throughout the network.  Neurons receiving these projections from the pulvinar, both directly and indirectly, learn locally based on the temporal difference in their activation states across this critical alpha-frequency time-cycle.  

\end{itemize}

In recognition of the critical predictive role of deep neocortical layers, and the ability to train deep hierarchical networks, we refer to this as the {\em DeepLeabra} learning algorithm, building on our earlier {\em Leabra} mechanism that performed the same temporal-difference-based error-driven learning in bidirectionally-connected networks modeled only on the superficial layers of the neocortex \cite{OReillyHazyHerd15,OReillyMunakataFrankEtAl12,OReillyMunakata00,OReilly96}.
A critical feature of Leabra is the ability to effectively and efficiently learn and process information using {\em bidirectional excitatory connectivity}, which introduces a number of significant computational challenges (but is clearly a major feature of the biology of the neocortex; \abbrevnopcite{RocklandPandya79,FellemanVanEssen91,MarkovVezoliChameauEtAl14}). In contrast, most existing deep backpropagation models are strictly feedforward, or only do bidirectional processing in a restricted manner.  Furthermore, Leabra incorporates both error-driven learning and a robust form of Hebbian learning based on the BCM algorithm \cite{BienenstockCooperMunro82,CooperIntratorBlaisEtAl04,ShouvalWangWittenberg10}, which is essential for successful learning in our model as explored below.  Thus, our current model builds directly on this earlier computational infrastructure.

\subsection{Predictive-Learning a Multi-level Generative Model of the Visual World}

As summarized above, there are several critical challenges that must be resolved to enable a general predictive-learning mechanism to develop systematic, high-level abstract representations.  To illustrate, our initial attempts to test the DeepLeabra framework followed the widely-adopted idea of a progressive development of hierarchically-organized neocortical areas, proceeding progressively from the bottom-up \cite{ShragerJohnson96,BengioYaoAlainEtAl13,Valpola14,RasmusBerglundHonkalaEtAl15,HintonSalakhutdinov06}.  Specifically, lower-level visual areas such as V1 and V2 develop their representations first, predicting whatever they can at this lower-level, and then higher areas are progressively added to build upon these lower levels and develop higher-level representations {\em learned from the residual prediction errors left over from the lower areas}.  However, we inevitably found that these models never really learned very well (i.e., they could not do a very good job of predicting what was going to happen in the next 100 msec), nor did we find evidence of useful abstract representations developing in higher areas.  The power of error-driven learning is predicated on the ability of the error gradient to accurately and adaptively reflect new aspects of the problem to be solved --- if the network just gets stuck at a high level of error, the error gradients may not be able to find a way out, and the network just thrashes around without really going anywhere.   Eventually we concluded that this approach may be entirely backward --- what if the residual error from relatively impoverished lower-level representations is {\em not} in fact a sound basis for the formation of useful higher-level abstractions?

Instead, we are now convinced that predictive learning must {\em start} with as much high-level abstract representation as possible, and focus on learning further such representations as quickly as possible thereafter, {\em because central, compact, abstract representations of things like spatial motion and object properties are essential for successful predictive models}.  Without these coherent, central, higher-level representations, the lower-level predictions are doomed to mediocrity --- they will learn a vague, muddled and incoherent predictive model, which does not then provide a good basis for developing higher-level abstract representations at a later stage of learning.

High-level abstract representations are essential because they consolidate and concentrate learning within a centralized set of representations (e.g., about the nature and relationship of different features of an object for the {\em What} pathway).  These central representations can much more easily maintain this essential information over time, to support consistent, stable predictions about how an object will appear in the next moment.  By contrast, lower-level areas such as V1 or V2 are huge and strongly retinotopically organized, such that any given set of neurons only encodes a relatively small portion of the visual world (e.g., around 1 degree of visual angle).  Therefore, the encoding of object properties and motion trajectories in such areas must inevitably be highly diffuse and disconnected, with entirely different populations of neurons representing an object at one moment to the next.  Such representations provide a poor basis for accurate predictions, given the underlying stability of object properties, and their current motion trajectories, over time (at least over the 100 msec alpha timescale of relevance here).

This general principle that abstract, high-level internal representations should project down to lower layers to generate more detailed, specific renderings of the visual world is central to the widely-advocated {\em generative model} framework \cite[e.g.,]{CarpenterGrossberg87,Mumford92,KawatoHayakawaInui93,Ullman95,DayanHintonNealEtAl95,RaoBallard99,LeeMumford03,Friston05,HintonSalakhutdinov06,YuilleKersten06,Friston08,Friston10,Lee15,Clark13,Valpola14,RasmusBerglundHonkalaEtAl15} (as we review in greater detail in the General Discussion).  This idea is easily stated and compelling, but notoriously difficult to achieve in practice, because of the intrinsic interdependencies among all the different levels of representation required, creating a form of catch-22 as noted in the introduction.  Indeed, avoiding this catch-22 circularity is exactly what makes the widely-adopted bottom-up approach so appealing.

Instead, we outline below our strategy for circumventing this catch-22 while still having a strong top-down influence of high-level abstract representations as early as possible, which reflects an opportunistic, progressive development of different visual pathways, along with the {\em emergent} bidirectional convergence of the final pathway.   We highlight how this overall strategy makes sense of many disparate properties of the development and function of the visual system.
\begin{itemize}
\item {\em First, it is relatively easy to form spatial abstractions, and learn about both externally-generated object motion, and internally-generated saccade motion.}  Unlike the formation of invariant object identity abstractions (in the {\em What} pathway), spatial location (in retinotopic coordinates at least) can be trivially abstracted by simply aggregating across different feature detectors at a given retinotopic location, resulting in an undifferentiated spatial {\em blob}.  These spatial blob representations can drive high-level, central spatial pathways that can learn to predict where a given blob will move next, based on prior history, basic visual motion filters, and efferent copy inputs of saccadic eye movement plans and motor actions.  We start our model off by learning these high-level representations, which correspond well with those in area LIP high in the dorsal visual stream, prior to any significant development of any of the rest of the model.  These high-level spatial representations then provide strong top-down drive to the lower levels of the model, giving them access to highly accurate spatial prediction signals.  This has the highly beneficial effect of partitioning off this spatial aspect of the overall prediction error, thereby concentrating the residual error signals around the remaining problems described next.

Biologically, there is increasing evidence that this dorsal spatial pathway develops first, and furthermore that there are specific developmental changes in connectivity in relevant areas including the pulvinar, V1, and LIP that specifically support this early development \cite{BridgeLeopoldBourne16}.  Furthermore, connectivity analyses show that one of the very rare asymmetric pathways in the visual system goes directly from V1 to LIP \abbrevcite{MarkovErcsey-RavaszGomesEtAl14}, providing a direct short-cut for high-level spatial representations in LIP.

\item \emph{There are \textbf{two} residual problems that need to be solved after the  spatial \emph{Where} problem has been factored out: the traditional \emph{What} problem of representing visual object properties in an invariant manner, and the problem of integrating both \emph{What} and \emph{Where} information for generating highly accurate visual predictions.}  Each of these problems presents its own distinct challenges, and each  benefits from having its own dedicated hierarchy of neural processing (which nevertheless need to interact extensively with the others).  In terms of further partitioning the residual prediction errors, we show that by including this \emph{What * Where} pathway, which we hypothesize may involve area MT (V5) and dorsal prelunate (DP) cortex (along with higher levels of that pathway, including MST and possibly area V6; \nopcite{FattoriPitzalisGalletti09,KravitzSaleemBakerEtAl11}), the residual error associated with developing a high-quality abstract representation of object features is thereby concentrated in the remaining \emph{What} pathway (areas V4 and TEO, feeding into higher IT areas) --- only then are these representations able to form.  This integration layer can develop representations that strongly mix spatial and object feature information, to solve the {\em binding problem} entailed by integrating these two separate factors \cite{OReillyBusbySoto03,CerOReilly06}.

Biologically, there has been considerable debate about the true extent of separation between the {\em What} vs. {\em Where} pathways \cite{FreudPlautBehrmann16,deHaanCowey11,SchenkMcIntosh10,SerenoMaunsell98,HongYaminsMajajEtAl16,ZoccolanKouhPoggioEtAl07}, and it is evident overall that there is considerable interconnectivity \abbrevcite{MarkovVezoliChameauEtAl14,MarkovErcsey-RavaszGomesEtAl14,FellemanVanEssen91}.  There are several proposals for dividing the dorsal pathway into two sub-pathways \cite{RizzolattiMatelli03,KravitzSaleemBakerEtAl11,HaakBeckmann17}, and the role of MT in particular has been recognized as highly ambiguous \cite{MilnerGoodale06}. By positing a third visual stream, whose job it is to integrate {\em What} and {\em Where} information, we can potentially make more sense of all this interconnectivity.  More generally, the overall objective of learning to accurately predict what will be seen next makes it clear that these areas must interact with each-other extensively, and our model requires extensive cross-stream connectivity, despite also exhibiting specialization within-streams.  As we expand the complexity of the environment, it is likely that additional areas will be useful for developing more abstract, compact representations relevant to things like object physics, 3D shape, motor action prediction at many levels (reach, grasp, etc), biological motion, etc, consistent with the above references on the diversity of visual areas and pathways.

\item \emph{The \emph{What} visual pathway takes a relatively long time to develop useful abstract representations, so leveraging its benefits requires a later developmental strengthening of top-down connections from this pathway.}  We were unable to find a way for this pathway to develop earlier, as it seems to be dependent on successful learning in the other pathways, consistent with the idea that the these other pathways partition and concentrate the residual error on object feature information.  Thus, these abstractions are not available early-on to support predictive learning, in violation of the principle that abstract high-level representations are essential.  We therefore need to posit a later developmental strengthening of these top-down connections, once the representations have sufficiently developed, and we show that this then results in a significant boost in overall prediction accuracy, which still requires a relatively long time period to fully develop.  This developing {\em What} pathway informs and reshapes the ongoing {\em What * Where} pathway learning --- there are considerable bootstrapping and emergent bidirectional dependencies between these pathways.  Biologically, there is various evidence for delayed development of the {\em What} pathway \cite{Rodman94,NishimuraScherfBehrmann09}.

\item \emph{The pulvinar (as a kind of projection screen) broadcasts the main prediction error signal throughout the \emph{What * Where} and \emph{What} streams, and structured interconnections among areas then result in the partitioning of the residual errors to develop specialized pathways.}  We have consistently found that our model depends critically on all areas at all levels receiving the main predictive error signal generated by the V1 layer 5IB driver inputs to the pulvinar in the plus phase.  This was initially quite surprising at a computational level, as it goes strongly against the classic hierarchical organization of visual processing, where higher areas form representations on top of foundations built in lower areas --- how can high-level abstractions be shaped by this very lowest level of error signals?  We now understand that the overall learning dynamic in our model is analogous to \emph{multiple regression}, where each pathway in the model learns by  absorbing a component of the overall error signal, such that the residual is then available to drive learning in another pathway.  Thus, each factor in this regression benefits by directly receiving the overall error signal, and the process of partitioning out the residuals across brain areas requires specific patterns of interconnectivity that we have managed to discover through a long process of experimentation, guided by various principles of connectivity that emerged in this process (as detailed below).  One clear such principle is that although the pulvinar connections have this unusual flat connectivity pattern, most other connections obey the standard hierarchical pattern of connectivity, and this is essential for supporting the development of increasingly abstract representations in higher areas.  Also, this wide broadcast from the pulvinar helps to coordinate all of the different layers and share the emerging prediction directly among them, which likely has important benefits as well.

Biologically, the pulvinar does indeed interconnect widely with all areas of the cortex, and there is strong evidence for the idea that the lowest-level V1-driven signal drives all the major areas in the {\em What * Where} and {\em What} pathways \cite{Shipp03,KaasLyon07}.  In particular, individual V1 5IB driver neurons have multiple (3-5 or so) strong driving synapses into the pulvinar at different levels, whereas other areas only seem to have a single such driving synapse \cite{Rockland98a,Rockland96}.  

\end{itemize}

In summary, we offer a complex, emergent, yet principled account for how the seemingly intractable problem of simultaneously learning concrete and abstract representations across multiple interconnected areas of the visual cortex can be solved.  The working computational model is essential here to demonstrate the success of this approach, given the complex and emergent nature of the learning process.  In the remainder of the paper, we present the model, the simple dynamic visual environment on which it is trained, and the way in which learning evolves over time in the model.  We then use a variety of techniques to probe the nature of what is learned, and the forces that shape this learning, corroborating the overall account just given.  Next, we explore the relevant biological data and provide detailed simulations of particularly relevant neural recording data.  Because this model simulates such a large portion of the posterior neocortex, the scope of potentially-relevant data is vast, so our treatment is necessarily selective and opportunistic --- subsequent work will go into further details.  Each of these explorations includes a number of testable predictions from our model, and more general such predictions are outlined in the General Discussion section.

\section{The DeepLeabra Predictive Learning Framework}

We begin with an overview of how DeepLeabra models the cortical and thalamic pathways to do predictive learning, in terms of differential functional roles for superficial and deep layers of the neocortex, and loops through the thalamus, and the temporal dynamics of information flow through this circuit.

\begin{figure*}
  \centering\includegraphics[width=5in]{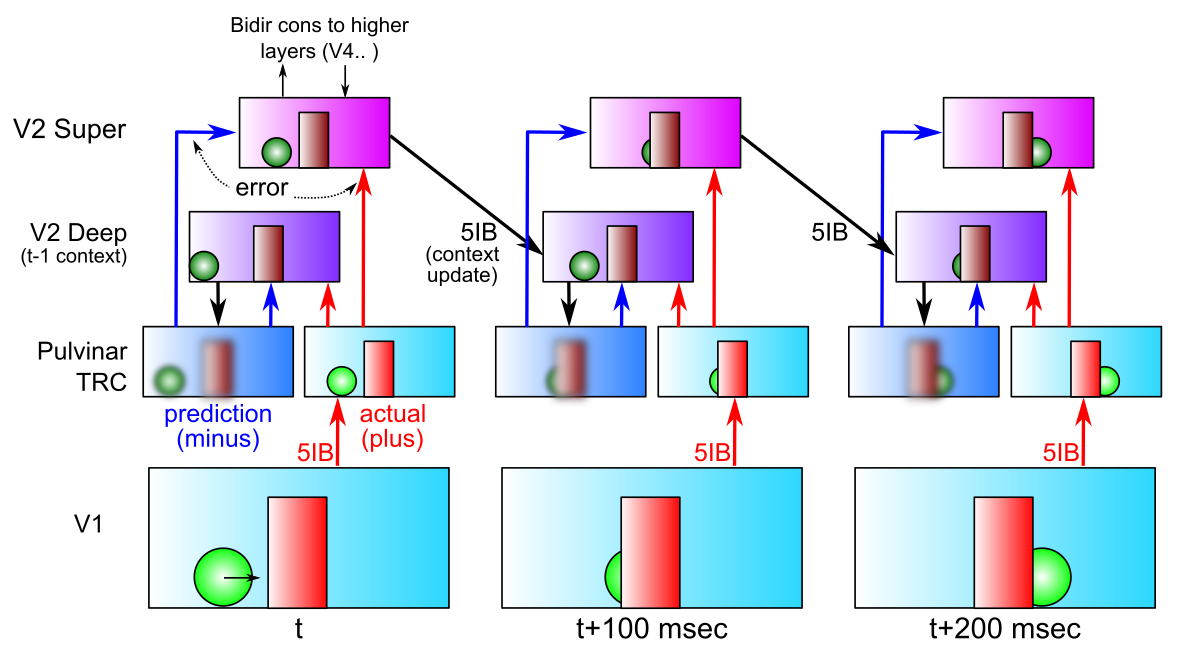}
  \caption{\footnotesize Schematic illustration of the temporal evolution of information flow in a DeepLeabra model predicting visual sequences, over a period of three alpha cycles of 100 msec each.  During each alpha cycle, the V2 Deep layer uses the prior 100 msec of context information to generate a prediction or expectation (minus phase) over the pulvinar thalamic relay cell (TRC) units of what will come in next via the 5IB strong driver inputs from V1, which herald the next plus or target phase of learning.  Error-driven learning occurs as a function of the temporal difference between the plus and minus activation states, in both superficial and deep networks, via the TRC projections into these networks.  The 5IB bursting in V2 drives an update of the local temporal context information in V2, which is used in generating the minus phase in the next alpha cycle, and so on.  These same 5IB cells drive a plus phase in higher area TRC's as well, which perform the same kind of {\em local} predictive auto-encoder learning as shown for V2 here.  This system is a predictive auto-encoder (generative model), because it is learning to generate a representation of the V1 inputs (as encoded via the relatively fixed V1 5IB to pulvinar projection).}
  \label{fig.pred_vis_motion}
\end{figure*}

Figure~\ref{fig.pred_vis_motion} provides an overall schematic for how predictive auto-encoder learning takes place in our framework, in terms of area V2 predicting the next pattern of activation on V1, over the period of three alpha-cycle ``movie frames'' (interestingly, actual film-based movies have a frame rate of 24 Hz, which is just over the 2x nyquist sampling limit for a 10 Hz process).  The V2 deep-layer neurons drive activation of a minus-phase prediction over the pulvinar, and then in the plus phase the 5IB neurons in area V1 drive the pulvinar with the actual sensory input state, and the temporal difference between the two represents the error signal that trains the superficial and deep layers to create better representations for making a more accurate prediction next time around.

Unpacking this and the prior summaries, here is the full set of explicit hypotheses and relevant biological data behind this predictive learning process in our model:
\begin{itemize}
\item The neocortex is composed of two separable but tightly interacting sub-networks, superficial and deep / thalamic (pulvinar).  The superficial-layer network consists of neocortical layers 4, 2, and 3, across different brain areas, with extensive bidirectional interconnectivity (feedforward going from 2/3 to layer 4 in the next area, and feedback coming from 2/3 in one area back to 2/3 in an earlier area; \abbrevnopcite{RocklandPandya79,FellemanVanEssen91,MarkovVezoliChameauEtAl14}). The deep / thalamic network starts in each area with the layer 5b intrinsic bursting (IB) neurons (5IB, \nopcite{ConnorsGutnickPrince82,LopesdaSilva91,ShermanGuillery06,FranceschettiGuatteoPanzicaEtAl95,FlintConnors96,SilvaAmitaiConnors91}), which receive inputs from local superficial neurons and top-down projections from other areas (e.g., higher-level task control signals).  These 5IB neurons then project to deep layer 6, which interconnects with the thalamus (which in turn projects back up to layer 4 of the superficial network and layer 6 in the deep network), and the 5IB neurons also provide a strong driving feedforward input to higher-area thalamic areas.

\item The superficial network represents the current state of the environment and internal state of the organism, at multiple different levels of abstraction, all mutually interacting.  It can be described computationally in terms of a classic Hopfield network / Boltzmann machine constraint satisfaction system \cite{Hopfield82,Hopfield84,AckleyHintonSejnowski85,RumelhartMcClelland82}, that settles over bidirectional activation propagation updates into a state (representation) that best satisfies the current bottom-up inputs and top-down knowledge / task-driven constraints.  This does not imply that the network converges fully to a stable settled attractor state --- just that it moves in that direction within the alpha-cycle time frame, after which changes in the deep / thalamic network (and in the sensory inputs) drive a new settling process under new constraints.

\item The deep / thalamic network in the posterior cortex is directly responsible for generating predictions over the pulvinar.  It must be phasically shielded from the current state information in the superficial layers, to be forced to generate a prediction as opposed to simply copying the current input state (in which case it would become a simple auto-encoder).  As such, it only phasically receives new bottom-up input about the state of the environment, triggered by alpha-frequency bursting of the layer 5IB neurons (which is also entrained via thalamocortical networks via various mechanisms \abbrevnopcite{LorinczKekesiJuhaszEtAl09,FranceschettiGuatteoPanzicaEtAl95,SaalmannPinskWangEtAl12}).  During the minus phase, when it is generating the next prediction, the deep state reflecting the last 5IB burst of activity is sustained and elaborated through regular spiking layer 6 neurons (i.e., layer 6CT corticothalamic neurons; \nopcite{Thomson10,ThomsonLamy07}) that project to the thalamic relay cells (TRC) of the pulvinar, which then project back to these same 6CT neurons (and up to the layer 4 inputs to the superficial network).  Computationally, we divide the 100 msec alpha cycle into 25 msec quarters, with the final quarter corresponding to the time of 5IB bursting and the plus phase (and the prior three quarters constituting the minus phase) --- these quarters are thus at the gamma frequency (40 hz), which is typically observed for superficial layer neural firing, and is thought to be modulated by the overall alpha frequency envelope \cite{DoughertyCoxNinomiyaEtAl17,vanKerkoerleSelfDagninoEtAl14,HaegensNacherLunaEtAl11,LakatosKarmosMehtaEtAl08,SpaakBonnefondMaierEtAl12,BollimuntaMoSchroederEtAl11,BollimuntaChenSchroederEtAl08}.

Extensive biological evidence supports the alpha-frequency dynamics of the deep layer network (and gamma for the superficial layers), including direct electrophysiological recording \cite{LuczakBarthoHarris13}, local-field-potential recordings from superficial vs. deep layers \cite{BuffaloFriesLandmanEtAl11,MaierAdamsAuraEtAl10,MaierAuraLeopold11,SpaakBonnefondMaierEtAl12,XingYehBurnsEtAl12,BastosVezoliBosmanEtAl15,MichalareasVezolivanPeltEtAl16}, and top-down-specific synchronization \cite{vonSteinChiangKonig00,vanKerkoerleSelfDagninoEtAl14}.  Furthermore, the pulvinar has been shown to drive alpha-frequency synchronization of cortical activity across areas in the alpha band \cite{SaalmannPinskWangEtAl12}.  Behaviorally, as reviewed below, there is extensive evidence of alpha-frequency effects on perception consistent with our framework  \cite{NunnOsselton74,VarelaToroJohnEtAl81,VanRullenKoch03,JensenBonnefondVanRullen12}.

\item Computationally, the deep / thalamic network activations encode temporal context information that reflects activations from the prior 100 msec period, in a manner similar to the simple recurrent network (SRN) model \cite{Elman90,Elman91,Jordan89}.  The SRN is so-named because it employs the {\em simple} trick of copying the current internal (hidden) layer representation to a context layer that then acts as an additional input to the hidden layer for generating a prediction of what will happen on the next time step.  In effect, we hypothesize that the time step for updating an SRN-like context layer is the 100 msec alpha cycle, and during a single alpha cycle, considerable bidirectional constraint satisfaction neural processing is taking place within a DeepLeabra network.  This contrasts with the standard SRN, which is typically implemented in a feedforward backpropagation network, where each time step and context update corresponds to a single feedforward activation pass through the network.  We discuss this and other relevant biological and computational issues in more detail in the Appendix.  Briefly, our model differs from a standard SRN by pre-computing the context-integrated net input, which deep layer neurons can maintain through bidirectional excitatory loops and longer-lasting channel dynamics, e.g., in NMDA and mGluR receptors.  But it fundamentally retains the copy-then-learn dynamic of an SRN, which we argue is essential because subsequent outcomes must be used to determine what is relevant from the past.

\item Biologically, there are two different types of cortical connections into pulvinar TRC neurons \cite{ShermanGuillery06}: strong, sparse {\em driver} connections originating from 5IB neurons (originally labeled R or type-2; \nopcite{Rockland98,Rockland96}), and weaker but much more numerous {\em modulatory} connections originating from 6CT neurons (E or type-1).  We depart from the modulatory notion of \incite{ShermanGuillery06}, and argue that these weaker 6CT inputs are capable of driving TRC activation by themselves, in the form of the minus-phase prediction representation.  Indeed, extensive {\em in vivo} electrophysiological recording data shows constant steady activation of pulvinar neurons across multiple alpha trials worth of time, suggesting that these projections are capable of driving TRC activation in between the 5IB bursting \cite{Bender82,PetersenRobinsonKeys85,BenderYouakim01,Robinson93,SaalmannPinskWangEtAl12,KomuraNikkuniHirashimaEtAl13}.  This minus phase is then followed by the strongly-driven 5IB plus-phase representation, which is essentially a copy of the sending layer activations (e.g., V1).  To generate the predicted minus-phase state, the layer 6CT neurons rely on integrated inputs from earlier 6 corticocortical (6CC) neurons and 5IB neurons, along with various other largely top-down inputs.  

\item In addition to the predictive learning functions of the deep / thalamic layers, these same circuits are also likely critical for supporting powerful top-down attentional mechanisms that have a net multiplicative effect on superficial-layer activations \cite{BortoneOlsenScanziani14,OlsenBortoneAdesnikEtAl12,BortoneOlsenScanziani14,OlsenBortoneAdesnikEtAl12}.  These attentional modulation signals cause the iterative constraint satisfaction process in the superficial network to focus on task-relevant information while down-regulating responses to irrelevant information --- in the real world, there are typically too many objects to track at any given time, so predictive learning must be directed toward the most important objects.  Indeed, there are well-established capacity constraints of around 2-4 objects (or ``fingers of instantiation,'' FINST's; \nopcite{Pylyshyn89}) that can be tracked at any given time, including during the predictive remapping process \cite{CavanaghHuntAfrazEtAl10}.  We are generally surprisingly unaware of how much we are {\em not} tracking, because typically we can just re-access the environment to encode any element we might have initially overlooked \cite{SimonsRensink05}.

Computationally, we show below that these deep / thalamic circuits produce attentional effects consistent with the abstract \incite{ReynoldsHeeger09} model, while the contributions of the deep layer networks to this function are broadly consistent with the folded-feedback model \cite{Grossberg99}.  Biologically, the layer 6CT neurons are known to exhibit a multiplicative influence over firing of superficial-layer neurons, in a manner consistent with the \incite{ReynoldsHeeger09} model \cite{BortoneOlsenScanziani14,OlsenBortoneAdesnikEtAl12}.  The importance of the pulvinar for attentional processing has been widely documented \cite[e.g.,]{LaBergeBuchsbaum90,BenderYouakim01,SaalmannPinskWangEtAl12}, and there is likely an additional important role of the thalamic reticular nucleus (TRN), which can contribute a surround-inhibition contrast-enhancing effect on top of the incoming attentional signal from the cortex \cite{Crick84,Pinault04,WimmerSchmittDavidsonEtAl15}.  We briefly elaborate on these ideas toward the end of this paper, and a subsequent paper will explore them in greater depth.

\end{itemize}

\section{A Comprehensive Model of Three Visual Streams}

The above DeepLeabra predictive auto-encoder learning mechanisms provide the core engine of our systems-level model of how the three different visual pathways ({\em Where}, {\em What * Where}, and {\em What}) work together to produce highly accurate visual predictions.  As summarized earlier, this model requires considerable additional structure and developmental organization to achieve fully successful learning, based on abstract high-level representations driving top-down inputs to the lower areas where the more detailed visual prediction is rendered.  The measure of success in this model is not just that it accurately predict the next sensory inputs, but, more importantly, that it develop these high-level abstract representations that can then provide a more systematic basis for intelligent behavior.  For example, by developing invariant object representations, an organism would be able to systematically respond appropriately to the presence of objects regardless of the perceptual details in which that object was viewed.

The strong correspondence between the specific computationally-motivated network properties and the known biology, reviewed in greater detail here, supports the idea that this model accurately describes how the actual mammalian visual neocortex learns.  We first provide an overview of the full model and the simple dynamic visual environment on which it is trained (including saccades), followed by basic computationally-oriented results demonstrating the key principles underlying its learning abilities.  Then, we provide detailed accounts of a range of different data of particular relevance to the model, followed by further testable predictions that the model could make.

\begin{figure*}
  \centering\includegraphics[width=4in]{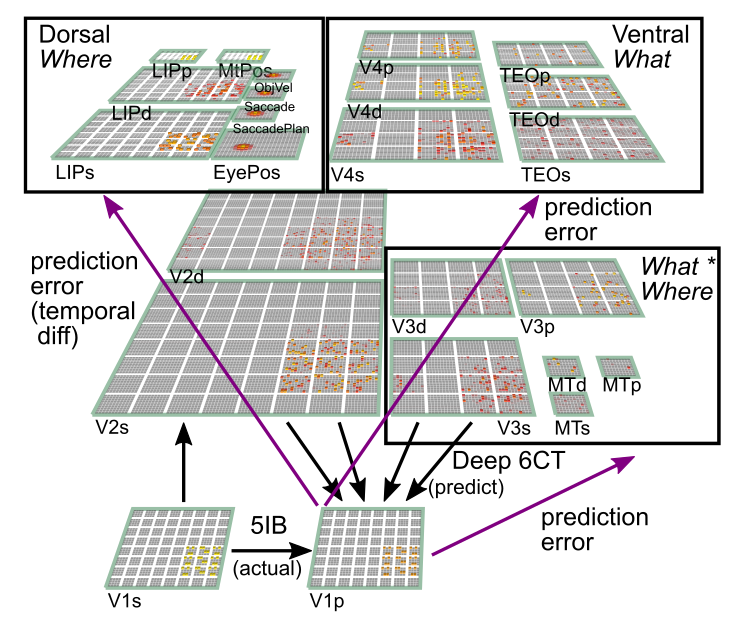}
  \caption{\footnotesize The three-visual-stream deep predictive learning model (What-Where-Integration or WWI model). The dorsal {\em Where} pathway learns first, using abstracted {\em spatial blob} representations, to predict where an object will move next, based on prior motion history, visual motion, and saccade efferent copy signals.  It then provides strong top-down inputs to lower areas to drive accurate spatial predictions, leaving the residual error to be more about {\em What} and {\em What * Where} integration information.  The V3 and MT areas constitute the {\em What * Where} integration pathway, sitting on top of V2 and learning to integrate visual features plus spatial information to accurately drive fully detailed predictions over the V1 pulvinar (V1p) ``projection screen'' layer (i.e., the cells distributed throughout the pulvinar that receive strong 5IB driver inputs).  V4 and TEO are the {\em What} pathway, and learn abstracted object feature representations, which uniquely generalize to novel objects, and, after some initial learning, drive strong top-down inputs to lower areas.  Most of the learning throughout the network is driven by a common predictive error signal encoded via a temporal difference over the pulvinar (V1p and other {\em p} layers), reflecting the difference between prediction (minus phase) and actual outcome (plus phase). {\em s} suffix = superficial  layer, {\em d} = deep layer.}
  \label{fig.wwi_model}
\end{figure*}

The model, which we refer to as the {\em What-Where-Integration} or {\em WWI} model, is shown in Figure~\ref{fig.wwi_model}, highlighting the three distinct visual streams ({\em Where, What,} and {\em What * Where}) all trained with a strong influence from a common predictive error signal represented as a temporal difference over the pulvinar.  The only external inputs to this model are the {\bf V1s} superficial layer activations, reflecting basic feature extraction (e.g., gabor oriented edge filtering) on retinal input signals, the saccade-related signals (anatomically in FEF) of current eye position ({\bf EyePos}), saccade motor plan and efferent copy of last saccade vector ({\bf SaccadePlan, Saccade}), and an object velocity representation reflecting output of known visual motion signals ({\bf ObjVel}) --- these last could be directly computed from the V1 inputs but it is simpler to provide as inputs.  There is no input of high-level category representations as are typically used in supervised backpropagation networks --- instead this model is entirely self-organizing and forms complex high-level representations without any explicit external shaping forces.  We also have a number of {\em decoders} (not shown in the figure) that receive inputs from various areas in the model, and attempt to decode things like object identity or position --- these provide one major means of understanding what these areas are representing (in a manner analogous to typical methods in neuroimaging of the brain).  Critically, these decoders do {\em not} feed back into the network and have absolutely no influence on learning in the model.

According to the known biology of the pulvinar, each of the different areas receives from its own subset of ventral pulvinar TRC neurons, but the wide distribution of V1 5IB driver inputs throughout the ventral pulvinar \cite{Shipp03} suggests that at least a portion of the pulvinar signal shares a common training plus-phase input across all the areas in the model.  This 5IB plus-phase input determines the resolution of the prediction that is learned --- biologically there may be only a few such 5IB neurons per microcolumn that present a kind of summary output for the entire microcolumn, and we just use a simple one-to-one mapping from our rate-coded microlumn-level superficial layer units.  Computationally, it was easier to represent this using a single {\bf V1p} layer that projects to all areas, and also receives deep-layer minus-phase prediction inputs from these same areas, such that predictions reflect the integrated best guesses from different areas and pathways in the model (i.e., a projection screen).  To measure network learning, we compute the cosine difference between the minus-phase prediction and plus-phase actual input over this V1p layer (cosine is computed as the normalized dot product between the two vectors, separately mean-normalized).  The full, trained model produces values around 0.9 or above on this measure, where 1.0 is perfect prediction.

The overall laminar structure and types of connectivity patterns in the model are based on our prior bidirectional object recognition model \cite{OReillyWyatteHerdEtAl13}, and follow general biological principles of higher areas being more compact and less retinotopically-distributed than lower layers, using convergent topographic projections to integrate over these lower layers. We did not use any non-biological weight sharing (convolution).  We extensively explored and optimized layer sizes and connectivity patterns for this model --- see Appendix for detailed parameters.

\subsection{The Dynamic Visual Environment}

\begin{figure}
  \centering\includegraphics[width=3in]{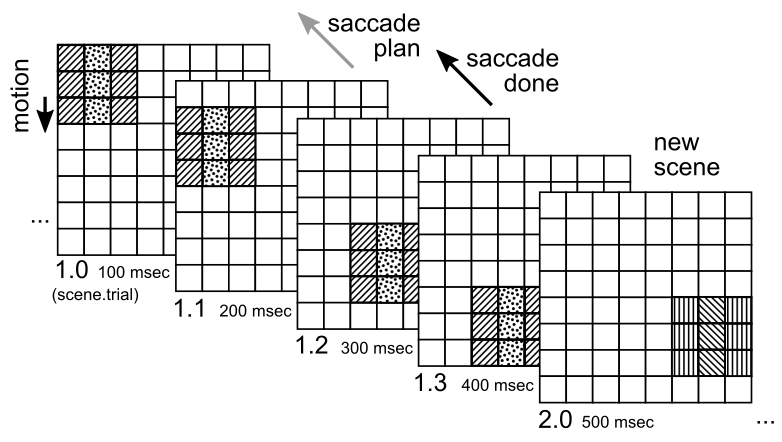}
  \caption{\footnotesize Dynamic visual environment, with 100 different objects composed of two independent sets of features (central column vs lateral flankers, 10 different patterns each), that have a constant motion vector (including the 0,0 no motion case) --- a 1 cell per trial downward motion is shown.  New scenes are rendered every 4 trials, and each trial represents one alpha cycle (100 msec, 10 Hz). A saccade is planned (i.e., a random vector generated) every 2nd trial, and executed between the 2nd and 3rd trial (note that trial index numbers start at 0).  The spatial {\em Where} pathway can accurately integrate object motion with saccade-generated displacements to predict where the object will appear on the 3rd trial.  The {\em What} pathway can maintain a representation of the object's visual features and apply them consistently across the scene in generating an expectation of what will be seen next.  Overall, the model can predict the next trial in this environment with high accuracy (except for the first trial, which is not predictable).}
  \label{fig.wwi_env}
\end{figure}

One critical requirement of a predictive learning model is an environment with sufficiently rich yet predictable dynamics over time to drive interesting learning --- one cannot use the kinds of randomly-ordered static images typically used with deep neural networks. The environment model that generates the V1 visual inputs (Figure~\ref{fig.wwi_env}) is designed to capture the most basic and essential features of our physical world: there are spatially contiguous objects with stable visual features over time, that can be moving relative to the observer in a stable manner over the period of roughly half a second.  Furthermore, the observer can move its eyes in a planned manner (saccades), which results in a discrete displacement of the visual input corresponding to the (opposite) vector of the saccade.  Saccades are the main reliable form of motor control that develops first, and including these in the model provides a template for how predictive learning can learn to anticipate the effects of motor actions more generally --- it is essential that the visual areas receive information about the {\em motor plan} (efference copy) in advance of the actual action, to be able to fully anticipate the effects \cite{vonHolst54,Wurtz08}.  This is a form of {\em forward model} \cite{KawatoFurukawaSuzuki87,JordanRumelhart92,MiallWolpert96}, as we elaborate in the General Discussion.

To keep things as simple and small as possible, we used an 8x8 grid of V1 hypercolumns (each hypercolumn having 4x4=16 feature bits), with an individual object subtending a 3x3 contiguous grid within that space, without going off the edge.  Thus, there are 6x6=36 different locations where the object can appear, and we randomly sampled the motion vector uniformly across the [-2,+2] range of integers (inclusive) separately along the horizontal (x) and vertical (y) dimensions, for a total of 25 different motion vectors.  The saccade vectors are drawn from the same distribution.  Both such vectors are constrained so as to keep the object fully visible.  There is an underlying ``world'' plane (16x16) where objects are allocentrically located, and eye positions reflect coordinates in this world plane --- objects are also constrained to lie entirely within this world plane.

Objects are constructed from two independent sets of features: one for the central vertical column, and the other for the two flanking columns.  These feature sets comprise 10 random bit patterns with 4 bits active and sharing at most 2 bits with any other such pattern, so there are 10x10=100 total objects under this scheme.  We trained the model with 90 of these objects, and reserved 10 for testing.  The combinatorial nature of these objects provides a good basis for generalization to the novel testing items.  In the real world, the generalization abilities of the human visual system, and large-scale deep neural networks, both support the existence of such a combinatorial (compositional) nature of objects' visual appearance, although the space is certainly much larger and less crisply defined --- typical deep neural networks train on 1,000 image categories with roughly 1,000 images per category, and are still likely significantly undersampling the relevant space.  Future work will explore scaling up our model to larger, real-object inputs, but the requirement of a dynamic physical simulation for predictive learning makes this much more challenging, as compared to using a large collection of static images.  We return to this issue in the discussion.

The temporal structure of the environment is organized into a sequence of {\em scenes}, with a new scene generated every 4 alpha-cycle {\em trials}, and a saccade takes place between the 2nd and 3rd trial, as well as between scenes (i.e., after the 4th trial and before the 1st trial of the next scene).  The object features remain consistent during a given scene, and change randomly for the next scene.  Thus, the first trial is unpredictable, and only on the second trial does the network have the ability to make an accurate prediction.  For this reason, the predictive learning framework in general requires at least 2 trials of processing for a novel visual input --- in combination with our hypothesis of alpha frequency predictive trials, we strongly predict that fixation durations should last at least 200 msec, which appears to be consistent with available data as reviewed below.  Another important reason for having 2 such trials is to allow for the planning of a new saccade on the 2nd trial, which is then executed prior to the start of the 3rd trial (i.e., the 3rd trial shows the post-saccade visual inputs).  The neural activity representing this planned saccade in the 2nd trial allows the model to accurately predict what the full visual input will be post-saccade.  We ignore the actual duration of the saccade, and assume that the system resynchronizes the alpha cycle post-saccade --- relevant data are discussed later.  There are 2 more trials to process the input post-saccade, and on the 2nd such trial (4th trial of the scene) the model makes a new saccade plan --- we assume that even though the object is new, its location is known and so an accurate saccade plan can be generated for the start of the next scene.

\subsection{Model Mechanisms}

The model uses standard {\em Leabra} equations \cite{OReillyHazyHerd15,OReillyMunakataFrankEtAl12,OReillyMunakata00}, detailed in the Appendix, for computing rate-coded activation states for each simulated neuron / unit, incorporating both excitatory long-range connections and local inhibitory currents that simulate the effects of inhibitory interneurons.  The rate-code activation function closely approximates the well-validated adaptive exponential spiking dynamics of neocortical pyramidal neurons \cite{BretteGerstner05}, and we assume that an individual simulated neuron in our model corresponds to a population of roughly 100 spiking neurons organized into microcolumns in the neocortex \cite{BuxhoevedenCasanova02,Mountcastle57,Mountcastle97,RaoWilliamsGoldman-Rakic99}.  Inhibition is computed as a simple linear proportion of both the {\em feedforward (FF)} excitatory net inputs to a given area, and the {\em feedback (FB)} overall activation level within a unit's layer --- this {\em FFFB} inhibition dynamic produces sparse distributed representations within each layer, which have long been shown to be computationally beneficial \cite{Kanerva88,Barlow89,Field94,OlshausenField97}.  Most of the layers have retinotopically-organized hypercolumn-level unit groups within a layer, and the same FFFB inhibitory dynamics operate simultaneously at both the layer and unit group level, with the overall inhibition for a unit being the MAX of each of these computations.  This ensures sparse distributed representations both within unit groups and across the entire layer.

\begin{figure}
  \begin{center}
  \begin{tabular}{ll}
    \parbox[b]{.1em}{a) \vspace*{1in}} &
    \includegraphics[width=3in]{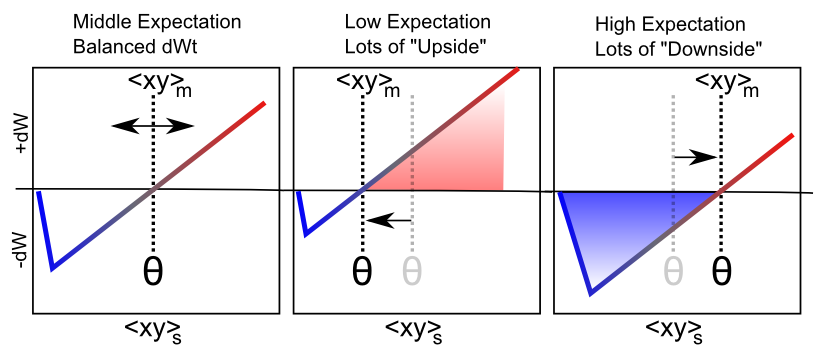} \\
    \parbox[b]{.1em}{b) \vspace*{1.4in}} &
    \includegraphics[width=3in]{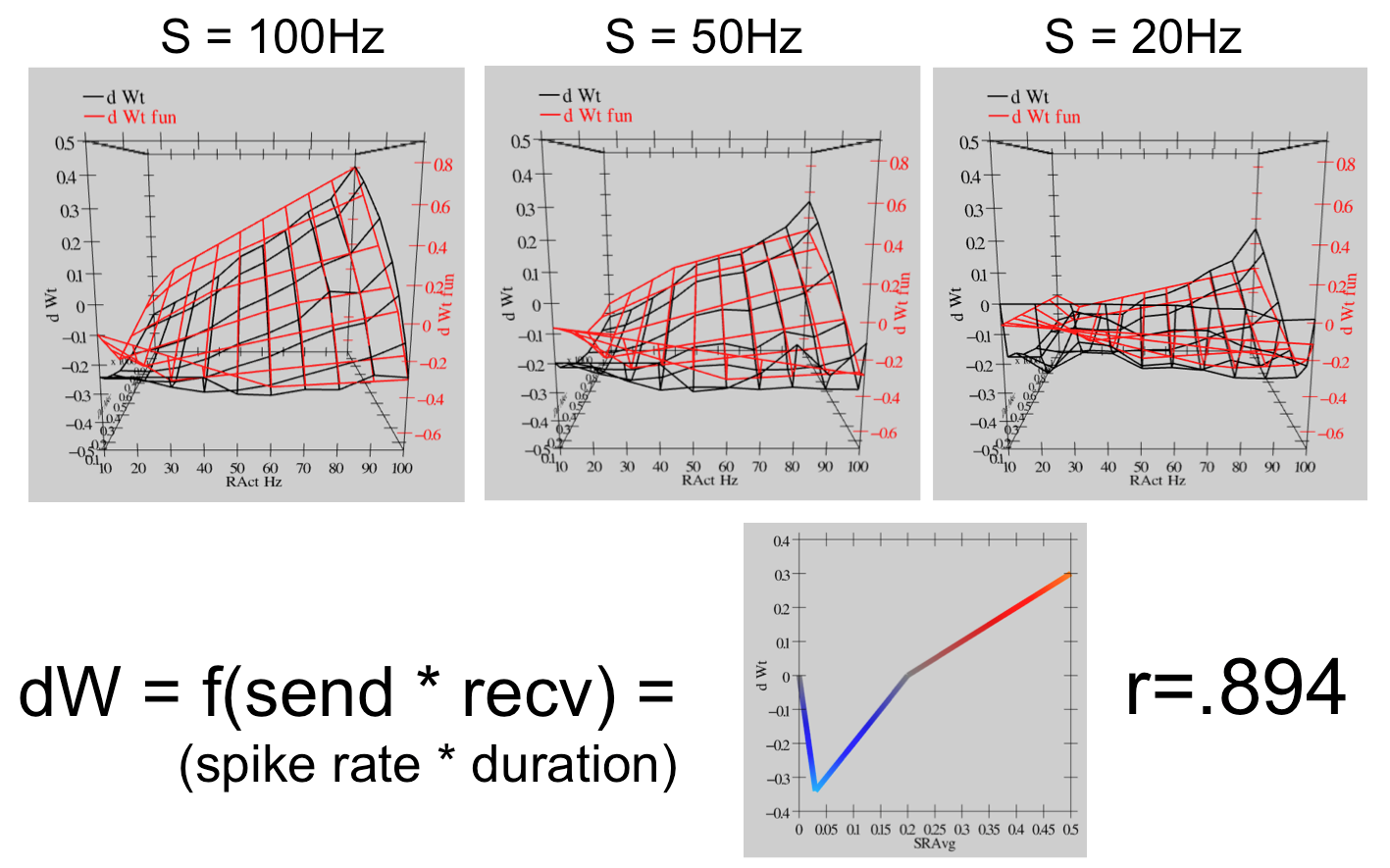}
  \end{tabular}
  \end{center}
 \caption{\footnotesize Error-driven synaptic plasticity in Leabra, using the {\em XCAL} function that is a linearized version of the BCM plasticity function, as derived from the Urakubo et al (2008) STDP model shown in panel (b).  a) The threshold $\theta$ between weight decrease (-dW, LTD) and weight increase (+dW, LTP) can adapt as a function of recent medium-time-scale average synaptic activity $<xy>_m$, which effectively captures the minus-phase expectation.  Learning is driven by the immediate short-term synaptic activity $<xy>_s$, reflecting the plus phase state, and the linear nature of the XCAL function results in an approximation to the CHL equation ($x^+ y^+ - x^- y^-$).  A more slowly-adapting threshold produces the BCM Hebbian learning dynamics (featuring a homeostatic negative-feedback mechanism that helps reduce hog units), and a mix of both such learning terms are used.  b) The fit to the Urakubo et al (2008) STDP model: a range of sending and receiving spiking frequencies were sampled, and net weight change from the model recorded (black lines).  A simple linear equation (the XCAL function) (red lines) fits the overall results well (although the best-fitting function has a small kink around the threshold, a straight line fits nearly as well, and computationally this kink does not affect learning if included).}
 \label{fig.xcal_learn}
\end{figure}

Synaptic plasticity in Leabra reflects a synthesis between computational and biological mechanisms.  Computationally, it performs both error-driven and Hebbian learning, and we'll see that both of these learning factors are essential for successful learning.  The error-driven learning arises from a temporal difference between plus (outcome) and minus (prediction) phases as noted above, approximately of the form of the Contrastive-Hebbian-Learning (CHL; \nopcite{Movellan90}) equations:
\begin{equation}
  \Delta w \approx \epsilon \left( x^+ y^+ - x^- y^- \right)
  \label{eq.chl}
\end{equation}
Where $+$ superscripts indicate plus phase, $-$ minus, and $x$ is the activation of the sending unit, while $y$ is that of the receiving unit.  This difference of sender-receiver products computes approximately the same gradient as error backpropagation, subject to symmetry constraints and a few other details \cite{OReilly96,XieSeung03,ScellierBengio17}.  Critically, each factor in this CHL equation is of a simple $x y$ Hebbian form, making the connection to biological mechanisms more straightforward.  We were able to enhance this biological connection significantly by deriving a CHL-like equation directly from a highly detailed biophysical model of spike-timing-dependent-plasticity (STDP;  \nopcite{UrakuboHondaFroemkeEtAl08};  Figure~\ref{fig.xcal_learn}).  Specifically, we found that the rate-code average behavior of this biophysical model, which accounts for a wide range of complex STDP data, can be accurately summarized with a simple linear function that resembles the BCM learning function \cite{BienenstockCooperMunro82,CooperIntratorBlaisEtAl04,ShouvalWangWittenberg10}.  This function (which we call {\em XCAL}: temporally eXtended Contrastive Attractor Learning) captures the well-established finding that low (but still elevated) levels of postsynaptic calcium (reflecting the Hebbian $x y$ product) drive a decrease in synaptic weights, while higher levels drive weight increases \cite{ArtolaBrocherSinger90,Lisman90,Lisman95,BearMalenka94}.

The essential feature of the BCM model is that the threshold crossover point between these two regimes can adapt over time, and by so doing, produce a homeostatic negative feedback mechanism that shifts the balance of weight increases and decreases as a function of how active a unit has been.  We realized that if such a threshold were to adapt on a rather more rapid timescale, it could reflect the minus-phase activation state as shown in the CHL equation above, and the linear nature of the learning function then produces the necessary subtraction of this dynamic threshold, with the basic Hebbian-style learning signal reflecting the calcium signal that drives plasticity (Figure~\ref{fig.xcal_learn}).  Interestingly, some recent data are consistent with more rapidly adapting thresholds \cite{LimMcKeeWoloszynEtAl15,JedlickaBenuskovaAbraham15,ZenkeGerstnerGanguli17}.  Furthermore, our model employs two timescales of threshold adaptation --- the shorter one reflecting the minus-phase expectation and a longer one reflecting overall activation levels over time --- thus achieving an elegant synthesis of  error-driven and BCM-like Hebbian learning.

In the current model, and most of our other large-scale deep visual models \cite{OReillyWyatteHerdEtAl13}, the BCM-like Hebbian learning plays a critical role in combating the {\em hog unit problem}, where a small subset of units takes over much of the representational space and are essentially always active.  This problem arises because of the presence of strong positive feedback loops in bidirectionally-connected networks, where units across bidirectionally connected areas can build up mutually reinforcing weights, causing these hogs to form and stabilize themselves.  Although error-driven learning should theoretically end up punishing these hog units if they are not contributing to solving the overall problem, it is often the case with challenging problems in deep networks that the error gradients are not very strong or clear at the start of learning, resulting in a kind of ``thrashing'' dynamic that is ineffective at combating these hog units (and indeed results in a reduction in overall variance in weight values, thereby reducing the random variability that drives exploration of different regions of the solution space).  In this context, the BCM Hebbian learning, by raising the learning threshold in proportion to overall unit activation levels, helps to push down the hog units.  In addition, we have found that using a normalized momentum learning factor (widely used in backpropagation networks) is helpful for reducing thrashing by driving synaptic weights more quickly along useful gradients, thereby combating hogging as well.

The above mechanisms are used for all neurons in the model, and sufficiently characterize the superficial layers (labeled with an {\bf s} suffix in Figure~\ref{fig.wwi_model}).  However, the deep layer and pulvinar neurons have a few special mechanisms to capture their unique functionality.  The deep layers in DeepLeabra (with a {\bf d} suffix) capture the firing of the final output stage of the deep neocortical layers, the layer 6CT corticothalamic neurons that project to the pulvinar (and top-down to other neocortical areas) \cite{Thomson10,ThomsonLamy07}.  As summarized above, these deep neurons receive a persistent excitatory input representing the SRN-like context information integrated over the superficial layer neurons from the prior alpha trial, and this input is updated as a result of simulated layer 5IB burst firing at the end of every trial.  Critically, this prior context state information is the {\em only} input these deep units receive about the sensory state as represented in the bottom-up feedforward pathways in the network --- this restriction is what forces the network to predict, as opposed to simply copy the current sensory input (which is impinging on the superficial layers during the current alpha trial).   The V4d and TEOd deep layers also receive a self-context projection, which integrates across the prior deep layer activations in addition to the superficial layers.  This supports more enduring activation states over time.  We tested this ``deeper'' context on all layers, but only found benefits for these higher {\em What} pathway layers, which is consistent with the idea that these areas have more sustained representations to support the development of more invariant representations \cite{Foldiak91,OReillyJohnson94,WiskottSejnowski02}.

The pulvinar neurons (with a {\bf p} suffix in Figure~\ref{fig.wwi_model}) are specialized to capture the strong driver effects of the 5IB driving inputs --- in the plus phase when these neurons fire, their input drowns out the signal from the layer 6CT prediction-generating inputs, and is used as the exclusive source of synaptic input for the pulvinar neurons.  Computationally, this is important because simply adding the drivers plus the existing 6CT inputs results in a constantly increasing error signal that drives synaptic weights ever upward (we refer to this as a {\em main effect} problem).  The driving input is computed directly from one-to-one connections from corresponding superficial layer neurons, which are subject to a thresholding process that we assume to be one of the major computational contributions of the 5IB stage.

\subsection{Connectivity Patterns}

Overall, the patterns of interconnectivity among the areas in our model largely follow known biological patterns \abbrevcite{RocklandPandya79,FellemanVanEssen91,MarkovVezoliChameauEtAl14,MarkovErcsey-RavaszGomesEtAl14,Thomson10,ThomsonLamy07,SchubertKotterStaiger07,ShermanGuillery06,DouglasMartin04}, but we also explored many other possibilities, to determine what works best computationally.  The resulting model only includes connections with a demonstrated computational value --- if adding a given connection made little overall difference, or made performance worse, it was left out of the default model.  Reassuringly, the computational benefits largely aligned with the known biology.  Below, we present results from manipulating a few particularly important connections, which provide key insights into how the model learns.

\begin{figure}
  \begin{center}
    \begin{tabular}{ll}
      \parbox[b]{.1em}{a) \vspace*{1.3in}} &
      \includegraphics[height=1.5in]{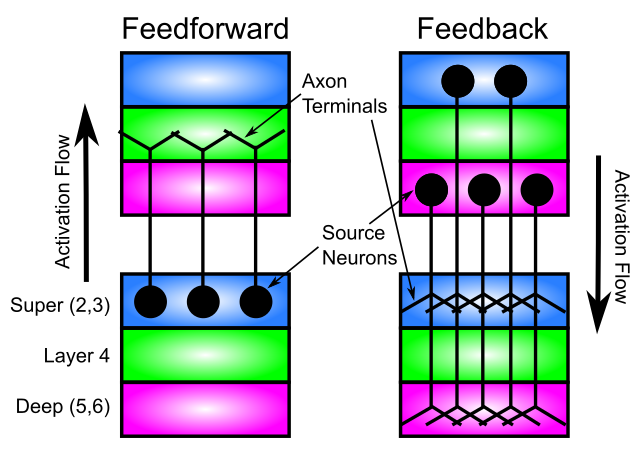} \\
      \parbox[b]{.1em}{b) \vspace*{.8in}} &
      \includegraphics[height=1in]{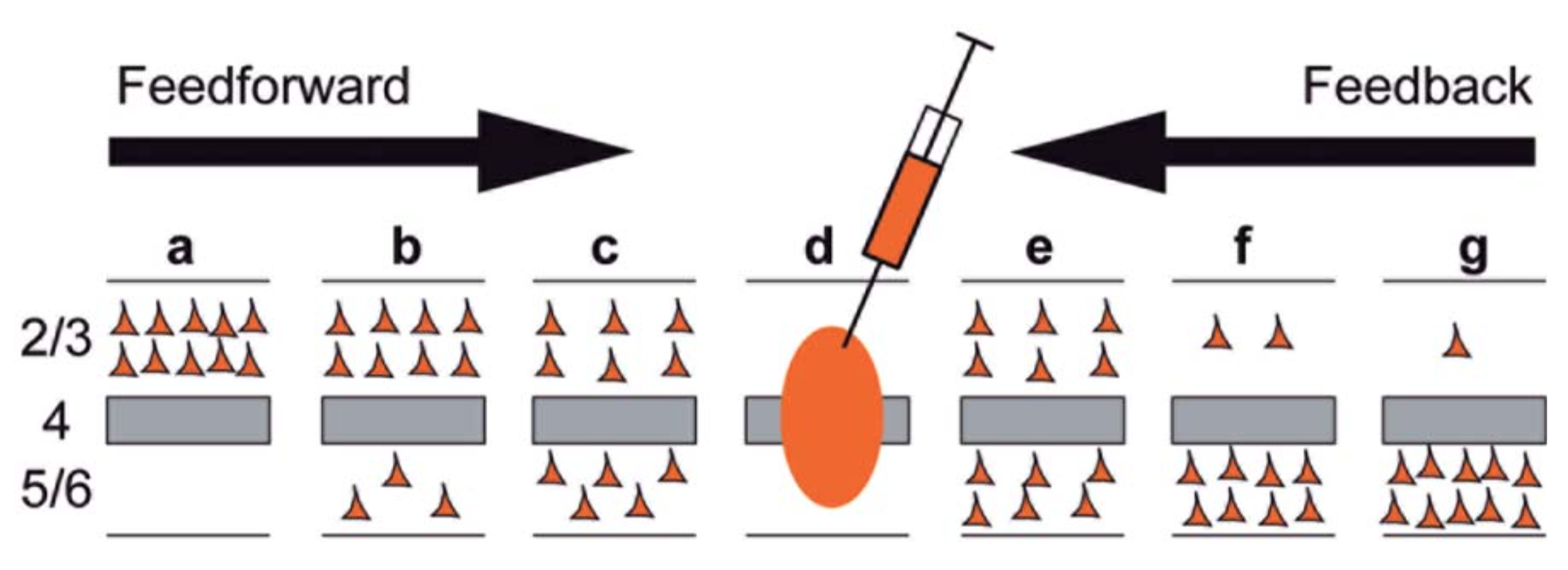} 
    \end{tabular}
  \end{center}
  \caption{\footnotesize Standard patterns of feedforward and feedback connectivity in neocortex.  a) Most feedforward connections originate in superficial layers of lower area, and terminate in layer 4 of higher area.  Feedback connections can originate in either superficial or deep layers, and in both cases terminate in both superficial and deep layers of the lower area. (adapted from Fellemen \& Van Essen, 1991). b) A more quantitative representation from Markov et al (2014), showing density of {\em retrograde} labeling from a given injection in a middle-level area (d) --- again, most feedforward projections originate from superficial layers of lower areas (a,b,c) and deep layers predominantly contribute to feedback (and more strongly for longer-range feedback).  However, there appears to be some feedforward contribution from deep-layers, which we did not find to be useful in our model.  Overall, these patterns are critical for the functioning of the predictive learning model as explained in the text.}
  \label{fig.ff_fb}
\end{figure}

Starting at the most general level, Figure~\ref{fig.ff_fb} (adapted from \nopcite{FellemanVanEssen91}) shows that feedforward connections originate in the superficial layers (2/3) in the lower area, and terminate in layer 4 of the higher area (i.e., the input layer of neocortex, where thalamic inputs from sensory areas terminate in primary sensory areas).  From layer 4, connections go straight up to the superficial layers, and in our model we combine the functionality of all of these layers (4,2,3) in the single superficial layer for a given area.  Completing the bidirectional loop of excitatory connections within the superficial layers, one type of feedback connectivity originates in the superficial layers of a higher area, and projects back to the superficial layers of a lower area.  This pattern of connectivity produces {\em bidirectional constraint satisfaction} dynamics, iteratively settling into {\em attractor states} that best represent the constraints present in the external inputs and internal learned synaptic weights \cite{Hopfield82,Hopfield84,AckleyHintonSejnowski85,RumelhartMcClelland82}.  Note that although \abbrevincite{MarkovVezoliChameauEtAl14} present evidence that the feedforward and feedback pathways in the superficial layers may be supported by separate populations of neurons (in layer 2 vs. 3B), both of these populations receive the same feedforward (via layer 4) and feedback (via layer 1 dendritic tufts) projections, so this may just be more of a wiring difference without strong functional implications --- we will explore these issues in later versions of our model.

As noted earlier, it is essential in the DeepLeabra model that the feedforward connections do {\em not} project directly to the deep layers (5,6), because that would give the predictive learning model direct access to the current sensory inputs, which is what it is trying to predict in the first place.  This would be analogous to a short-circuit in electrical terms.  Furthermore, as we demonstrate below, it is very important that the feedback connections from superficial layers {\em do} drive the deep layers directly --- we found that the deep layers benefit considerably from top-down connections from higher areas, both from other deep layers and from higher-order superficial layers.  Computationally, there is the possibility that superficial information from these top-down super-to-deep projections, reflecting current inputs, could short-circuit the predictive learning process.  However, because this information is coming only from areas higher in the network, it is already contingent on the quality of the lower-level area in question, and thus is not capable of short-circuiting the learning process.  More generally, it seems that the deep layers in our model only benefited from top-down projections, not bottom-up ones (which could only be from other deep layers, due to the short-circuit problem).  The fact that the deep layers only seem to receive direct feedback is a basic feature of the neocortical connectivity that also makes sense in terms of generative predictive models, where the best source of predictive information comes top-down from compact, high-level representations (as discussed earlier).

\begin{figure}
  \begin{center}
    \begin{tabular}{ll}
      \parbox[b]{.1em}{a) \vspace*{1.6in}} &
      \includegraphics[width=3in]{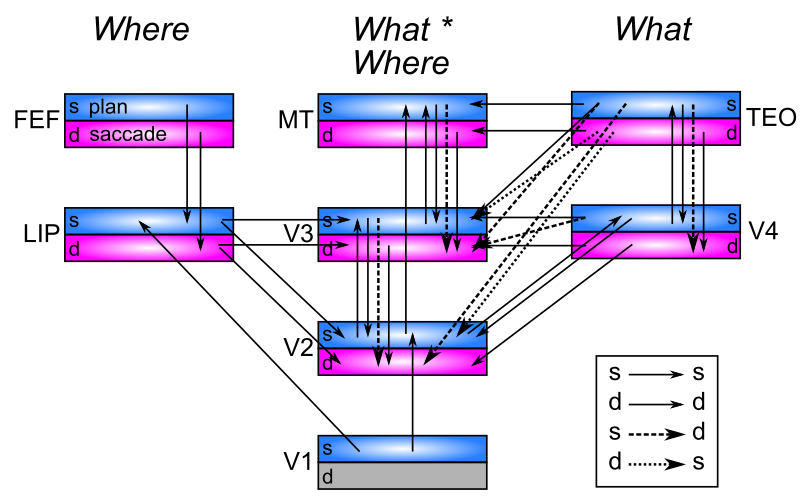} \\
      \parbox[b]{.1em}{b) \vspace*{1.8in}} &
      \includegraphics[height=2in]{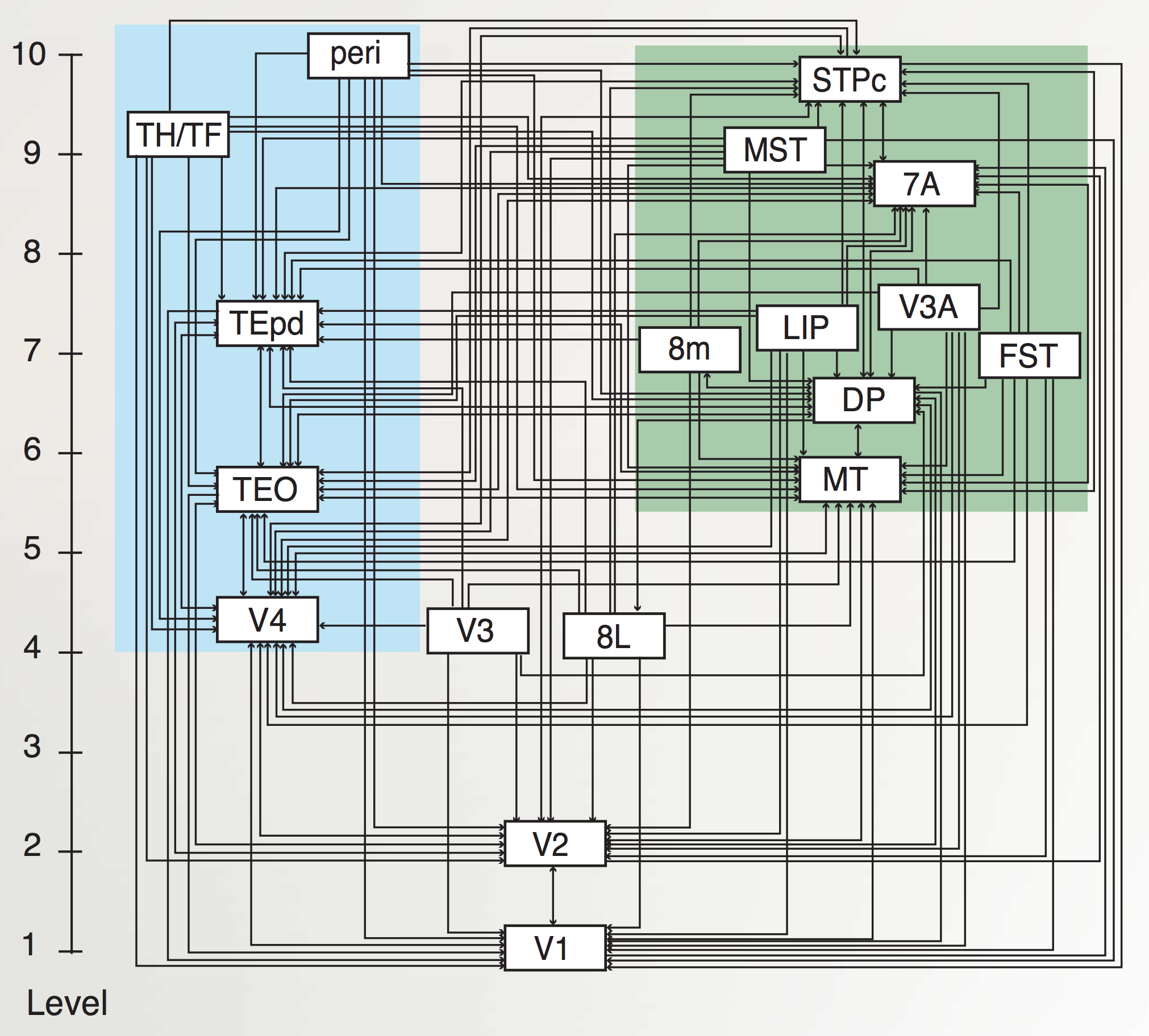}
    \end{tabular}
  \end{center}
  \caption{\footnotesize a) Superficial and deep-layer connectivity in the model.  Note the repeating motif between hierarchically-adjacent areas, with bidirectional connectivity between superficial layers, and feedback into deep layers from both higher-level superficial and deep layers, according to canonical pattern shown in previous figure.  Special patterns of connectivity from TEO to V3 and V2, involving crossed super-to-deep and deep-to-super pathways, provide top-down support for predictions based on high-level object representations (particularly important for novel test items).  b) Anatomical hierarchy as determined by percentage of superficial layer source labeling (SLN) by Markov et al (2014) --- the hierarchical levels are well matched for our model, but we functionally divide the dorsal pathway (shown in green background) into the two separable components of a {\em Where} and a {\em What * Where} integration pathway.  It is likely that area DP is also part of this integration pathway.  8L = FEF for small-displacement saccades, while 8m = FEF for large-displacement saccades.}
  \label{fig.model_cons}
\end{figure}

Figure~\ref{fig.model_cons} shows the full pattern of superficial and deep layer connections among all the areas in our model, in comparison to the cortical hierarchy of the macaque from \abbrevincite{MarkovVezoliChameauEtAl14}.  For the hierarchically adjacent levels outside of the {\em Where} pathway, the characteristic pattern shown in Figure~\ref{fig.ff_fb} is present: standard bidirectional excitatory connectivity among superficial neurons, together with top-down projections from both superficial and deep into the deep layers (note that V1 is strictly an input layer in this model, so all top-down and deep-layer connectivity was omitted).  The most interesting connections concern the way that the {\em What} pathway influences the {\em What * Where} pathway, which involved the only instances of deep-to-superficial connections (from TEOd to V3s \& V2s), in addition to the opposite crossing of superficial-to-deep (from TEOs to V3d \& V2d).  These connections are essential for allowing more abstract, high-level TEO representations to positively influence the low-level predictions generated over V1p -- especially for the novel untrained items.

Next, we consider the interconnectivity with the pulvinar.  Biologically, the pulvinar has long remained a bit of a mystery, in part because its obvious anatomical divisions do not appear to coincide with its functional organization --- there are coherent retinotopic maps that spread across multiple anatomical divisions, at odd angles, which makes analysis difficult.  \incite{Shipp03} provides an impressive synthesis of the literature, building on the pioneering work of \incite{Bender81}, and clarifies various points of confusion, such that we were able to build our model on the foundation of this synthesis.  The major conclusions are that there are four major retinotopically-organized maps in the pulvinar, three corresponding to the ventral cortical pathway, and one for dorsal, and that these maps also have a coarse hierarchical topography, but also considerable levels of intermixing across hierarchical levels.

The first two major ventral pulvinar maps (VP1, VP2) were first characterized by \incite{Bender81} as being {\em first-order} and {\em second-order}, while \incite{Shipp03} also refers to them as $1^\circ$ and $2^\circ$ (confusingly suggesting a difference in visual angle size of receptive field, which is {\em not} the case).  As \incite{Bender81} emphasizes, these two maps have highly similar properties overall (electrophysiology and patterns of connectivity with cortex), and one primary difference lies in the nature of their topographic organization in the brain, mirroring that of V1 and V2 respectively (where V2/VP2 are wrapped around the central core of V1/VP1).  Another major difference is that VP1 (located in inferior pulvinar) receives direct projections from the superior colliculus, while VP2 (in lateral pulvinar) does not.  We are excited to explore possible contributions of collicular inputs in future models --- they may serve as another source of plus-phase training signals, and could have important implications for spatial attention maps, saccade signals, and also subcortical object / pattern recognition signals (e.g., low-level face detector cells; \nopcite{MortonJohnson91}). For the present model, we use a single common VP substrate.  The third ventral pulvinar map, VP3, appears to be dedicated to MT (V5) --- we will see below that this may be a separate map because it has a unique developmental trajectory, consistent with the early development of a spatial {\em Where} system in our model \cite{BridgeLeopoldBourne16}.  The single dorsal pulvinar map (DP) interconnects with higher-level dorsal pathway areas, including LIP as represented in our model.  \incite{Shipp03} argues that overall the VP3 map can really be considered a part of the DP map --- this straddling of ventral and dorsal pathways fits well overall with it playing a key {\em What * Where} integration role in our model.

All of these pulvinar maps have a third dimension of organization beyond their 2D retinotopic maps, the {\em axis of iso-representation} (AIR), which {\em roughly} reflects the corresponding cortical hierarchy (although it is inverted relative to cortex in the caudal-rostral dimension).  The lower visual areas, V1, V2, and V3, project extensively across the AIR dimension (with the densest projections in the most rostral region), and V1 in particular has multiple branches of driving (R, type-2) projections along this dimension \cite{Rockland98a,Rockland96,ShermanGuillery06}.  By contrast, higher areas send these driving projections more caudally along the AIR dimension (corresponding to higher-level areas), while also sending weaker (E, type-1) projections to the more rostral, lower areas.  Overall, the connectivity from pulvinar to cortex tends to be reciprocal (symmetric) to the connectivity from cortex to pulvinar.

Our overall conclusion from this biological data is that the pulvinar serves as a kind of {\em shared projection screen} (similar to the {\em blackboard} proposal of \nopcite{Mumford91}) where multiple different cortical areas can provide convergent input to shape an overall integrated representation.  The projections from pulvinar to cortex then share this converged information broadly back to the same areas that provided input in creating it.  As \incite{Mumford91} emphasized, there is a fundamental puzzle about the pulvinar: it lacks any interconnections among its principal TRC neurons, and therefore does not appear to be capable of doing any processing.  This fact is precisely what makes it so attractive as a substrate for projecting representations on to.  Furthermore, the massive projection from pulvinar to cortex, targeting the layer 4 {\em input} neurons, suggests that the pulvinar is somehow involved in representing the sensory input to the brain.  In addition to this projection-screen-like aspect, there is also a rough hierarchical gradient, so the higher-level cortical areas participate more strongly with shaping the more caudal, higher-level representations in the pulvinar, but there is still plenty of mixing here with lower-level cortical areas providing input into these caudal pulvinar areas, and higher-level cortical areas also providing plenty of input into the rostral, lower-level pulvinar areas.

Our model then goes beyond these basic characterizations to further specify that the convergent, integrated representations in the pulvinar are actually {\em predictions} about what state the strong driving inputs will generate at the next interval of alpha-cycle 5IB burst firing.  And the projections from pulvinar back to cortex then carry the critical error signal, in the form of a temporal difference between the prediction and driven states, to train the cortex to produce better such predictions over time.  This account helps to make sense of the otherwise somewhat puzzling roles of the two types inputs to the pulvinar \cite{ShermanGuillery06}, and why the strong driver inputs appear to obey the hierarchical topographic organization somewhat more strongly than the other inputs \cite{Rockland98a,Rockland96}: this establishes a spectrum of increasingly abstract {\em ground truth} driver inputs to be predicted.  Thus, the ``cartoon'' of a single projection screen in the pulvinar is inaccurate (but a useful first approximation) --- it is really a number of different screens at various levels of abstraction.

\begin{figure}
  \centering\includegraphics[width=3in]{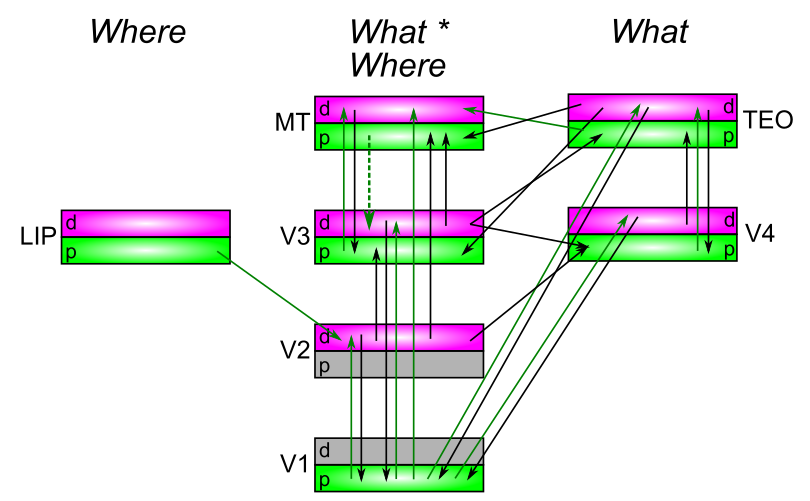}
  \caption{\footnotesize Connectivity for deep layers and pulvinar in the model, which generally mirror the corticocortical pathways (previous figure).  Each pulvinar layer (p) receives 5IB driving inputs from the labeled layer (e.g., V1p receives 5IB drivers from V1).  In reality these neurons are more distributed throughout the pulvinar, but it is computationally convenient to organize them together as shown.  Deep layers (d) provide predictive input into pulvinar, and pulvinar projections send error signals (via temporal differences between predictions and actual state) to {\em both} deep and superficial layers of given areas (only d shown).  Most areas send deep-layer prediction inputs into the main V1p prediction layer, and receive reciprocal error signals therefrom.  The strongest constraint we found was that pulvinar outputs (colored green) must generally project only to higher areas, not to lower areas, with the exceptions of MTp $\rightarrow$ V3 and LIPp $\rightarrow$ V2.  V2p was omitted because it is largely redundant with V1p in this simple model.}
  \label{fig.model_cons_pulv}
\end{figure}

Figure~\ref{fig.model_cons_pulv} shows the connectivity of deep layers and pulvinar areas in our model.  The overall patterns of connectivity generally mirror those of the corticocortical pathways (Figure~\ref{fig.model_cons}) --- obeying the general {\em replication principle} of \incite{Shipp03}.  Note that the V1d deep layers (6CT) generally project down to the LGN, not the pulvinar, so the next-higher layer, V2d, provides the primary detailed, retinotopically-organized predictive input to the V1p (interestingly, the pulvinar receptive field sizes match those of V2; \nopcite{Bender81}).  Thus, the extensive top-down corticocortical pathways target V2d, to drive V1p predictions (and we omit V1d from our model).  One could label V1p as V2p to align those functions, but there are also distinct pulvinar neurons (anatomically intermixed with V1p neurons) that receive V2 5IB driver inputs, and have similar inputs and outputs as V1p, so we reserve the term V2p for that population of neurons.  However, we did not implement V2p in the current model because it was largely redundant with V1p --- in the future we plan to add binocular vision and real-world 3D objects, at which point the V2p layer should contain important distinct shape information beyond that in V1p.

The higher-level areas also have their own associated pulvinar layers, which again anatomically are intermixed with V1p, but there is a gradient of the distribution that overall mirrors the caudal-rostral hierarchy of visual areas \cite{Shipp03}.  These pulvinar layers receive a variety of deep-layer inputs, mostly from neighboring areas, to predict their plus-phase firing patterns.  Interestingly, we found a strong constraint on the outputs of these pulvinar areas: they were only beneficial when they projected to higher-level areas.  This makes computational sense in terms of the overall generative, auto-encoder framework, where the higher-level areas are learning to be able to reconstruct lower-level representations.  It does not make sense that lower-level areas would have the representational abstractions necessary to accurately drive higher-level representations.  Nevertheless, the deep-layer inputs from these lower-level areas can still provide useful information for helping drive the prediction, even though it is not by itself sufficient.  This overall constraint is potentially consistent with the patterns of pulvino-cortical connectivity reviewed in \incite{Shipp03}, which appears to be more strongly hierarchically organized compared to the cortico-pulvinar direction.  However, more detailed examination of connectivity patterns relative to the strong intermixing of information across the entire pulvinar axis would be necessary to clearly evaluate the validity of this constraint in the biology.

Overall, we argue that the close fit between the characteristic patterns of neocortical /  pulvinar connectivity, and the specific, detailed demands of our WWI predictive learning model provides support for the notion that these patterns have evolved to support this functionality.

\subsection{Early Development of Predictive Spatial Maps in the \emph{Where} Pathway}

A central principle of our overall framework is that high-level abstract representations are important for driving lower-level predictions via strong top-down connections.  In the case of the dorsal {\em Where} pathway, it is relatively straightforward to create the relevant spatial abstractions directly from the V1 inputs, and drive predictive learning of object and self-motion (including saccades) on these abstracted spatial {\em blob} representations at the high levels of the dorsal pathway.  The higher levels (e.g., LIP) are compact enough to be capable of remapping saccades over the full span of visual space, whereas in lower levels the degree of interconnectivity across areas would be impossible given the size of the areas.  This is consistent with the framework of \incite{CavanaghHuntAfrazEtAl10} (building on \nopcite{Wurtz08}), who argue that predictive remapping across saccades is performed at the high levels of the dorsal stream, and it then drives top-down activation in lower areas.  Later, we apply our model to account for specific data in the predictive remapping literature.

The two essential features that must be extracted from V1 inputs to make this work are just the retinotopic location irrespective of features (i.e., the spatial blob), and the visual motion vector.  Based on a wide range of data discussed next, we hypothesize that area MT (V5) extracts both of these features.  The LIP area in our model then integrates these MT inputs together with the saccade plan and actual saccade vector representations (from area FEF and/or superior colliculus) to generate a prediction of where the spatial blob will appear on the next alpha trial, projected onto the LIPp pulvinar.  The LIPp is then driven in the plus phase by 5IB bursting output of area MT, providing the ground truth for where the object actually did move.

Due to the relative simplicity of this spatial prediction task, we hypothesized that the brain should learn it {\em first}, before anything else of significance is attempted, to absorb as much of the predictive error associated with the spatial aspect, and thereby drive other areas to take on the remaining {\em What * Where} and {\em What} components.  Biologically, this appears to be a well-supported hypothesis.  \incite{BridgeLeopoldBourne16} review a range of data showing that area MT and its associated VP3 pulvinar area do indeed develop very early, in part through a unique pathway of strong connections from the retina to VP3 (medial inferior pulvinar) that is present early in life, and then is significantly reduced a few months later in development. There is also evidence of direct LGN to MT projections \cite{SincichParkWohlgemuthEtAl04}.  Neurally, area MT matures earlier than other visual areas, at the same time as V1 \cite{BourneRosa06}, and behaviorally motion sensitivity develops before form sensitivity in macaques \cite{KiorpesPriceHall-HaroEtAl12}.  \incite{BridgeLeopoldBourne16} also argue that this early development of MT then drives early learning in other dorsal-stream pathways, and that after this early developmental phase, MT shifts over to being driven more strongly by direct V1 inputs and other cortical inputs, as the unique retino-pulvinar pathway retreats.

\begin{figure}
  \centering\includegraphics[width=2in]{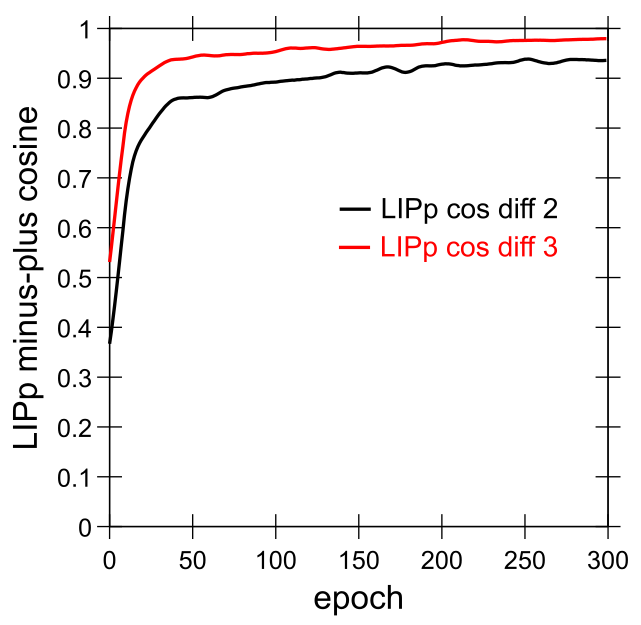}
  \caption{\footnotesize Learning curves for LIP spatial prediction accuracy, measured as cosine between minus and plus phase representations over the LIPp pulvinar layer (perfect accuracy is 1.0).  Trial 2 (LIPp cos diff 2, which is the 3rd trial of the sequence) is right after the saccade and thus requires integrating saccade motion plus intrinsic object motion.  This curve achieves high levels of predictive accuracy, demonstrating that our model is indeed successfully doing predictive remapping, at least within this {\em What} pathway.  Trial 3 (LIPp cos diff 3; 4th trial) only requires tracking intrinsic object motion, and is thus easier than the full saccadic remaping task.  One epoch = 512 alpha cycles = 51.2 seconds of real time, so this total training period represents approximately 5 hours of real time learning. }
  \label{fig.lip_pretrain}
\end{figure}

In our model, we simplify this overall developmental dynamic in several ways.  First, we turn off the entire rest of the model for the initial training of the {\em Where} pathway.  Second, we use a separate {\bf MTPos} layer as a proxy for the direct retino-pulvinar pathway, which just collapses all the feature distinctions within a given 8x8 spatial location from the V1 input, producing an entirely spatial input to the LIP.  We also use an {\bf ObjVel} input that encodes the visual velocity vector based on object motion, which we assume this early MT layer also provides.  Instead of phasing these early drivers out and shifting over to a more cortically integrated MT later, we just add a new MT layer as shown in the {\em What * Where} pathway of our model (Figure~\ref{fig.wwi_model}).  A later model could explore a more realistic developmental transition of a common MT area, potentially revealing interesting benefits from the early developmental phase.

We initialized the connectivity of LIP with random weights shaped by topographic sigmoidal and gaussian basis function representations, as has long been recognized as a theoretically-important feature of parietal processing \cite{ZipserAndersen88,PougetSejnowski97}.  This improved the learning time compared to purely random weights (see the Appendix for details).  The learning curves for this {\em Where} pathway are shown in Figure~\ref{fig.lip_pretrain}, for both the post-saccade trial and the trial thereafter.  This graph demonstrates that the model is indeed capable of successful predictive remapping using a representation of the saccade plan, integrated with the current object location.  Interestingly, as explored later, our model predicts that this predictive remapping happens first in the superficial layers of LIP, and then later and more fully in the deep layers --- and these deep layers actually benefit from receiving the actual saccade command, instead of the planning inputs which drive initial updating of the superficial layers.  The total training time is approximately 5 hours simulated real-time, with 512 100 msec alpha cycles per epoch, and 300 epochs, which is clearly well within realistic limits.  The more complex, higher-resolution learning in the human brain would likely take significantly longer.

Again, we argue that the particular computational demands of our generative predictive learning model align well with the unique developmental trajectory of area MT and associated pulvinar, providing further support for the overall framework.

\subsection{Later Development of TEO Top-Down Pathway}

Another developmental aspect of our model concerns the TEO top-down projections into V3 and V2 --- we found small but significant benefits in overall predictive accuracy and ability to decode object information from TEO from delaying the point at which these projections actually influence these lower areas.  Computationally, this makes sense because it allows the more fully developed TEO object representations to drive these lower areas, instead of the rapidly changing and initially quite noisy representations from the start.  Overall, this reflects an attempt to find a good compromise for the difficult co-dependency problem in the {\em What} pathway, where high-level abstract representations take a while to develop, and yet are needed for improved prediction performance at the lower levels, which in turn drives better learning of these lower level representations, upon which the TEO representations themselves depend.

Biologically, we were unable to find directly relevant data specifically about the development of top-down projections from TEO, but more general data suggest that IT overall develops relatively slowly compared to other visual areas \cite{Rodman94} and that the visual functions associated with IT emerge relatively later in development and continue to develop over a relatively long timecourse \cite{NishimuraScherfBehrmann09}.  Thus, this particular feature of our model is overall plausible but not directly supported, and it is quite likely that various other developmental manipulations could have similar benefits, so this remains an area for future exploration.

\section{Results: Understanding how the Model Learns}

The first set of results are focused on various tests, manipulations, and analyses that show how the model learns, and how the different pathways and mechanisms interact to produce its overall high levels of predictive learning and development of abstract object representations in the {\em What} pathway, which are documented first.  The subsequent results section then explores how the model accounts for some detailed empirical data of particular relevance.

\begin{figure}
  \begin{center}
    \begin{tabular}{ll}
      \parbox[b]{.1em}{a) \vspace*{2in}} &
      \includegraphics[width=2.7in]{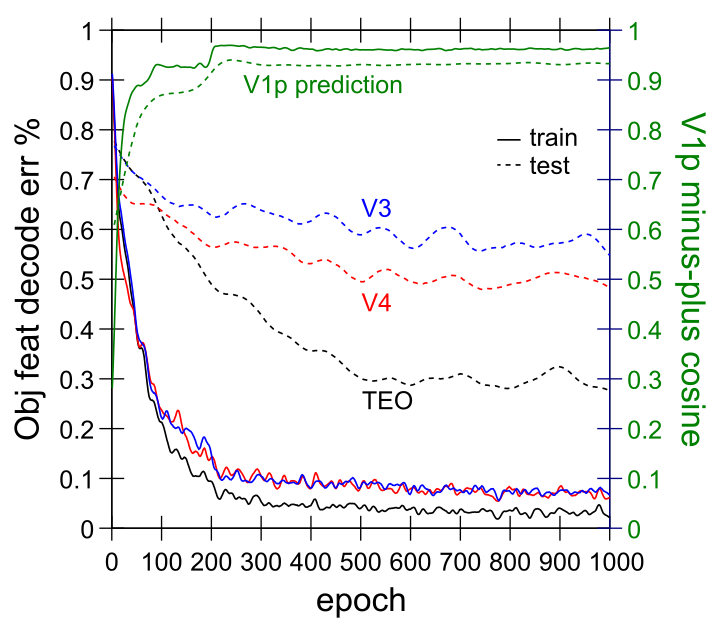} \\
      \parbox[b]{.1em}{b) \vspace*{2in}} &
      \includegraphics[width=2.5in]{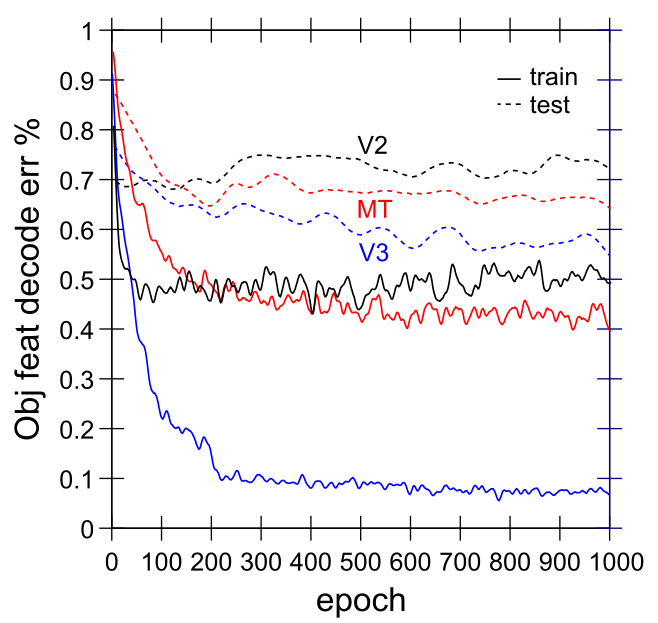}
    \end{tabular}
  \end{center}
  \caption{\footnotesize a) Learning curves for full model, showing accuracy (proportion error) in decoding the object features from each of 3 different layers (V3, V4, TEO), and overall prediction accuracy in terms of minus vs. plus phase cosine over the V1p pulvinar layer, at trial 3 (the last trial), which is nearly perfect.  Note the discrete jump in prediction accuracy when we turn on the top-down weights from TEO, at epoch 200.  The decoding shows a roughly 2x reduction in error for TEO vs. V3, and is especially evident in raw terms for the 10 novel untrained testing items.   This shows that TEO has developed much more systematic object representations than those in other layers.  b) Object feature decoding in layers V2 and MT versus V3, showing that MT specifically seems to learn in the {\em opposite} direction compared to TEO, producing significantly worse object decoding accuracy compared to V3, which serves as its input.  Nevertheless, MT does have slightly better object representations compared to V2.  Training curves are bumpier than testing curves because testing occurs only every 5 epochs, and all curves are smoothed with a gaussian filter to remove high-frequency trial-to-trial variance due to differences in environmental inputs.  One epoch = 512 alpha cycles = 51.2 seconds of real time, so this total training period represents approximately 16 hours of real time learning. Due to the time required (12 hrs using 64 processors in parallel on our cluster), results are from single runs, but we did run multiple replications of several key conditions and they were very reliable.}
  \label{fig.train_lcurve}
\end{figure}

The learning curves for the full intact model are shown in Figure~\ref{fig.train_lcurve}, showing that the model achieves high levels of predictive accuracy in terms of the cosine difference between the minus and plus phase activation states over the V1p pulvinar layer (green lines, 1.0 is perfect, model achieves roughly .96 on training and .93 on testing).  Furthermore, the TEO layer develops a much more systematic, generalizable representation of objects compared to other layers.  This is evident in the ability of the {\em decoder} (trained using the standard Leabra error-driven learning algorithm, but critically not interacting at all with the model via reciprocal connections) to decode both of the object feature dimensions accurately (each has 10 features, so chance is 1/10 per dimension, or 1/100=.01 for both).  The decoding of TEO is roughly 2x better (i.e., a 2x reduction in error) compared to the V3 layer.  Numerically, this is particularly evident for the 10 novel testing objects, suggesting that the TEO layer has developed a largely systematic encoding of the object dimensions, supporting roughly 70\% accuracy at decoding the object dimensions.

This measure of systematic object feature decoding is not just of computational interest: ecologically, it supports the ability of an organism to accurately and consistently identify objects in the environment, and respond appropriately.  Thus, we regard this measure as the most important indicator of overall function in the model: while predictive accuracy is the engine that trains everything, the essential product of this is developing a high-level abstract understanding of the environment that then provides a strong basis for adaptive behavior.  Anatomically, TEO provides the input to the higher areas of IT, medial temporal lobe, and ventral and medial prefrontal cortex, all of which build upon these basic invariant object representations to guide goal-driven behavior and high-level memory encoding.

As Figure~\ref{fig.train_lcurve}a shows, some of the improved TEO object decoding performance is due to improvements made by V4, indicating the need for multiple processing layers in the {\em What} pathway, consistent with the biology and recent deep neural network models.  V2 has very low object decoding accuracy, so V3 produces large gains in object decoding accuracy, but mainly for the trained items --- the novel test items show only a modest improvement.  Thus, the trained-object decoding accuracy measure does not necessarily indicate that V3 has invariant or compact object representations --- just that the information can be extracted by the decoder in any way (albeit within the constraints of a single-layer set of weights).  The test-object decoding performance is really the best measure of how systematic and invariant the object representations are, as is evident in the direct analysis of the representations shown next.

Interestingly, the MT layer shows {\em worse} object decoding accuracy compared to its input layer, V3 (Figure~\ref{fig.train_lcurve}b), indicating that it has learned in the opposite direction from V4 and TEO, in terms of extracting invariant object representations.  This oppositional dynamic between MT (i.e., the {\em What * Where} pathway) and IT (the {\em What} pathway) reflects the critical contributions of the these two pathways in enabling each other to partition distinct parts of the overall prediction problem, and it is evident in many of the other results below.

We also examined the ability to decode object position information from various layers, and found that TEO, V4, and MT all had essentially ceiling levels of decoding accuracy.  Because we used a gaussian blob spatial representation for spatial location, we measured decoding accuracy in terms of a cosine difference between the target location representation and that produced in the minus phase over the decoding layer (which again had no interaction with the rest of the network), and these cosines were at 0.995 for these layers for the testing items, and interestingly, somewhat lower for the training items (0.99 for TEO and V4, and 0.98 for MT).  Thus, TEO not only encodes abstract object identity, but also spatial location information, consistent with available empirical data \cite{MajajHongSolomonEtAl15}.  The differences in accuracy between MT and TEO may reflect the comparatively smaller size of MT --- when we used a larger MT layer, it started to take on more of the object identity encoding job and this interfered with learning of these object representations in TEO.  We hypothesize that the early developmental engagement of the MT more strongly biases it toward spatial representations, which could have the same overall effect as constraining its size as we do here.  This more complex developmental dynamic will be explored in subsequent work.

\subsection{Nature of TEO vs. MT Representations}

\begin{figure*}
  \centering\includegraphics[width=6in]{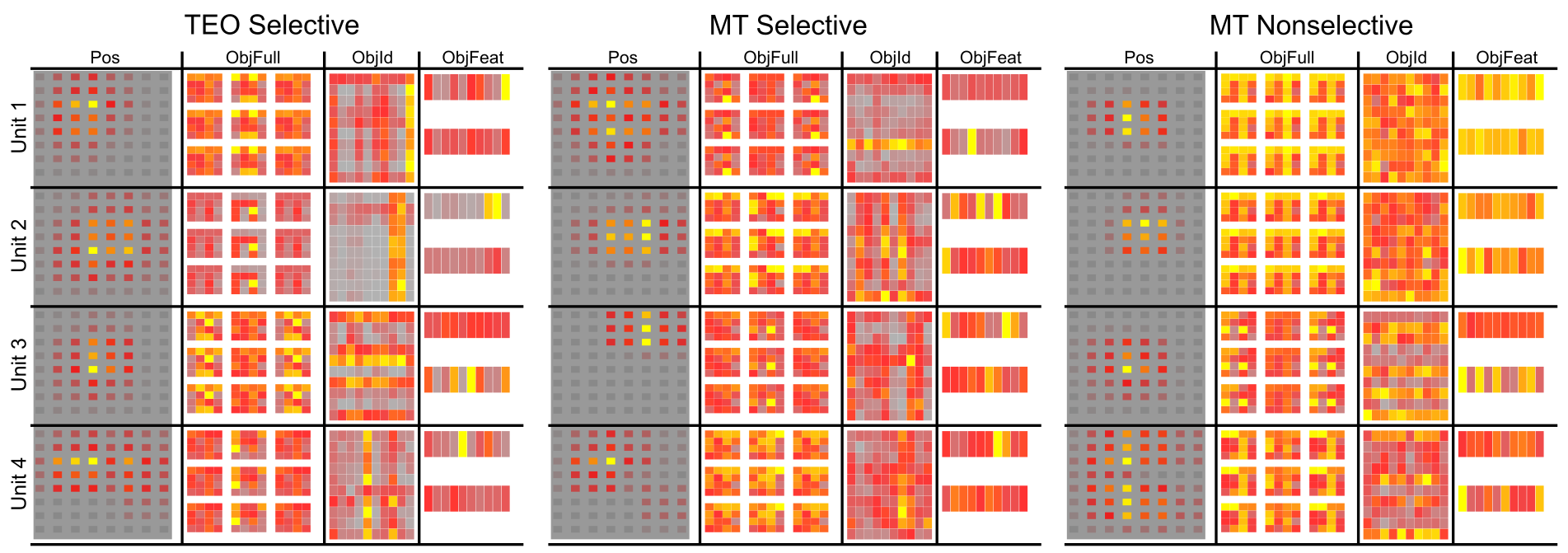}
  \caption{\footnotesize Activation-based receptive fields for TEO vs. MT (superficial layers), selected for relative feature selectivity, and MT non-selective cases.  Each cell shows weighted average activation across position and object decoding patterns as a function of unit activity, for 4 target units from each layer (large-scale rows).  Pos: position of object, showing large receptive fields in both TEO and MT (the center of the field is sampled more frequently due to nature of sampling constraints, so it is emphasized).  ObjFeat: 10 features x 2 dimensions (rows) of the object that was present --- e.g., the 2nd TEO unit from top selectively and strongly encodes two of the features from one of the dimensions (top row). ObjId: localist encoding (1 out of 100) of the object identity --- due to combinatorial nature of objects, those sharing the same features are aligned vertically or horizontally for the two dimensions, providing a fuller picture of the degree of feature selectivity (i.e., how solid and consistent are the lines).  ObjFull: the full rendered object pattern.  Overall, TEO has cleaner, more selective ActRF's compared to MT, even in the selected sample (see table 1 for selection details).  The non-selective patterns tend to have tighter spatial position coding, and very broad / distributed object coding.}
  \label{fig.actrf}
\end{figure*}

\begin{table*}
  \centering
  \begin{tabular}{l|rrrrr}
    & & \multicolumn{2}{c}{{\bf Spatial RF Size}} &
    \multicolumn{2}{c}{{\bf Cos Trial 2-3 Consistency}} \\
    {\bf Area} & {\bf \% Selective} & {\bf All} & {\bf Selective}  & {\bf All} & {\bf Selective} \\ \hline
    TEO & 60\% & 64\% & 71\% & 0.73 & 0.80 \\
    MT  & 30\% & 57\% & 67\% & 0.60 & 0.71 \\
  \end{tabular}
  \caption{\footnotesize Quantitative analysis of selectivity, stability, and receptive field size for ActRF representations in TEO vs. MT.  Selectivity was cheaply determined by thresholding average activation in the ObjId ActRF --- by experimentation, a threshold of 0.4 (on max-normalized 0-1 data) did a good job of separating the feature-selective (having clear lines in the Id ActRF) vs. more complex non-selective units.  There were twice as many such selective units (\% Selective) in TEO compared to MT, and the majority of TEO units were selective.  The next two columns show the average percent of object position cells that units responded to, for All units and for the selectively responding ones, showing that the feature-selective units had larger receptive fields, and that these fields on average covered a large portion of the spatial locations.  MT receptive fields were smaller overall.  The final two columns measure the consistency (cosine similarity) of the ActRF's computed on trial 2 (immediately post-saccade) vs. trial 3 --- the selective ones are more consistent across time, and TEO is more consistent than MT over time.}
  \label{tab.actrf_stats}
\end{table*}

To better understand the nature of the representations that developed in the high layers of the model, we used a form of the {\em spike triggered averaging} technique that computes a weighted average of the activation state across the network, weighted by the activation level of a given {\em target} unit (we refer to this as an activation-based receptive field, or ActRF).  When the target unit is off, then those network states are effectively ignored (they are multiplied by 0 in the weighted average).  And to the extent it is on, the result is an average, weighted by strength of activation, of the activation states correlated with the activity of the target unit.  In other words, it gives you a pretty clear picture of what the activation patterns in the rest of the network are like when this unit is responding.  Furthermore, it can be used with any kind of pattern, even ones not directly connected to the target unit --- including the decoder patterns which provide a very clear analysis of the unit's response profiles.

Figure~\ref{fig.actrf} shows the ActRF patterns for a sample of more feature-selective TEO and MT units, and non-selective MT units (which were a majority in MT, while the feature-selective ones were a majority in TEO; Table~\ref{tab.actrf_stats}).  As explained in the figure, the object ID and feature decoder layers allow us to see how consistently the TEO units respond to a subset of feature values, across a range of different spatial locations.  This clearly shows that TEO units have developed the characteristic invariant object recognition property of actual TEO neurons, responding systematically to subsets of object features across a range of locations.  Table~\ref{tab.actrf_stats} shows that 60\% of the TEO units had this object-feature selectivity, while only 30\% of MT neurons did (and even with those, the tuning was less clear and consistent than in TEO).  This table also shows the percent of all 64 spatial locations where units responded, showing that TEO had larger receptive fields than MT, and that the feature-selective receptive fields are larger on average than the non-feature-selective ones.

The non-feature-selective receptive fields in MT and TEO (Figure~\ref{fig.actrf}) tended to have more focal spatial coding, and broader distributed object feature tuning (including cases with essentially no feature selectivity at all).  These are clearly going to be more useful for the {\em What * Where} integration process, and their prevalence in MT supports this functional role for this area.  Nevertheless, these unit types also developed in TEO --- as is typical in neural network models, and in the brain, a full distributed spectrum of neural coding types tend to emerge over learning across all areas --- there are no truly representationally {\em pure} areas \cite{BehrmannPlaut13}.  This is overall consistent with available data on TEO neurons, which also encode spatial location along with many other properties, and have a broad range of selectivities \cite[e.g.,]{HongYaminsMajajEtAl16,MajajHongSolomonEtAl15,ZoccolanKouhPoggioEtAl07,Tanaka96,LogothetisSheinberg96}.  More generally, these results are consistent with coarse-coded distributed representations of high-dimensional data (also known as {\em mixed selectivity} \nopcite{FusiMillerRigotti16}), which are useful for efficiently binding multiple features into a coherent object representation \cite{HintonMcClellandRumelhart86,OReillyBusby02,OReillyBusbySoto03,CerOReilly06}.  The greater complexity and higher-dimensionality of the {\em What * Where} pathway reflects their particular specialization for this kind of binding, but the differences are clearly quantitative, not qualitative.

One further analysis we performed was to compare the consistency (cosine similarity) of ActRF patterns based on activity on trial 2 (immediately post-saccade) to those from trial 3.  This provides an indication of how temporally stable these representations are over the 4 trial scene where a single object is present.  Table~\ref{tab.actrf_stats} shows that again TEO had overall more such consistency compared to MT, and that the feature-selective units were more consistent than the non-selective ones.

Taken together these analyses strongly show that, consistent with the decoding results, the model's TEO has developed systematic invariant object representations, without any external pressure to do so.  This purely self-organized learning, in an environment with a relatively large number (100) of highly overlapping and confusable objects, goes beyond existing auto-encoder neural network models, that tend to extract broad central tendencies across the inputs (e.g., the famous Google auto-encoder network that extracted a blurry cat face from millions of images from the internet; \nopcite{LeMongaDevinEtAl12}).  Success in these auto-encoder models is instead typically measured in terms of reductions in number of supervised training trials required on top of the auto-encoding pre-training \cite{Valpola14,RasmusBerglundHonkalaEtAl15}.

\subsection{Importance of a Deep Hierarchy: Testing Flatter Models}

\begin{figure}
  \begin{center}
    \begin{tabular}{ll}
      \parbox[b]{.1em}{a) \vspace*{2in}} &
      \includegraphics[width=2.5in]{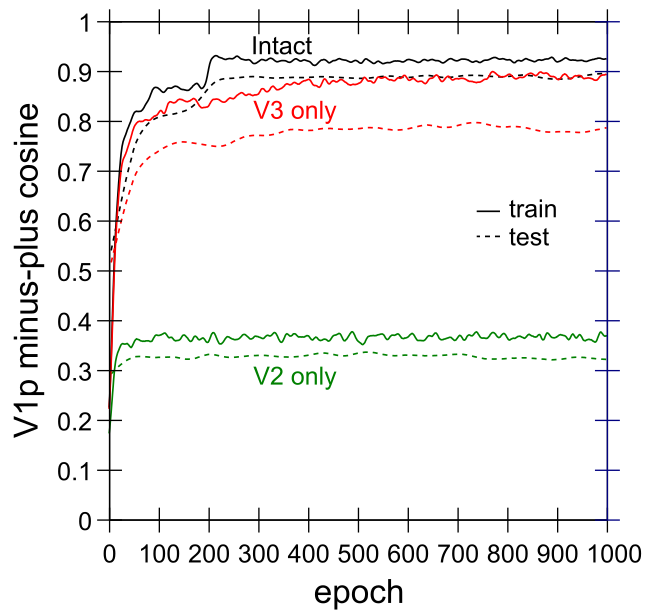} \\
      \parbox[b]{.1em}{b) \vspace*{2in}} &
      \includegraphics[width=2.5in]{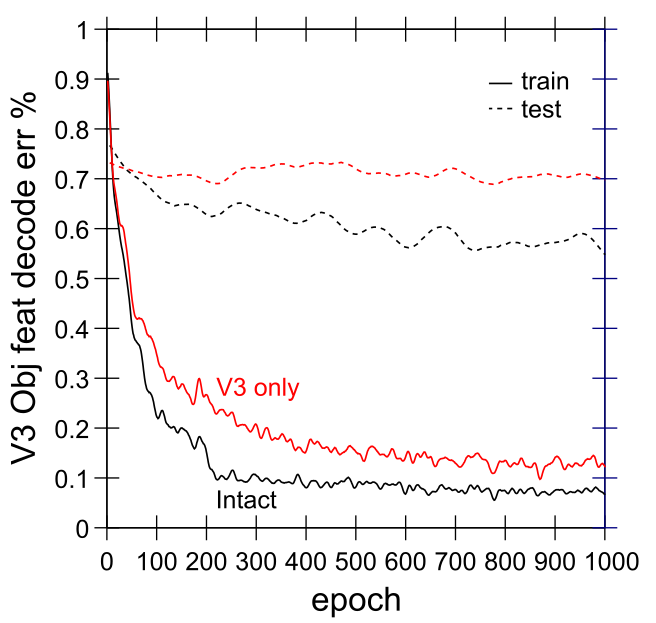}
    \end{tabular}
  \end{center}
  \caption{\footnotesize a) Prediction accuracy (minus vs. plus phase cosine over the V1p pulvinar layer), at trial 2 (the post-saccade trial) for model with only V2 (no V3, MT, V4, TEO) or only V3, compared to the full Intact model.  A single layer alone (V2 only) cannot do very well, despite getting nearly-perfect spatial inputs from the pre-trained LIP {\em Where} network.  Adding V3 on top of V2 produces a dramatic improvement, but the novel testing patterns are notably worse than the trained ones.  b) Object feature decoding accuracy from layer V3 in V3 only vs. Intact model, showing that the top-down projections from higher layers play a significant role in shaping the object encoding in V3 in the Intact model. }
  \label{fig.v2v3only}
\end{figure}

Figure~\ref{fig.v2v3only}a shows the effects of removing the higher levels of the network, demonstrating that a deep hierarchy of layers is important for achieving high levels of predictive accuracy in this task, particularly with respect to the novel test items.  Performance on these test items indicates to what extent the model is shaping predictive mappings specifically around the trained objects (resulting in poor testing performance), versus having a more generalized, abstract capability of mixing independent {\em What} and {\em Where} pathway information (resulting in good testing performance).  With only V2, prediction accuracy on V1p is dramatically worse, with cosine levels between .3 and .4 and not much sign of learning progress overall.  Adding V3 improves training performance dramatically --- the more compact representations and integrative connectivity of V3 adds considerably more systematicity and power.  Nevertheless, the performance on the testing items remains differentially lower compared to the training performance, suggesting that the V3-only network is missing the ability to more systematically represent objects.  Figure~\ref{fig.v2v3only}b reinforces the importance of yet higher layers above V3: these higher layers (MT, V4, TEO) provide a top-down shaping influence on the V3 representations that makes it easier to decode the object features from V3.

\begin{figure}
  \begin{center}
    \begin{tabular}{ll}
      \parbox[b]{.1em}{a) \vspace*{2in}} &
      \includegraphics[width=2.5in]{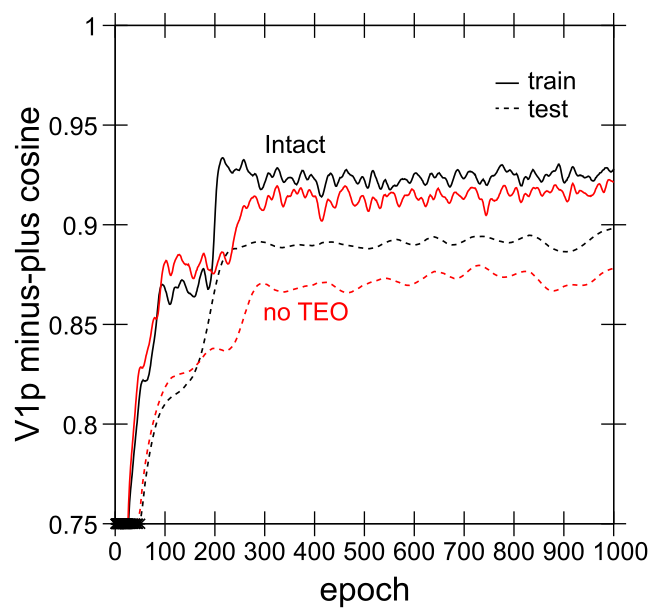} \\
      \parbox[b]{.1em}{b) \vspace*{2in}} &
      \includegraphics[width=2.5in]{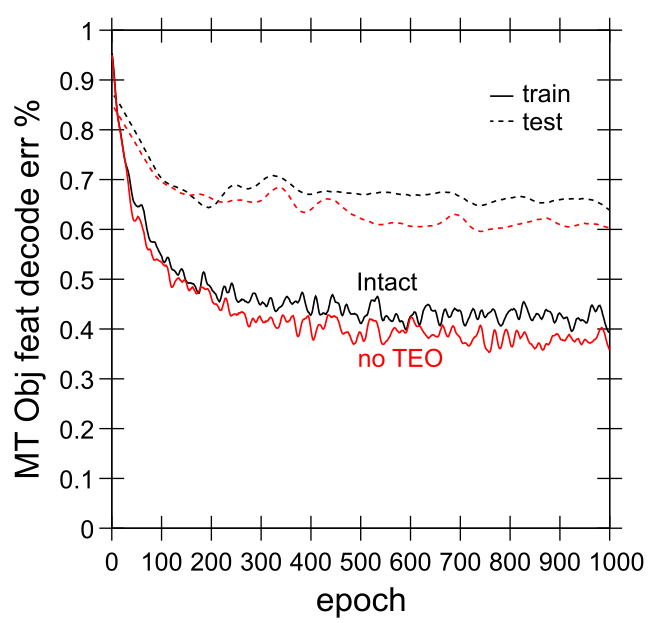}
    \end{tabular}
  \end{center}
  \caption{\footnotesize a) Prediction accuracy (as in prior figure) for Intact versus model with no TEO area, showing small but reliable impairment, more for test than trained objects.  b) Object feature decoding accuracy from layer MT for Intact vs. no TEO model, showing {\em improvement} in object detection in MT when TEO is lesioned, consistent with opponent relationship between these pathways. }
  \label{fig.noteo}
\end{figure}

Figure~\ref{fig.noteo} shows effects of only removing the TEO area, with everything else as in the full Intact model.  This results in a small but reliable impairment in prediction accuracy, more for the novel testing objects than the trained objects, consistent with the importance of the abstract high-level TEO representations providing top-down drive into the lower-layer predictions.  Here you can also more clearly see (due to the use of a more restricted vertical range in the graph) the significant bump in prediction accuracy in the Intact model right after the top-down connections from TEO are turned on at epoch 200, reflecting the hypothesized delay in maturation of these projections.  Interestingly, the no-TEO model also shows a bump, but at epoch 250, which is when we drop the learning rate on our standard learning rate schedule, which overall produces better learning results and reflects a likely developmental slowing of effective learning rate.  Overall, we anticipate that with more complex, high-dimensional real-world objects, this high-level TEO contribution to overall prediction accuracy will be significantly more important, compared to the relatively simple objects used here.  Nevertheless, even in this simple case, and especially in the novel testing objects, we obtain an indication of these top-down effects.

Another manifestation of the opponent-dynamics between MT and TEO is evident in Figure~\ref{fig.noteo}b, showing the object decoding accuracy in area MT for both the Intact and no-TEO models.  Interestingly, the ability to decode objects actually {\em improves} in MT with the TEO removal, suggesting that it is partially taking on some of the {\em What} pathway function that TEO otherwise dominates in the intact model.   We also tested the removal of MT --- in earlier versions of the model this consistently produced major reciprocal impairments on object encoding in TEO, as TEO took on more of the {\em What * Where} integration task from the missing MT.  However, due to various improvements in the V4/TEO pathway parameters, it became more robust and the removal of MT only had relatively small (but reliable) effects on TEO object decoding (not graphed).

\subsection{Developmental Timing: Early {\em Where} and Late {\em What} Pathways}

\begin{figure}
  \begin{center}
    \begin{tabular}{ll}
      \parbox[b]{.1em}{a) \vspace*{2in}} &
      \includegraphics[width=2.7in]{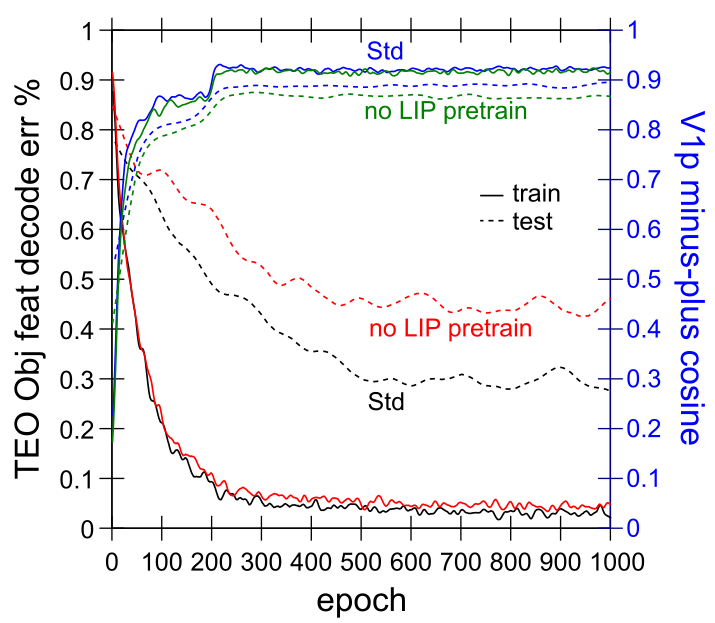} \\
       \parbox[b]{.1em}{b) \vspace*{2in}} &
     \includegraphics[width=2.5in]{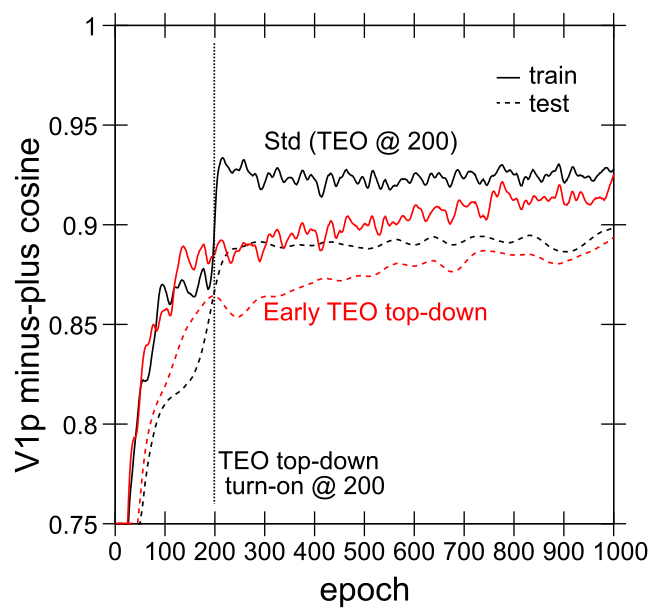}
    \end{tabular}
  \end{center}
  \caption{\footnotesize a) Learning without first pretraining the {\em Where} LIP pathway compared to the standard (Std) training, this has a significant impact on the development of systematic TEO object representations, particularly for the testing items.  This has corresponding effects on V1p prediction accuracy (top lines), again particularly on the testing items (the size of these effects is roughly proportional to the relatively small overall impact of TEO on prediction error as shown in earlier figures).  Overall, this again supports the importance of partitioning the prediction error so that the TEO can focus on learning more directly about object features. b) Prediction accuracy effects of having top-down TEO to V2,V3 projections effective right from the start of learning, as opposed to coming on after 200 epochs as in the standard model.  The delayed engagement of TEO allows overall predictive performance to improve significantly earlier.}
  \label{fig.devel}
\end{figure}

The importance of the early development of the LIP spatial prediction pathway on subsequent learning in the full network is shown in Figure~\ref{fig.devel}a.  The main effects from not using the pretrained LIP pathway weights are on the development of systematic object feature representations in TEO, reflected in significant reduction in object decoding accuracy on test items, and a corresponding impact on V1p prediction accuracy specifically for these test items.  The relatively large impact on testing object decoding is interesting given that the LIP trains quite quickly (a majority of the learning takes place within the first 10 epochs; Figure~\ref{fig.lip_pretrain}).  This again suggests that the partitioning of the spatial component of prediction error is important for allowing the TEO to develop more systematic object encodings, and that doing so before the TEO has any significant learning pressure is critical.  With larger more complex spatial and object representational spaces in the real system, these effects would likely be magnified considerably.

Figure~\ref{fig.devel}b shows the advantages of a developmental delay in the strengthening of the top-down projections from TEO to lower areas (V2, V3).  By waiting until the TEO area has had a chance to develop more abstract object representations, the impact of these more systematic representations produces an immediate bump in predictive accuracy, whereas when these lower layers have first learned to incorporate the less systematic initial TEO representations, it takes much longer to overcome that initial learning and begin to incorporate the more systematic top-down inputs.

\subsection{Limitations of Outside-In Progressive Learning}

\begin{figure}
  \centering\includegraphics[width=2.7in]{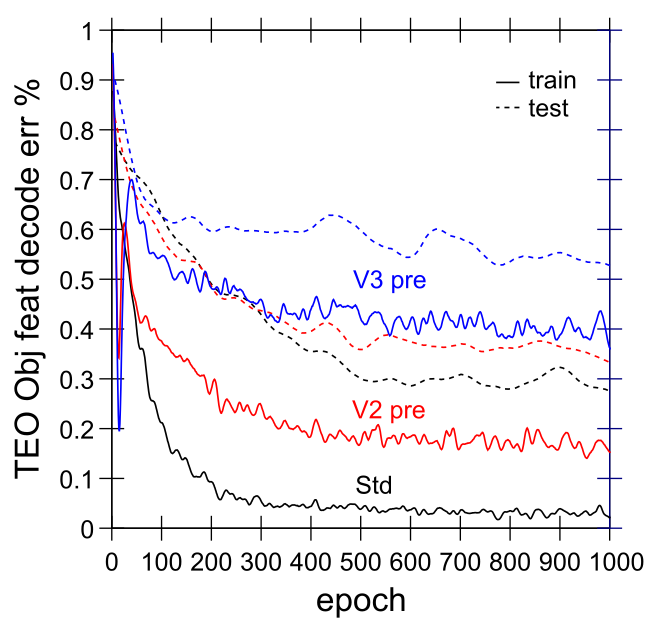}
  \caption{\footnotesize Effects of pretraining using weights from V2 only or V3 only model on object decoding accuracy from the TEO area, as a test of the standard outside-in developmental training approach.  This significantly impairs the development of systematic invariant object representations in TEO, presumably by interfering with the prediction error partitioning process, and the top-down influence of more abstract object representations during learning.}
  \label{fig.v23_pre}
\end{figure}

Next we tested the standard approach of training deep hierarchical auto-encoders and related models, where progressively higher layers are added after earlier layers have had a chance to develop their initial representations.  We did this by using the weights from the V2 only and V3 only cases described above as initial starting weights for training the full standard model.  Figure~\ref{fig.v23_pre} shows that this significantly impaired the ability to decode object features from the TEO area of the model.  We argue that this resulted from these models developing representations that tried to solve all aspects of the prediction problem without the benefits of more abstract higher-level representations driving top-down input into these lower layers.  Interestingly, the V3-only case was significantly worse here compared to the V2-only, even though V3-only did a better job overall of prediction (Figure~\ref{fig.v2v3only}).  This suggests that the representations developed during this initial pretraining fused the {\em What} and {\em Where} aspects of the prediction problem in a way that made it difficult to then extract a more pure object-invariant representation.  Instead, we argue that our standard version of the model depends critically on the interactions between MT and TEO pathways {\em from the very start of the learning process} for partitioning the prediction problem, allowing TEO to more fully develop its more pure {\em What} representations.

Also, these pretrained models actually did relatively well at the V1p prediction learning task, with the V2-pre case even doing slightly better than the default model, suggesting that prediction error in this simple model may not fully reflect the beneficial contributions from high-level abstract representations.  We anticipate that with more complex, high-dimensional real-world objects, these high-level representations will be essential for accurate prediction.

\subsection{Importance of V1p for Higher Areas}

\begin{figure}
  \centering\includegraphics[width=2.7in]{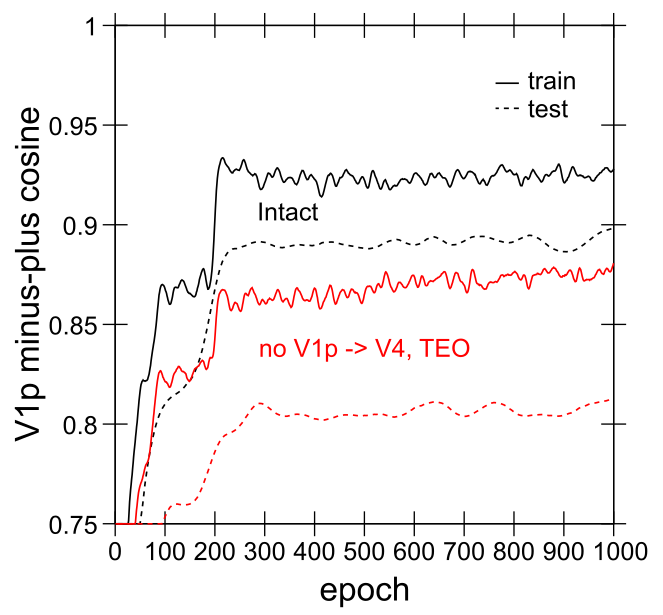}
  \caption{\footnotesize Effects of removing the V1p to V4,TEO projection on overall V1p prediction accuracy, showing similar effects to a TEO lesion, indicating that the {\em What} pathway is essentially non-functional without this V1p pulvinar projection.  Consistent with this, object decoding accuracy in TEO was also completely abolished (not shown).}
  \label{fig.nov1p_to_it}
\end{figure}

One of the potentially puzzling aspects of the pulvinar connectivity is that it appears to route information from low levels of the visual hierarchy (V1, V2) into the higher-level areas such as V4 and TEO.  How could such a low-level signal, reflecting detailed prediction errors in our model, be beneficial for shaping higher-level representations?  As we have argued above, we think this signal is useful in the context of interactions with other areas, to help partition the overall prediction error signal, such that the {\em What} pathway ends up being able to focus on improving the prediction accuracy specifically for the object features component.  In other words, this shared projection-screen-like representation enables the different areas to effectively coordinate and specialize on specific aspects of the overall prediction task.  Throughout the development of our model, we consistently found that removing the V1p projections to TEO or V4 impaired performance (object decoding and prediction error) significantly.  And in the final model, removing this projection from {\em both} V4 and TEO results in a {\em complete failure} to be able to decode object features from TEO or V4.  These layers instead develop some entirely different form of representations, and prediction accuracy also suffers significantly (Figure~\ref{fig.nov1p_to_it}).  However, there are only relatively minimal effects in the final model of only removing V1p projections to TEO, increasing the object decoding error for trained objects from around .05 to .1, and, surprisingly, having no effect on test objects.  Thus, we think that TEO can largely receive the relevant V1p error signals indirectly through its interconnections with V4, but removing this signal from both V4 and TEO is catastrophic.

Also, it is worth noting that throughout most of our model development, we had a small bug in the environment program, which resulted in occasionally unpredictable input sequences being presented.  It is possible that the magnified effects of the V1p to TEO projection in these earlier models may reflect its importance for more robust, fault-tolerant learning.  We plan to explore this idea in future research.

\subsection{Importance of Temporal Context, Hebbian Learning, Momentum}

\begin{figure}
  \begin{center}
    \begin{tabular}{ll}
      \parbox[b]{.1em}{a) \vspace*{2in}} &
      \includegraphics[width=2.5in]{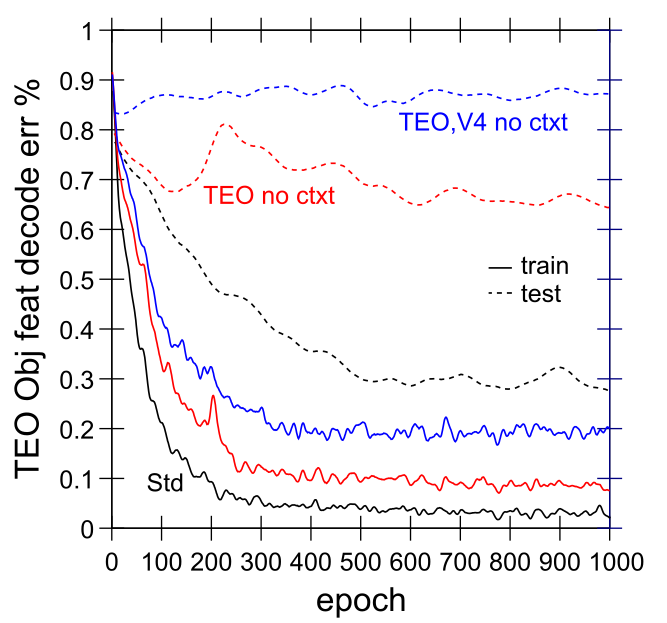} \\
      \parbox[b]{.1em}{b) \vspace*{2in}} &
      \includegraphics[width=2.5in]{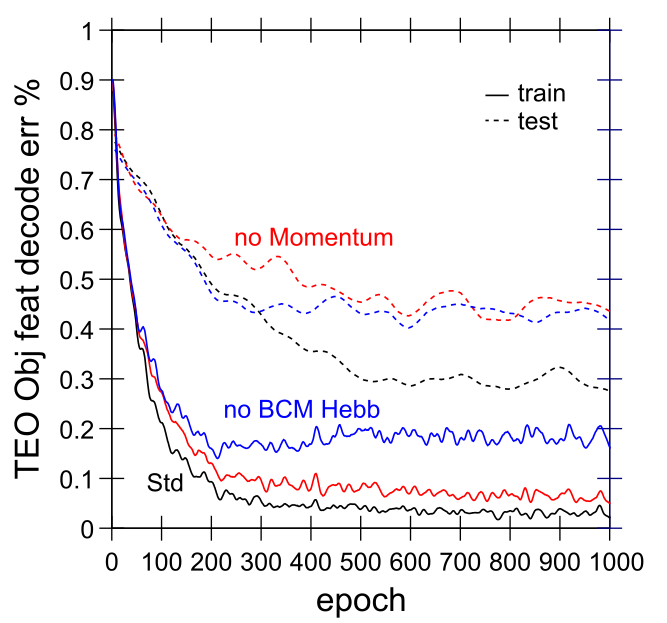}
    \end{tabular}
  \end{center}
  \caption{\footnotesize a) Effects of removing the deep-layer context inputs into TEO and V4 + TEO together --- this has a major impact on ability to decode object features from TEO, particularly in the case of the novel testing items. b) Effects of not using momentum or BCM-like Hebbian learning.}
  \label{fig.ctxt_moment_hebb}
\end{figure}

Finally, we report the effects of various important elements of the DeepLeabra computational framework, including the deep-layer temporal context mechanism, the combination of BCM-like Hebbian learning along with error-driven learning, and the effects of using momentum in the learning rule.   Figure~\ref{fig.ctxt_moment_hebb} shows that each of these factors plays an important role in contributing to the overall performance of the intact network.  For the Hebbian and momentum factors, both of these produced more ``dead'' units (the flip-side of the hog units mentioned above --- these are easier to quantify), particularly in the higher layers, with hebbian being particularly important for TEO while momentum was more important for V4.

\subsection{Summary}

The above results, which represent a small subset of the extensive explorations we performed over the development of the final model (1,160 different model runs, requiring over 45 CPU-{\em years} of computation on our 576 CPU cluster), together support a consistent overall picture of how it learns over time.  The three different pathways of the model, {\em Where}, {\em What}, and {\em What * Where}, interact in important ways to enable the joint goals of highly accurate prediction generation, and the development of invariant, systematic object representations in the ventral {\em What} pathway.  This latter outcome depends on the other sources of prediction error being managed by other areas, and represents an important new way of understanding how a purely self-organizing learning system can develop these essential high-level abstract representations.  In other words, this is a case where ``it takes the whole network to raise a model'' --- the entire predictive learning problem must be solved with a complete, interacting network, and cannot be solved piece-wise.  Furthermore, the entire network must be interacting bidirectionally, with top-down excitatory connections playing a critical role in shaping the overall learning process in lower layers, which then feed back up into the higher layers, etc.  Thus, this model represents a truly {\em emergent} system.

\section{Results: Accounting for Empirical Data}

In this section, we apply our model to a set of important empirical phenomena that directly relate to predictive learning, starting with the case of predictive remapping, which is perhaps the most iconic example of a predictive phenomenon in the brain.  We then simulate key data from monkey electrophysiology showing top-down effects emerging after roughly one alpha cycle, shaping lower-level representations according to higher-level interpretations of the overall scene.  Finally, we simulate data that has been interpreted as supporting an alternative explicit-error-coding framework for generative models, showing that it emerges naturally from our model.  Although these are but a small subset of the possible data within the scope of such a comprehensive model, they address some of the most important and relevant data.  Future work will explore many other such phenomena.

\subsection{Predictive remapping}

\begin{figure}
  \begin{center}
    \begin{tabular}{ll}
      \parbox[b]{.1em}{a) \vspace*{1.4in}} &
      \includegraphics[width=3in]{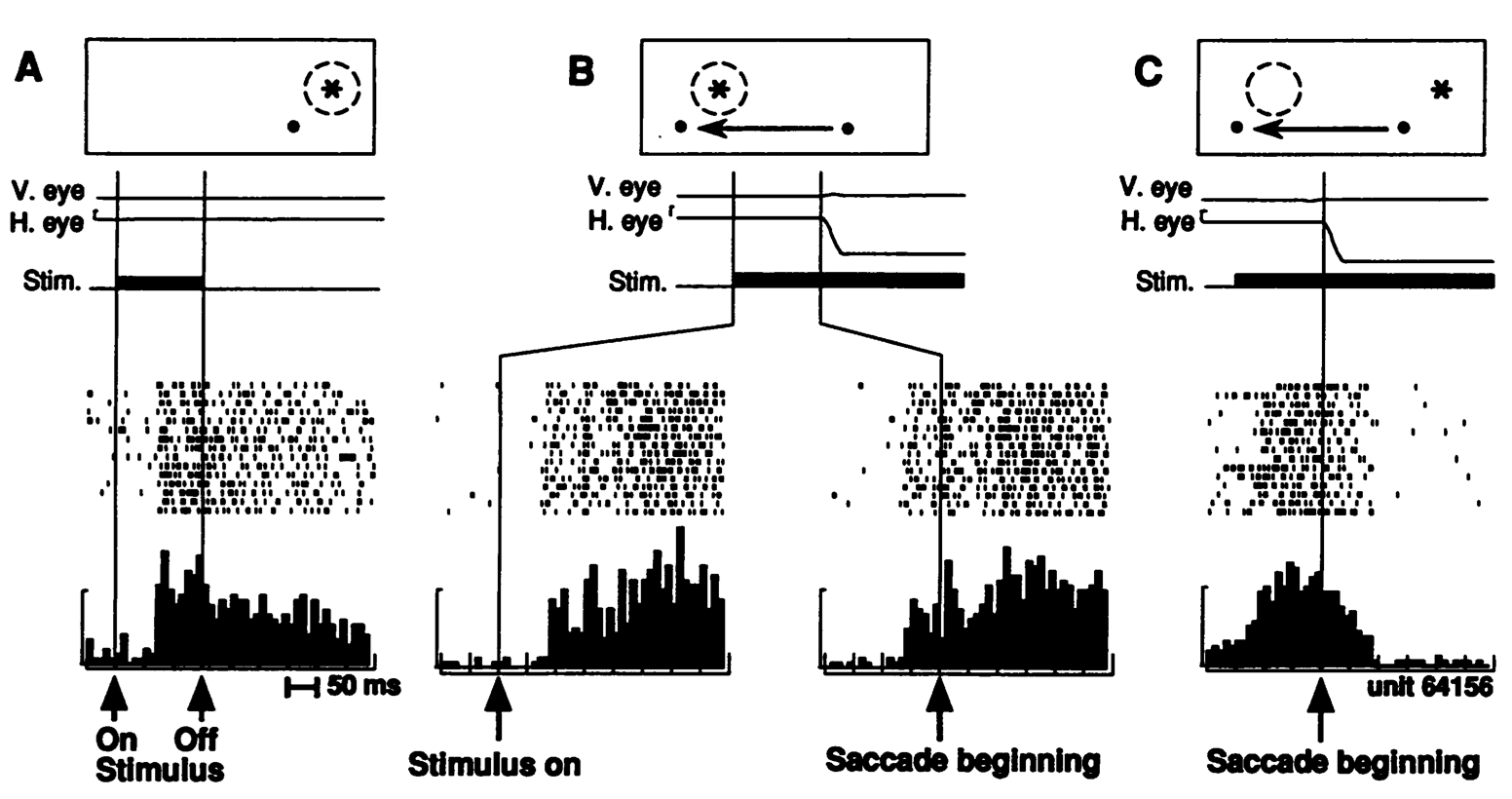} \\
      \parbox[b]{.1em}{b) \vspace*{1.4in}} &
      \includegraphics[width=3in]{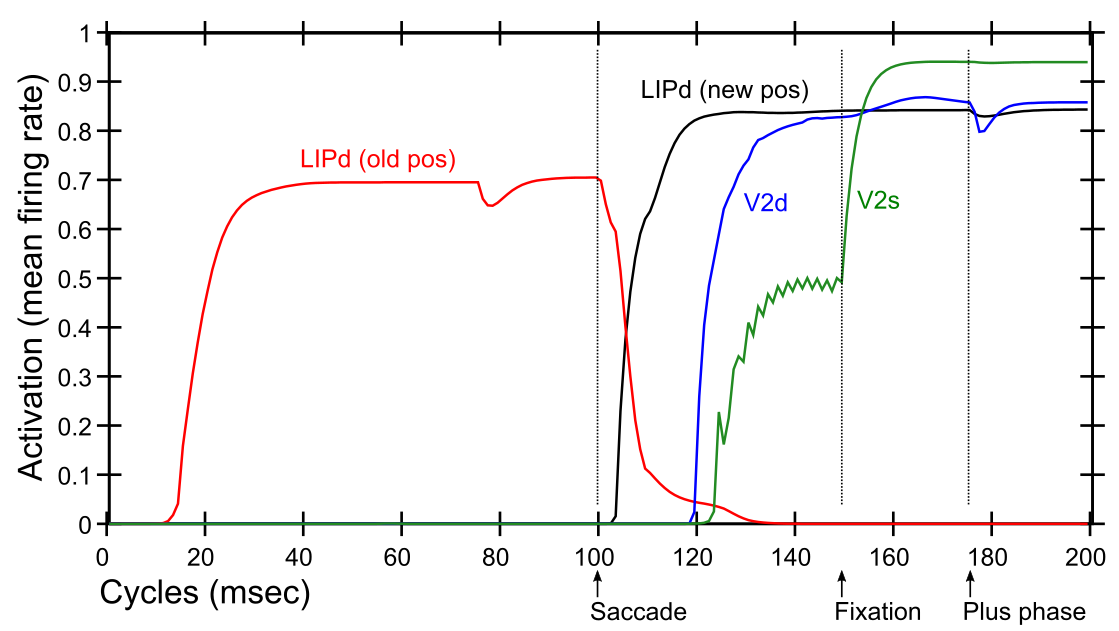}
    \end{tabular}
  \end{center}
  \caption{\footnotesize a) Original remapping data in LIP from Duhamel et al (1992).  A) shows stimulus (star) response within receptive field (dashed circle) relative to fixation dot (upper right of fixation).  B) Just prior to monkey making a saccade to new fixation (moving left), stimulus is turned on in receptive field location that {\em will be} upper right of the new fixation point, and the LIP neuron responds to that stimulus in advance of the saccade completing.  The neuron does not respond to the stimulus in that location if it is not about to make a saccade that puts it within its receptive field (not shown).  This is predictive remapping.  C) response to the old stimulus location goes away as saccade is initiated.  b) Data from our model, from individual units in LIPd, V2d, and V2s, showing that the LIP deep neurons respond to the saccade first, activating in the new location and deactivating in the old, and this LIP activation goes top-down to V3 and V2 to drive updating there, generally at a longer latency and with less activation especially in the superficial layers.  When the new stimulus appears at the point of fixation (after a 50 msec saccade here), the {\em primed} V2s units get fully activated by the incoming stimulus.  But the deep neurons are insulated from this superficial input until the plus phase, when the cascade of 5IB firing drives activation of the actual stimulus location into the pulvinar, which then reflects up into all the other layers.}
  \label{fig.remap_units}
\end{figure}

\begin{figure*}
  \centering\includegraphics[width=6in]{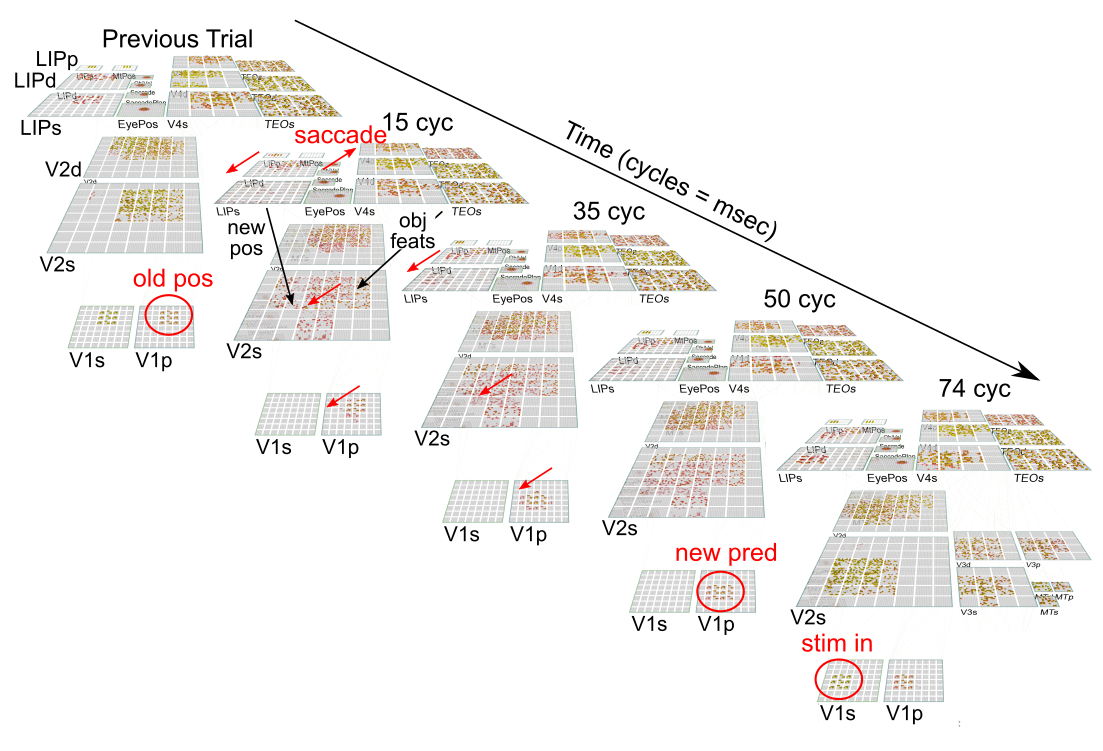}
  \caption{\footnotesize Predictive remapping in the entire model, from the prior trial state through the minus phase (75 cycles) of the post-saccade fixation trial.  Though hard to see, the LIPd deep-layer activation state moves first within the first 15 cycles, which then drives LIPp (which then updates LIPs as well), and sends top-down input to V3 (not shown) and V2, which ultimately consolidates on a new predicted V1p state by 50 cycles.  On cycle 74 (75th cycle), the new sensory input appears, matching the prediction.  To make this accurate prediction, these lower layers receive top-down input from TEO providing a representation of object features, and these streams are combined (with considerable help from the V3/MT {\em What * Where} integration pathways) to drive an accurate prediction on the pulvinar (V1p) about what the visual input will look like when it arrives, after the saccade fixation.}
  \label{fig.remap_net}
\end{figure*}

The remarkable phenomenon of predictive remapping, where neurons in the visual stream appear to remap their spatial receptive field in anticipation of the effects of a saccade \cite{DuhamelColbyGoldberg92,ColbyDuhamelGoldberg97,GottliebKusunokiGoldberg98,NakamuraColby02,NeupaneGuittonPack16}, is the exactly what one would expect if the brain is performing predictive learning.  And indeed, our model was designed specifically to capture this effect, using saccades as one of the major sources of spatial prediction that the model needs to learn (the other being intrinsic motion of the object itself).  Predictive remapping was initially described in area LIP \cite{DuhamelColbyGoldberg92}, but it has also been found as low as V2 in the early visual stream, but, interestingly, not in V1 \cite{NakamuraColby02}.  In LIP, around the time of the saccade, neurons fire for stimuli that will appear in the new retinotopically-defined receptive field location, in anticipation of the effects of the saccade (Figure~\ref{fig.remap_units}a).

Figure~\ref{fig.remap_units}b shows the activity profiles of characteristic units in our model from LIP and V2 layers, providing a clear match to the observed data.  Importantly, our model predicts that the remapping starts in LIP, which has direct input from the relevant eye movement signals, and this then drives top-down updating of activations in lower layers (V3, V2).  Figure~\ref{fig.remap_net} shows this same trial in terms of full network activation patterns.  This is consistent with the theoretical frameworks of \incite{CavanaghHuntAfrazEtAl10} and \incite{Wurtz08}, who strongly emphasize that this remapping must occur in these higher layers first, and then drive a top-down attentional signal to lower layers.  It is simply not possible for lower layers to remap across the relevant visual angle of saccades, which can be quite far, and would require massive interconnectivity in these lower layers.  Instead, it makes much more sense for a compact, high-level spatial layer like LIP to do the essential spatial remapping, and then send the result down to lower layers.  Critically, our model predicts that this top-down remapping largely stops at V2, because that is the first layer that is driven by the predictive signals from the pulvinar --- V1 is largely driven by LGN thalamus, and does not engage in this same kind of predictive learning process.  This is consistent with available data \cite{NakamuraColby02}, which also supports our prediction that V2 remapping is weaker and slower than that in LIP.

Our model makes some testable predictions about the relationship between saccades and the alpha cycle.  For example, depth-electrode recording in LIP should be able to distinguish between a predictive representation emerging in the deep layers, strongly synchronized with the alpha cycle, and a more fluid superficial-layer representation reflecting current attentional foci, which is then updated via the predictive signals from the deep layers around the time of a saccade.  We also predict that the pulvinar plays a critical role in broadcasting the predicted saccade outcome information to superficial LIP and other areas (along with LIP deep-layer top-down projections).  Indeed, very recent data appears strongly consistent with these predictions, showing a strong alpha-frequency coherence between the current and predicted receptive fields in V4, which they speculate to be driven by top-down and pulvinar-driven alpha dynamics \cite{NeupaneGuittonPack2017}.  This appears to be a very strong confirmation of a major prediction from our model.

In a future, larger-scale model, we plan to address the potentially important differences between microsaccades (less than 1 degree) and full saccades \cite{Martinez-CondeOtero-MillanMacknik13,Martinez-CondeMacknikHubel04}.  Unlike full saccades, microsaccades {\em can} be predicted within the typical receptive field sizes of V2 neurons, and there is evidence that visual motion signals are also used to predict the outcome of such saccades (along with passive visual drift which is also prevalent at these small scales).  Interestingly, the new cortical hierarchy analysis by \abbrevincite{MarkovVezoliChameauEtAl14} (Figure~\ref{fig.model_cons}b) separates the frontal eye fields (FEF, area 8) into two parts, at different locations in the hierarchy.  The part of FEF responsible for small-displacement saccades (8L) is located at the same level as V3, while the large-displacement part (8m) is higher up at the level of LIP and is assumed to provide the saccade signals in our current model.  Thus, this microsaccade system involving 8L, V3, and V2 may provide a rich source of additional predictive learning training for shaping these high-resolution, lower areas of the visual system.

\subsection{Top-down Activation of V1 from Higher-Levels}

\begin{figure}
  \begin{center}
    \begin{tabular}{ll}
      \parbox[b]{.1em}{a) \vspace*{1.4in}} &
      \includegraphics[width=3in]{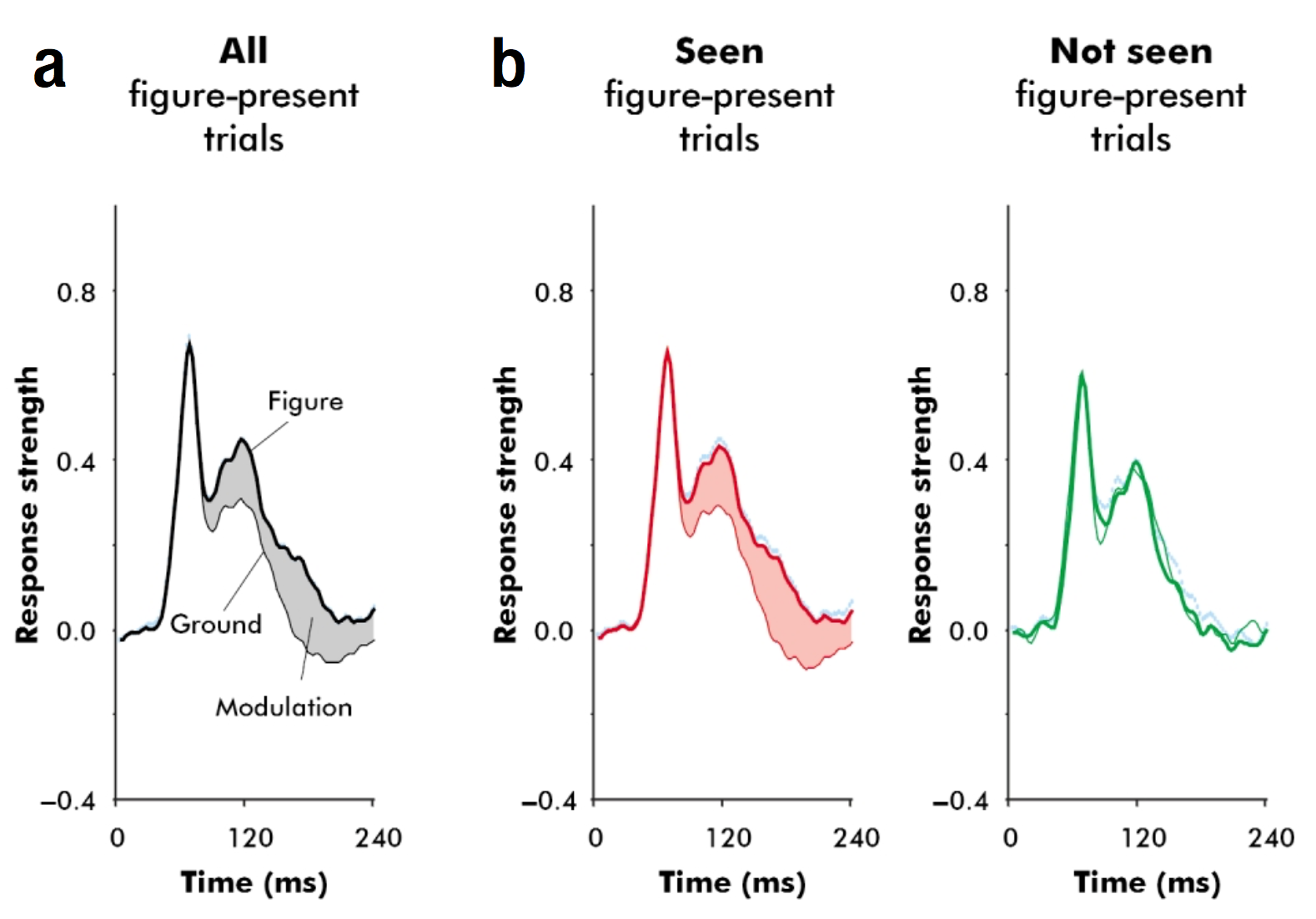} \\
      \parbox[b]{.1em}{b) \vspace*{1.4in}} &
      \includegraphics[width=3in]{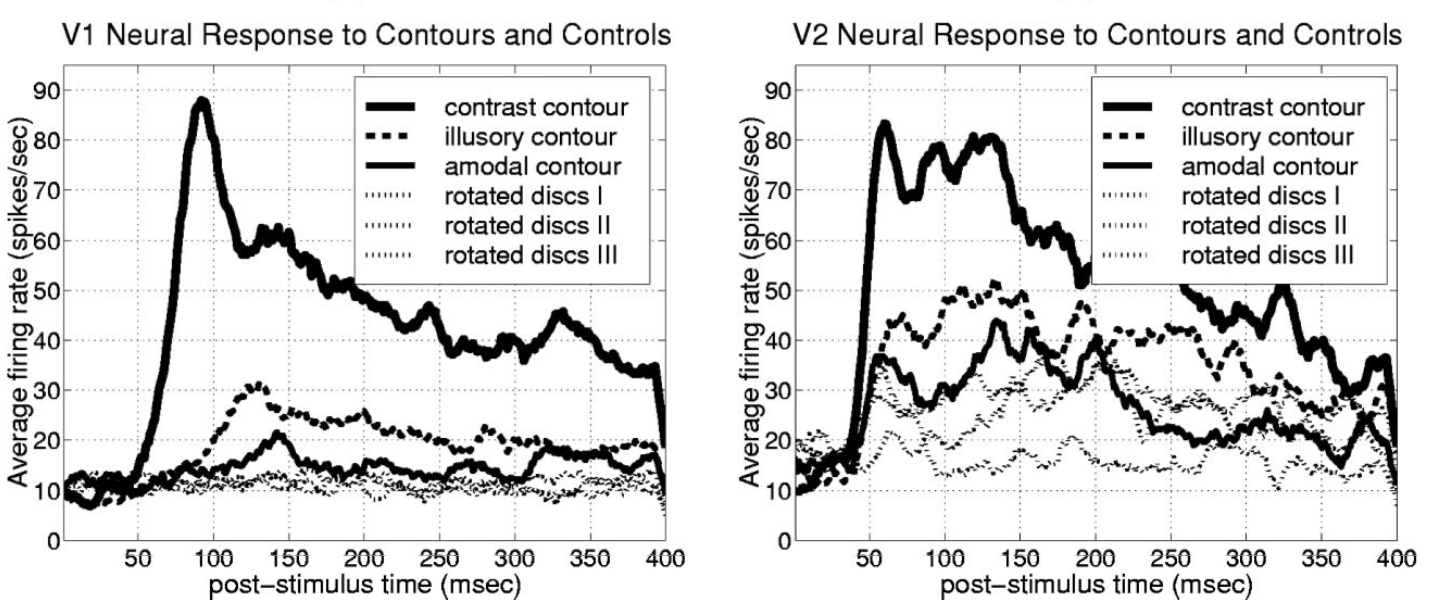} \\
      \parbox[b]{.1em}{c) \vspace*{1.4in}} &
      \includegraphics[width=3in]{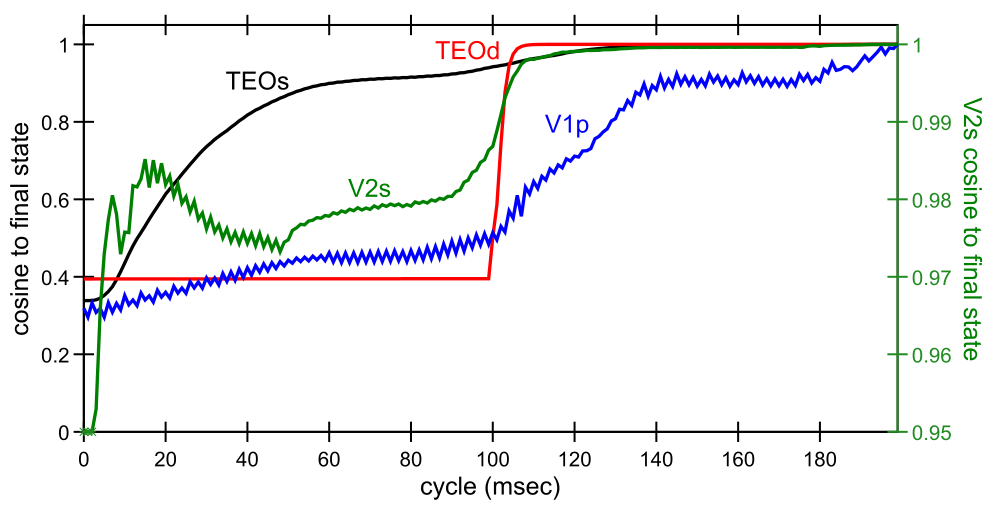}
    \end{tabular}
  \end{center}
  \caption{\footnotesize Top-down effects on lower-level neural firing.  a) Top-down modulation of V1 firing as a function of a texture-defined figure/ground stimulus, emerging after 100 msec (one alpha trial) in monkeys, specifically as a function of whether the monkey makes a behavioral response indicating that the figure was seen (Super et al, 2001, reprinted with permission).  b) Emergence of V1, V2 neural firing to illusory and amodal contours, suggesting earlier V2 responding driving top-down V1 responses that emerge after 100 msec (Lee et al., 2002, reprinted with permission).  c) Top-down driven activation in V1 and V2 of our model, using partially-occluded stimuli, showing the cosine of the current activity pattern on a layer in comparison to the final activation state at the end of the 200 msec window.  TEOs (superficial neurons) converges on its final state the most quickly, and drives top-down updating of V2s and V1p (pulvinar) representations, which are then more strongly driven when the TEO deep-layer (TEOd) updates after one alpha cycle.  The final V1p state reflects a largely accurate prediction of the complete object features (see supplemental information for a video of actual network states) --- the remaining change at the very end reflects plus-phase signal driving back to partial input, which does not perturb higher layers.  Note that V2s is plotted on a separate scale (shown at right) because it is a very large activation pattern that doesn't change as much as the others.}
  \label{fig.top_down}
\end{figure}

There have been a number of important demonstrations that neurons in lower visual areas (V1, V2) reflect higher-level interpretations of a visual display, with this top-down signal emerging typically after around 100 msec \cite{SuperSpekreijseLamme01,FahrenfortScholteLamme08,LeeNguyen01,LeeYangRomeroEtAl02} (Figure~\ref{fig.top_down}).  Importantly, these effects depend on the animal being awake, and on having indicated that the higher-level percept was actually formed \cite{SuperSpekreijseLamme01}, and other factors such as context that shape the nature of the high-level interpretation \cite{LeeMumford03}.  Given the importance of top-down activation from higher layers in our model, we tested for the presence of similar such effects.  Because of the simplicity of our visual environment, we could not directly replicate the existing experiments (which involve 2D-cues for depth perception), but instead used a simple proxy, where the object inputs were partially obscured (11\% of active features turned off), such that higher-level representations were needed to complete the original full pattern.

As Figure~\ref{fig.top_down} shows, our model shows the same kind of top-down effects in lower layers as have been observed in monkeys (and in our prior bidirectional object-recognition model; \nopcite{OReillyWyatteHerdEtAl13}).  The consistent observation that these top-down effects emerge just after 100 msec is consistent with the importance of deep-layer updating at the alpha rhythm (and the relative importance of deep-layer projections for top-down activation), which is an essential property of our model.

\subsection{Activation Differences between Predicted and Unpredicted Inputs}

As we review more extensively in the General Discussion section, there is an important difference between our model and many other types of generative models, which postulate the presence of neurons that explicitly code for the mismatch error between the top-down generated model and the bottom-up sensory input \cite{Mumford92,RaoBallard99,KawatoHayakawaInui93,Friston05}.  Under these frameworks, top-down pathways have a net inhibitory effect on lower-level neurons, subtracting away predicted aspects of the signal.  This is the opposite of the excitatory top-down effects just shown above, where top-down excitation can fill in missing elements and shape the representation to accentuate lower-level elements that are consistent with the higher-level interpretation of a scene.

Nevertheless, there are various sources of evidence that have been seen to support these explicit error-coding models, principally the finding of relatively less activation for predicted versus unexpected outcomes \cite[e.g.,]{SummerfieldTrittschuhMontiEtAl08,TodorovicEdeMarisEtAl11,MeyerOlson11,BastosUsreyAdamsEtAl12} (sometimes the opposite result is found; \nopcite{AndersonSheinberg08}).  However, there are a number of alternative mechanisms that can account for this same pattern, and various attempts to systematically evaluate the available evidence have been inconclusive and somewhat mutually contradictory \cite{KokLange15,KokJeheedeLange12,SummerfieldEgner09,LeeMumford03}.  None of these reviews concludes that there is any solid {\em direct} evidence for explicit error coding, including the most recent one \cite{KokLange15}, but they nevertheless reach different overall conclusions based on the overall body of indirect evidence, much of which comes from human neuroimaging studies and is subject to various forms of alternative explanations.

Here, we explore the extent to which our model, which definitely lacks any such explicit error coding neurons, can account for some of the observed patterns of data.  First, to review some of the major alternative explanations, there are well-established temporal dynamics of neural firing that naturally cause neurons to reduce their firing level over time, lasting for different time scales.  As is evident in just about every electrophysiological recording in neocortex (e.g., Figure~\ref{fig.top_down}a,b) neurons typically exhibit a large initial transient burst of activation, followed by a slower decrease in firing rate over the next several hundred milliseconds.  Some of the initial burst may be due to delay in onset of inhibitory feedback mechanisms, and there are also well-documented rapid-onset, transient spike frequency adaptation mechanisms that are essential for accurately capturing pyramidal cell firing patterns \cite{BretteGerstner05,GerstnerNaud09}.  Lasting slightly longer are synaptic depression effects \cite{MarkramTsodyks96,AbbottVarelaSenEtAl97,Hennig13} which can account for several important aspects of neural adaptation \cite{MullerMethaKrauskopfEtAl99}.  At a yet longer-lasting time-scale, fast synaptic plasticity interacting with inhibitory dynamics can account for an overall {\em sharpening} phenomenon across distributed neural representations, where the tuning of active neurons becomes narrower and more selective, while weak, broadly-tuned neurons drop out, resulting in an overall net reduction in neural activation \cite{Desimone96,WiggsMartin98,NormanOReilly03}.  This sharpening dynamic is considered likely to underlie many aspects of the {\em repetition suppression} effect widely-observed in human neuroimaging studies \cite{Grill-SpectorHensonMartin06}, and many of the phenomena typically offered in support of explicit error-coding are also consistent with a sharpening-based account \cite{KokJeheedeLange12,LeeMumford03}.

One clear way in which the above mechanisms could produce a seeming inhibition of inputs that are consistent with a prediction, is if the prediction process drives top-down activation of relevant neural representations {\em in advance of stimulus input}, such that these representations are {\em already} adapted / depressed / sharpened by the time the stimulus arrives.  It is unclear why this kind of effect would {\em not} arise, and it should account for all of the same prediction-dependent phenomena as the explicit error-coding account.  However, our current model does not have any of the above basic adaptation, synaptic depression, or fast synaptic plasticity mechanisms turned on (although all of them are available in our simulator) --- we will more systematically investigate this type of explanation in future work.

Instead, we investigated another possible mechanism behind relatively higher activation levels for unpredicted outcomes, that might help to explain why these effects are more easily seen in human neuroimaging: {\em representational churn}.  When something unexpected happens, a given layer will transition from representing the predicted outcome to then representing what actually happened.  This ``churn'' through representational states, when imaged using something like fMRI which has a long time constant of signal integration, or even faster ERP imaging along with typical aggregation and smoothing processing, can show up as a net overall increase in neural activation, even without instantaneous activation increasing at any given point.  There is a larger ``smear'' of neural activation over time in the unpredictable case compared to a case where a single stable representation is active over time (i.e., the predicted outcome actually occurs).  Any additional suppression of these stable representations over time would only accentuate the magnitude of the difference between unpredicted and predicted, as it would differentially affect the stable predicted representations.

\begin{figure}
  \begin{center}
  \begin{tabular}{ll}
    \parbox[b]{.1em}{a) \vspace*{1.8in}} &
    \includegraphics[height=2.2in]{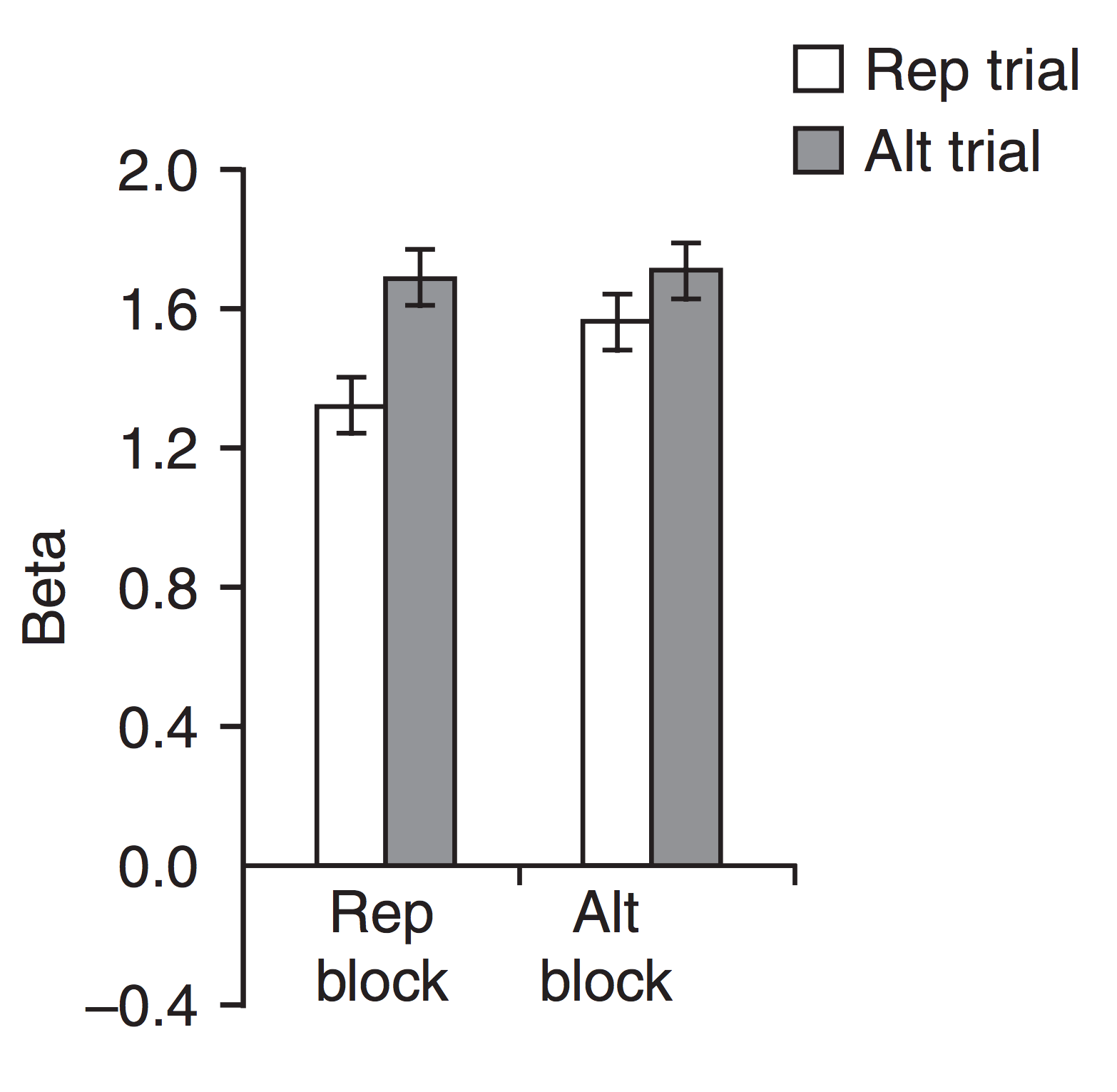} \\
    \parbox[b]{.1em}{b) \vspace*{1.8in}} &
    \includegraphics[height=2.2in]{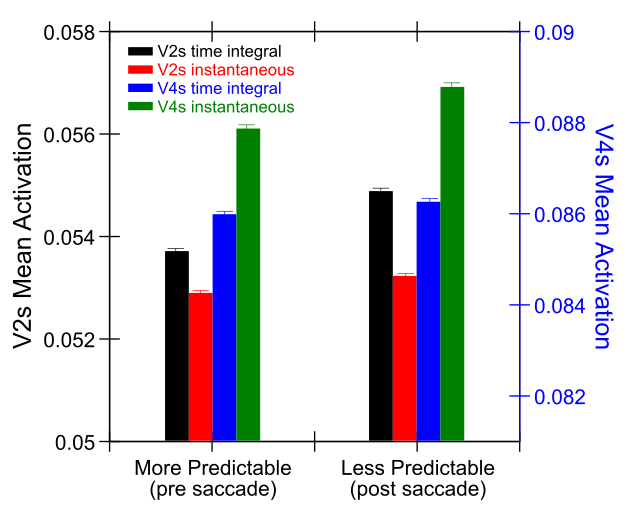}
  \end{tabular}
  \end{center}
  \caption{\footnotesize Activation reductions for more vs. less predictable trials  a) Data from Summerfield et al. (2008) fMRI study, comparing a block-wise manipulation of probability of repetition (75\% for Rep block, 25\% for Alt block).  Repetition suppression is enhanced when repetitions are more expected (Rep block).  b) Results from our model, on the trial before saccade (2nd trial) which is more predictable based on first trial inputs compared to the immediate post-saccade trial (2nd trial), which is less predictable due to the residual difficulty in fully predicting saccade outcome.  The V2s layer shows a significant increase in time-averaged activation across the trial for the less predictable case (black bars), even though this is not seen in instantaneous activations (red).  Higher up in V4 we see the reverse pattern, where instantaneous activation (green) is higher for the less predictable case, but the time-average does not differ --- this is because there is much less churn in V4, but it does perform its own time integral over V2.}
  \label{fig.act_churn}
\end{figure}

Figure~\ref{fig.act_churn}b shows this churn-based effect for layer V2s comparing the more predictable pre-saccade trial (2nd trial) with the post-saccade trial (3rd trial), which is less predictable due to residual difficulty in fully predicting the outcome of the saccade.  Because the V2 layer is highly retinotopically organized, it experiences this representational churn when predictions do not quite align with the new inputs, and the time-averaged activation over this third trial is higher than when it is relatively more stable in the second trial.  This is even though the instantaneous activation (recorded at the end of the trial) is essentially the same.  The same patterns were seen in the V2s (superficial) and V2d (deep) layers (not shown).  In contrast, at the higher, less topographically-organized V4 layer, there is much less difference in churn across the two trials, and the time-averaged activations do not differ.  However, by the end of the third trial, the instantaneous activation is somewhat higher, presumably because V4 is itself integrating over the V2 layer.  This effect is not present in the next higher (TEO) layer (not shown).  

For comparison, Figure~\ref{fig.act_churn}a shows fMRI data from \incite{SummerfieldTrittschuhMontiEtAl08} that has been interpreted as supporting the existence of explicit error coding neurons.  They compared cases where stimulus repetitions (faces) were more or less predictable, and found more of a repetition suppression effect in the more predictable case.  In the context of our model and the churn effect, we would say that people formed stronger predictions of the face repeating in the 75\% repetition block, and when it actually did repeat there was thus less churn compared to when repetitions were less frequent and predictions were weaker.  Interestingly, there was no effect of reducing activation in the alternating case when alternations were 75\% of trials, even though the alternation was more ``predictable'' --- it is impossible to form a concrete prediction for the alternation case, so whatever face does show up there is a surprise from a visual prediction standpoint, and results in equivalent amounts of representational churn.

\begin{figure}
  \centering\includegraphics[width=3in]{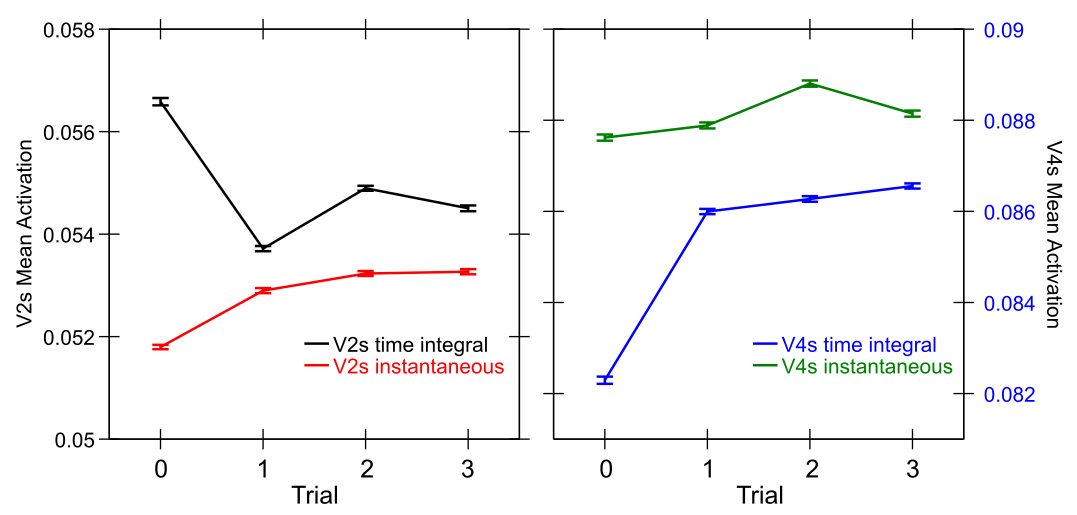}
  \caption{\footnotesize Time-integral and instantaneous activation across all 4 trials for V2s and V4s layers (data in previous graph comes from Trials 1,2).  Overall, trial 3 should be similarly predictable as trial 1, and the activations are consistent with this.  Trial 0 is highly unpredictable, and shows even higher levels of churn in V2s time-integral, while overall having lower instantaneous activity.  V4 is ramping up to its final activation during this trial and thus the time integral is lower.}
  \label{fig.act_churn_all}
\end{figure}

Finally, Figure~\ref{fig.act_churn_all} shows all four trials to give a fuller picture of the activation dynamics, and further evidence that the activation increases selectively in the more unpredictable post-saccade trial (Trial 2, the 3rd trial) compared to both of the surrounding more-predictable trials.  Also, we separated the data according to trials where the object was correctly decoded from TEO from those where it was not (all of the above data are from correct trials).  The error trials overall showed similar patterns of activation, but, interestingly, exhibited a consistent and sizeable reduction activation overall across all the trials (a difference of about .02 in V2s).  This is consistent with the idea that overall network coherence and representational strength is important for accurate performance, as is often found in electrophysiological correlates of behavior.

In summary, these analyses demonstrate a novel origin for observed relative reductions in (time-averaged) activation for more predictable vs. more unpredictable trials.  We anticipate that adding the various forms of repetition suppression mechanisms mentioned above will only increase the strength and robustness of these basic effects, and then it would be appropriate to make a number of more strongly testable predictions from the model.  One clear prediction from the model is that higher brain areas can integrate over ``churn'' present in lower areas, to produce in instantaneous activation that is only present in time-averaged activation at the lower level.  While any small set of data points may be consistent with a variety of models, comparing error vs. correct performance across a variety of trial types, layers, and neural measures should prove strongly constraining.

\section{Attention Mechanisms in Deep / Thalamic Networks}

Finally, although the focus of this paper is on predictive learning, there is another side to the DeepLeabra framework involving the ability of the very same deep / thalamic networks to modulate cortical activation, focusing attention on some elements of a scene and downregulating others.  Biologically and computationally these circuits are synergistic, in that the same mechanisms serve both predictive learning and attentional functions.  More generally, we think there is a larger underlying synergy, where predictions are only made about attentionally-selected objects, and, to perhaps a lesser extent, vice-versa. 

Consistent with this attentional aspect of the model, there is a rapidly-growing literature on the behavioral correlates of alpha-frequency EEG power in humans, along with many demonstrations of alpha-frequency entrainment and phase effects on perception \cite{NunnOsselton74,VarelaToroJohnEtAl81,VanRullenKoch03,KlimeschSausengHanslmayr07,BuschDuboisVanRullen09,MathewsonFabianiGrattonEtAl10,JensenMazaheri10,VanrullenDubois11,PalvaPalva11,RohenkohlNobre11,JensenBonnefondVanRullen12,JensenGipsBergmannEtAl14}.  The working hypothesis for most researchers in the field at this point is that there is a modulation of cortical inhibition at the alpha frequency, and top-down attentional mechanisms can selectively lift this inhibition, resulting in the robust finding of reduced alpha power in brain areas that are under the spotlight of attention, relative to higher alpha power in unattended areas.  Most of this work has taken place in humans, and until recently the detailed biological basis for these effects have been elusive.

There is now a clear biological account emerging, based on careful laminar depth electrode recordings in monkeys, showing that alpha-frequency bursting driven by deep layer (5IB) neurons has a modulatory effect on inhibition throughout the corresponding cortical column \cite{DoughertyCoxNinomiyaEtAl17,vanKerkoerleSelfDagninoEtAl14,BortoneOlsenScanziani14,OlsenBortoneAdesnikEtAl12}.  Specifically, multiple researchers have found that deep-layer alpha sources of local field potential (LFP) modulate spiking of superficial-layer neurons \cite{DoughertyCoxNinomiyaEtAl17,vanKerkoerleSelfDagninoEtAl14,HaegensNacherLunaEtAl11,LakatosKarmosMehtaEtAl08,SpaakBonnefondMaierEtAl12,BollimuntaMoSchroederEtAl11,BollimuntaChenSchroederEtAl08}, consistent with the well-characterized effects of layer 6CT neurons \cite{BortoneOlsenScanziani14,OlsenBortoneAdesnikEtAl12}.  

Thus, the alpha cycle appears to organize both the predictive learning and attentional update dynamics, in a synergistic fashion, with the deep / thalamic network providing an outer loop to the inner-loop of superficial layer constraint-satisfaction processing.  These nested loops can be thought of in terms of the widely-used expectation-maximization (EM) algorithm \cite{DempsterLairdRubin77}.  As elaborated below, the diffuse integrative {\em context} connections within the deep layer (e.g., supported by the 6CC neurons and other broad corticocortical connectivity among these deep neurons; \nopcite{Thomson10,ThomsonLamy07}), are as important for the attentional computation as they were for SRN-like temporal context as described above.

\begin{figure}
  \begin{center}
  \begin{tabular}{ll}
    \parbox[b]{.1em}{a) \vspace*{1.8in}} &
    \includegraphics[width=3in]{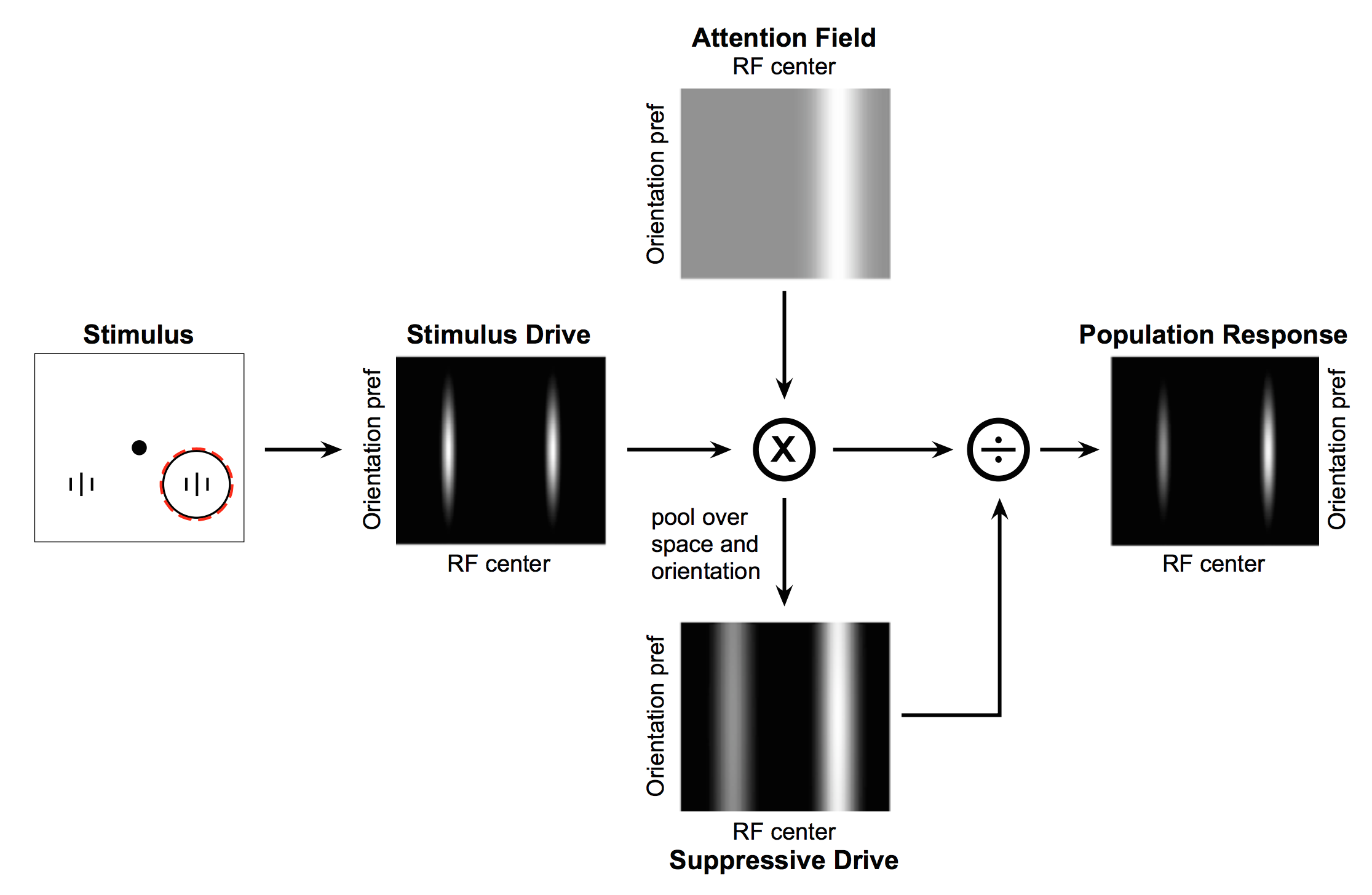} \\
    \parbox[b]{.1em}{b) \vspace*{2in}} &
    \includegraphics[width=2.5in]{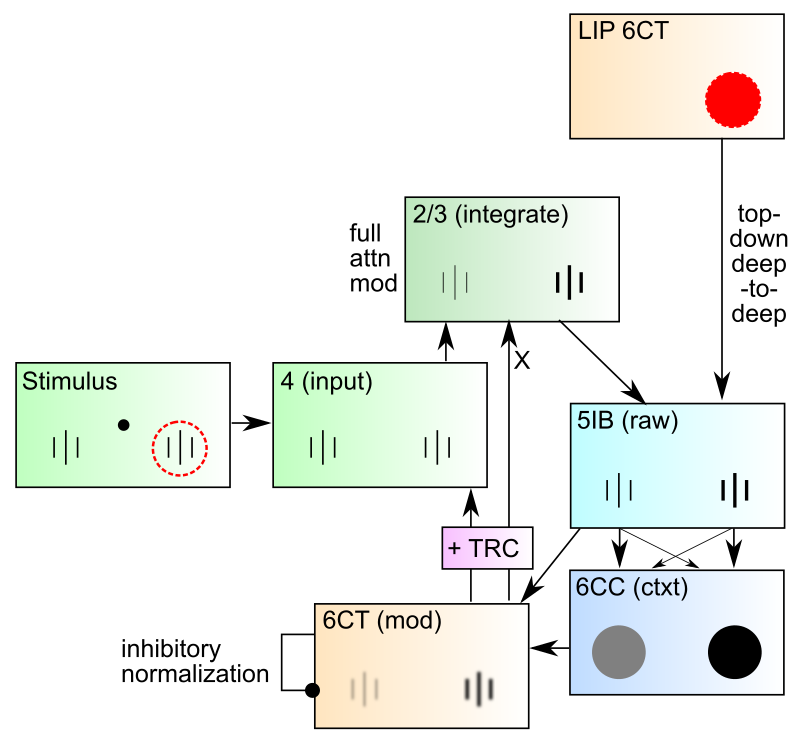}
  \end{tabular}
  \end{center}
  \caption{\footnotesize a) The Reynolds \& Heeger (2009) computational model of pooling and normalization processes in attention.  b) How attentional modulation is computed across the deep layers in DeepLeabra, in response to a top-down attentional focus (as encoded in LIP of parietal cortex).  Layer 4 receives bottom-up sensory input (initially equally weighted), which then drives superficial layers (2/3), which initially do not reflect the attentional modulation (not shown).  The deep 5IB neurons integrate deep-to-deep top-down attentional inputs from LIP plus the local stimulus features from 2/3, to produce the {\em raw} deep output, prior to the contextual normalization process.  The 6CC neurons integrate across the 5IB activations (context integration or pooling).  6CT then integrates this contextual and direct activation from 5IB, to produce, for the first time in the circuit, a properly renormalized multiplicative gain-field activation pattern, with surround inhibition both within the 6CT layer and further downstream in the TRN and TRC circuit providing the critical renormalization process.  These 6CT activations then modulate (multiply) the superficial-layer activations to produce {\em both} an increase the attended location, and a decrease for the unattended location, as shown.  In the biology, this modulation affects the layer 4 inputs (not shown) as well as 2/3.  Our model subsumes layer 4 into layer 2/3 neurons. }
  \label{fig.attn_compute}
\end{figure}

We found that our model captures the essential computations of the well-validated, abstract mathematical model of attention from \incite{ReynoldsHeeger09}, as shown in 
Figure~\ref{fig.attn_compute}.  Working backward from the 6CT modulatory layer, we posit that this layer encodes a final normalized attentional mask that has an overall multiplicative or gain-field effect on neural activations in the superficial network, which is consistent with relevant data \cite{BortoneOlsenScanziani14,OlsenBortoneAdesnikEtAl12,DoughertyCoxNinomiyaEtAl17,vanKerkoerleSelfDagninoEtAl14}.  Thus, where activations are strong in this layer, the corresponding superficial layer activations will remain strong, but where they are weaker, the superficial layer activations will be reduced.  The normalization in 6CT occurs via inhibitory feedback circuits, both locally within layer 6 and through the TRN and TRC circuits of the thalamus (which then feed back into 6CT as well).  This normalization process is affected by the 6CC layer prior to 6CT, which does the pooled integration over space and features, and then feeds into 6CT.  One step prior, area 5IB combines local stimulus features and the top-down attentional inputs from higher-level areas (e.g., LIP in this case, which has been shown to support spatially-organized attentional activations; \nopcite{BisleyGoldberg10}).  Thus, all of the same essential computations from the \incite{ReynoldsHeeger09} model can be performed across these different deep layers.

\begin{figure}
  \begin{center}
  \begin{tabular}{ll}
    \parbox[b]{.1em}{a) \vspace*{2in}} &
    \includegraphics[height=2.2in]{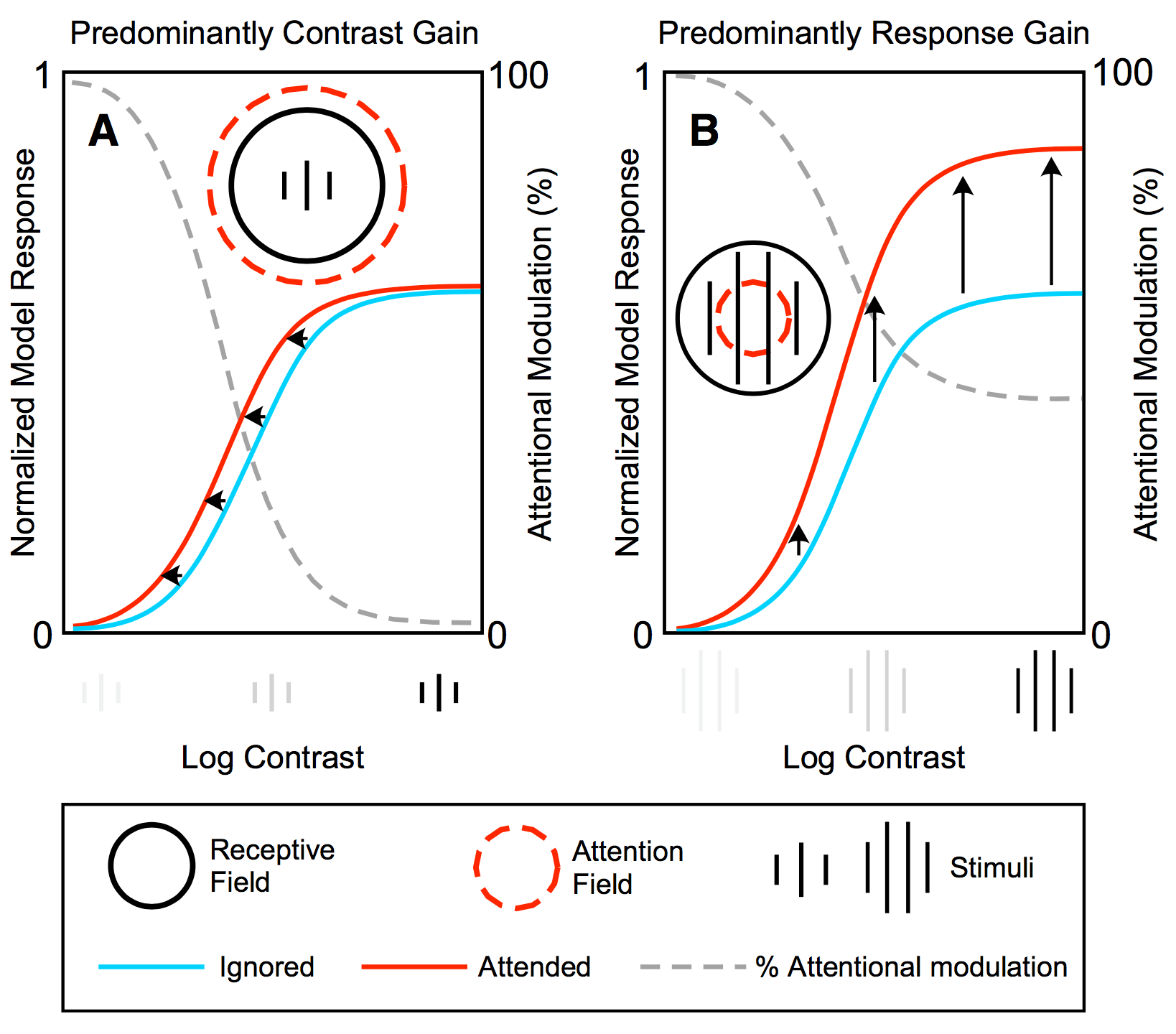} \\
    \parbox[b]{.1em}{b) \vspace*{2in}} &
    \includegraphics[height=2.2in]{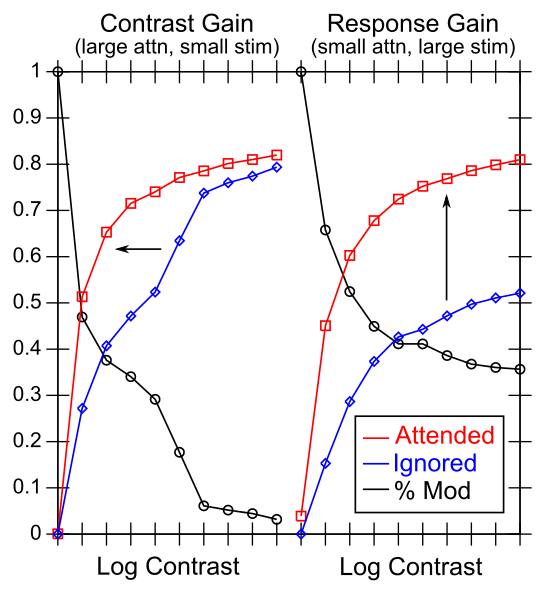}
  \end{tabular}
  \end{center}
  \caption{\footnotesize a) Key data accounted for by Reynolds \& Heeger (2009) model, showing two qualitatively different types of attentional modulation that can emerge from the same model, as a result of differences in size of attentional spotlight relative to stimulus size.  These different effects, which fit experimental data, result directly from the pooling and normalization processes, and are thus a key test of the model dynamics.  b) Results from a DeepLeabra model driven by large LIP attentional top-down spotlight relative to a small bottom-up stimulus (left) versus a small LIP spotlight relative to a larger stimulus, reproducing same qualitative effects. }
  \label{fig.attn_data}
\end{figure}

Figure~\ref{fig.attn_data} shows that our model captures the same key data as the \incite{ReynoldsHeeger09} model, where the relative balance of the enhancing vs. suppressive effects of attentional modulation can shift depending on the relative sizes of the attentional spotlight and the stimulus input (and as a function of stimulus contrast), producing the shift from contrast gain to response gain effects of attention.  Thus, although there is much more work to be done here to explore the full range of attentional dynamics, this provides a solid foundation building on the well-established \incite{ReynoldsHeeger09} model.  Furthermore, our model is related to the {\em folded-feedback} model of \incite{Grossberg99} (see \nopcite{RaizadaGrossberg03} for a more elaborated version), which also posits this same kind of attentional modulation dynamic between layer 6 and the superficial layers.  Interestingly, top-down attentional signals, like those coming from LIP down to lower-level visual pathways, are preferentially communicated via a network of deep-to-deep projections \abbrevcite{MarkovVezoliChameauEtAl14,vonSteinChiangKonig00,vanKerkoerleSelfDagninoEtAl14}.

In a future paper, we plan to apply our model to a wide range of alpha-frequency effects on perception and attention \cite{NunnOsselton74,VarelaToroJohnEtAl81,VanRullenKoch03,KlimeschSausengHanslmayr07,BuschDuboisVanRullen09,MathewsonFabianiGrattonEtAl10,JensenMazaheri10,VanrullenDubois11,PalvaPalva11,RohenkohlNobre11,JensenBonnefondVanRullen12,JensenGipsBergmannEtAl14}, to better understand how deep, thalamic, and superficial-layer dynamics interact to produce these effects, and how predictive learning and attention interact as well.  Is it possible that some of these effects could be driven just by the alpha-frequency context updating in the predictive learning aspect, or are they all due to attentional modulation effects?  What causes the alpha phase to reset \cite{CalderoneLakatosButlerEtAl14}, and how does the interplay between intrinsic oscillatory dynamics and external driving stimuli work?   What about the effects of saccades, which also appear to reset the alpha phase  \cite{MelloniSchwiedrzikRodriguezEtAl09,ParadisoMeshiPisarcikEtAl12,MaldonadoBabulSingerEtAl08,RajkaiLakatosChenEtAl08,ItoMaldonadoSingerEtAl11}?

One major goal of this work would be to provide a more satisfying integration of the inhibitory versus excitatory effects of alpha modulation \cite{PalvaPalva11,PalvaPalva07,GulbinaiteIlhanVanRullen17}.  In our model (and \incite{ReynoldsHeeger09}), the final modulatory signal carried by layer 6CT neurons is excitatory (these are excitatory pyramidal neurons), and its multiplicative effect on other neurons is hypothesized to result from an interaction between excitatory and inhibitory circuits.  \incite{BortoneOlsenScanziani14} clearly demonstrate that 6CT neurons strongly activate inhibitory interneurons in layer 6 that synapse throughout the cortical column, providing a strong overall background of inhibition.  However, they are also careful to emphasize that there are many excitatory synapses onto other pyramidal and thalamic TRC neurons, that can have the opposite effect, and the net overall effect is likely to depend critically on spatial topography (e.g., surround inhibition with central excitation) and also the local activity levels of the receiving neurons.  Furthermore, it is not clear how the known biological mechanisms would cause the level of inhibition of superficial spiking to be a direct function of overall EEG-level alpha power, as many theories assume \cite[e.g.,]{KlimeschSausengHanslmayr07,JensenGipsBergmannEtAl14}.  Instead, it is certain that gamma power increases when superficial neurons are disinhibited, which may directly remove some alpha power that they would otherwise have been contributing, and these superficial neurons may also have a consequent impact on the level of alpha synchrony in the deep layers (e.g., by affecting the timing of 5IB bursting).  Thus, the causal arrow may go the other way, consistent with various issues raised by \incite{PalvaPalva11,PalvaPalva07}.

\section{General Discussion}

We have presented a comprehensive model of the visual system that demonstrates how predictive learning within a generative framework leads to high-level invariant object representations without any external training signal.  The model follows known biology and accounts for data across many levels of analysis, from low-level synaptic plasticity to systems-level organization and connectivity of the areas and pathways of the visual system, including the development of these pathways.  The pulvinar nucleus of the thalamus plays a central role as a kind of projection screen, upon which the different visual areas across levels of abstraction collaboratively project their predictions for what the visual input will look like when the next alpha-frequency (100 msec) 5IB driver inputs provide their ground truth plus-phase training signal.  The pulvinar broadcasts back out to all the areas that contribute to it, enabling neurons everywhere to learn based on the temporal difference between the minus-phase prediction and plus-phase target.  Synaptic plasticity mechanisms capable of using this temporal difference were derived directly from a biophysically detailed model of spike-timing dependent plasticity \cite{UrakuboHondaFroemkeEtAl08}.  Computationally, the direct and indirect propagation of this prediction error signal produces powerful error-backpropagation learning, capable of shaping deep hierarchies of representations to minimize the prediction error.

The collective prediction error signal from the pulvinar is partitioned into three separable components by three different visual pathways: {\em Where}, {\em What}, and {\em What * Where} integration, through a combination of developmental sequencing and emergent dynamics of learning shaped by specific patterns of interconnectivity.  This allows compact, high-level, abstract representations at the top of each of these pathways to drive low-level predictions, which is essential for successful predictive learning, as the lower-level areas are too retinotopically diffuse to provide effective predictive representations over time.  The particular developmental and connectivity constraints that emerge from these principles, along with the results of extensive experimentation in our model, align remarkably well with available data on the primate visual system.  

To summarize, here are some of the major, well-established biological properties that are central to our model (along with many other details enumerated throughout the paper):
\begin{itemize}
\item The existence of a strong synchronized, low-frequency modulation of cortex (at the alpha frequency).
\item Specificity of this alpha modulation to deep layers and thalamus, as opposed to superficial layers.
\item Nature of deep-layer connectivity to pulvinar, specifically having {\em both} a numerous, weaker, plastic pathway (for generating a prediction) and a sparse, strong, fixed pathway (for providing a {\em ground truth} target).
\item Synchronization of this strong pathway input with the alpha cycle.
\item Broad connectivity of pulvinar with different visual pathways (afferent and efferent).
\item Lack of direct bottom-up superficial projections into the deep layers, but presence of these projections top-down.
\item Bidirectional (top-down and bottom-up) connectivity between superficial layers.
\item Early development of the {\em Where} (MT, LIP) pathway.
\item Organization into three separable (yet highly interconnected) visual pathways, particularly a third putative {\em What * Where} integration pathway.
\end{itemize}

While there are various other theoretical interpretations of each of these different phenomena, we are not aware of another framework that ties together all these different elements under an overarching computational model.  Furthermore, we argue that our model provides a theoretical continuity between levels of analysis that have previously not been well-aligned.  For example, biologists tend to think that the brain learns using Hebbian learning mechanisms, but computationally these are very limited, and computer scientists have overwhelmingly embraced error-driven backpropagation models instead.  However, error-backpropagation is widely regarded as biologically implausible for a variety of reasons \cite[e.g.,]{Crick89}, not all of which are resolved by local, activation-based versions \cite{OReilly96,Movellan90,XieSeung03,ScellierBengio17}.  One of the most important unresolved such issue is the question of where the error signals actually come from to drive backpropagation --- current models rely extensively on large human-labeled datasets.  Thus, the ability of our model to provide a biologically-sound framework for powerful error-backpropagation learning using only raw sensory streams, through the principle of predictive learning, establishes a clear theoretical continuity between levels.

As such, we offer it as a possible answer to the longstanding mystery of how the neocortex develops and learns over the first several months of life to produce the foundations of all our high-level cognitive abilities.  In particular, the finding that this purely self-organizing predictive learning process, in combination with all the systems-level structure in which it is embedded, can form systematic invariant object representations characteristic of those found in TEO and other IT areas, provides a foundation for subsequent word learning and language development.  We are excited to extend our model with auditory pathways, to understand how combined multi-modal predictive learning across vision and audition interact in this next level of cognitive learning (which also likely shapes the nature of visual learning in important ways not captured in the present model).  Preliminary work in this direction using earlier versions of our predictive learning framework suggests that the auditory pathway is highly amenable to predictive learning approaches in general, due to the intrinsically temporal nature of auditory signals, consistent with the success of predictive learning frameworks in linguistic datasets \cite{Elman90,Elman91,MikolovSutskeverChenEtAl13}.

In the remainder of the discussion, we compare this framework with other related frameworks, consider some broader implications of our approach, and then highlight a few of the many central testable predictions from our model, followed by a further discussion of a number of unresolved questions for future research.

\subsection{Comparison with other Frameworks}

\subsubsection{Generative Models}

Our framework fits within the broader context of {\em generative models} in psychology and neuroscience, which embody the principle of {\em recognition by synthesis}, which goes back at least to Helmholtz in 1867 \incite{Helmholtz13}.  This idea was advanced by a number of different researchers in various ways in the 1990's as a possible way of understanding neural function \cite{Mumford92,KawatoHayakawaInui93,Ullman95,DayanHintonNealEtAl95,RaoBallard99}, with \incite{CarpenterGrossberg87} having a somewhat different but related earlier framework.  Common to most of these frameworks is the notion of a hierarchy of areas stacked upon each other, with higher layers having more abstract, compact internal models of the environment, and some kind of interplay between a feedforward pathway of sensory information flowing up this hierarchy, and a feedback pathway driving top-down signals based on internal generative models.

Most of these models \cite{Mumford92,KawatoHayakawaInui93,DayanHintonNealEtAl95,RaoBallard99} adopt an {\em explicit error-coding} framework, where certain neurons explicitly subtract the top-down model-based signals from the bottom-up sensory-driven signals, to represent the mismatch between the two (while another population represents the accumulated top-down prediction itself).  This error signal is typically fed forward to higher layers, which then use it to adjust their current model parameters to better fit with the sensory inputs, in an iterative process.  Somewhat confusingly, these error signals are sometimes referred to as ``prediction errors'' but this sense of the word prediction does not typically include the critical ``about the future'' aspect --- they are usually just static ``predictions'' of the current sensory inputs, from the generative model (a more appropriate term would be {\em generative errors} or something to that effect).  \incite{Mumford92} hypothesized that the neocortical superficial layer neurons encode this error signal and project it feedforward, while the deep layers transmit the model-based predictions top-down --- this same idea was also advocated by others \cite{RaoBallard99,KawatoHayakawaInui93}.  \incite{CarpenterGrossberg87} adopted a more discretized, localist version of this process, where a single upper-layer neuron is activated (representing the internal model), and the degree of mismatch between its top-down weights and the current stimulus is used, with a sensitivity threshold, to determine whether to keep that neuron active, or select a new one to encode the current input stimulus.  

The hierarchical generative model idea was embraced and further developed with the subsequent popularity of the Bayesian framework, where it has a direct and clear relationship to the key Bayesian twist  \cite[e.g.,]{LeeMumford03,Friston05,YuilleKersten06,Friston08,Friston10,Lee15}.  This Bayesian twist turns a question about how likely various hypotheses (models) are given some observed data, into the question of how likely the {\em data} is given various hypotheses (i.e, the generative model).  The latter form is typically much easier to compute, and inference (going from the data to the model) can be performed by adapting the model to more closely generate the observed data, as proposed in these early neural models.  In machine learning and statistics, a generative model has a more formal definition in terms of capturing the full probability distribution of the data, and a well-defined formal probabilistic structure such that it truly can generate plausible data {\em de novo}.  In contrast, our use of the term as a model of brain function is much looser, including all such models that include any aspect of a generative process, such as neural network auto-encoders.

In contrast to the above models, the counter-streams model of \incite{Ullman95} holds that the feedforward and feedback pathways are collaborative and amplify areas of congruence or match between top-down and bottom-up pathways.  This is more in the spirit of the bidirectional constraint satisfaction framework that is a foundation of our approach, based on earlier frameworks developed in the 1980's \cite{Hopfield82,Hopfield84,AckleyHintonSejnowski85,RumelhartMcClelland82}.  In this overall framework, the activation states for both the superficial and deep layers {\em always represent the best guess internal representation of the sensory inputs}, not a difference or error signal.  This allows both top-down and bottom-up signals to converge on shaping these internal representation states in a collaborative way (i.e., bidirectional constraint satisfaction), instead of positing a fundamentally subtractive or contrastive relationship between the bottom-up and top-down pathways.  As we have demonstrated, this excitatory, collaborative influence of top-down inputs is critical for allowing high-level abstract representations to shape accurate low-level predictions in our model.

In sum, there is a fundamental division between frameworks based on the principle that bottom-up and top-down streams have a net subtractive, mismatch-coding relationship, versus those based on a more collaborative, match-amplification dynamic between the two streams (the deep layers in the mismatch-coding generative models do exhibit this match-amplification property, so the contrast here is focused specifically on the hypothesized superficial error-coding neurons).  Computationally, there may be a critical difference between these approaches in terms of how effectively they converge on an interpretation of the current sensory input.  Intuitively, this difference corresponds to the difference between the ``Yes, and..'' approach to collaborative problem solving, versus the ``No, but..'' approach, as highlighted in a popular book by comedy writers \cite{LeonardYorton15}.  The collaborative, positive approach brings {\em all} of the available constraints (top-down and bottom-up) to bear on rapidly converging on a reasonable interpretation.  In contrast, the error-based generative models are dominated by critical negative input from the top-down pathway, which is great for eliminating bad interpretations but not for collaboratively finding good ones.  Also, the strictly hierarchical nature of most generative models, where each layer serves exclusively as the model for the layer below it, may make the inference process more difficult.  In contrast, all of the different levels of abstraction in our model collaborate together to produce a single integrated prediction, projected onto the pulvinar ``silver screen of the Cartesian theater.''  The broad projections from pulvinar back to cortex then share this developing prediction with all the relevant contributing layers, helping to coordinate all levels together simultaneously, instead of each working separately on their own relatively isolated problem.

Instead of using error signals during the online inference process, we think they are more effectively used to guide the learning process, which takes place over a much longer time period, and only needs to converge once.  Here, the stochastic gradient descent process embodied by the error backpropagation algorithm has consistently proven its value as a way of optimizing learning in deep hierarchical networks.  

Biologically, we reviewed above the evidence bearing on whether superficial layer neurons in the neocortex encode prediction errors, and showed that our model can account for the key finding of reduced activation for predicted relative to unpredicted events.  This and other alternative accounts of the main indirect evidence for explicit error-coding neurons, together with the notable lack of any solid direct evidence for this central hypothesis of most generative model frameworks, should be sufficient to render such a framework biologically implausible at best.  More generally, there are so many detailed electrophysiological recordings of neurons throughout the cortex showing that neural firing positively encodes representations of the current environment, that it seems rather unlikely there could be a large population of explicit error-coding neurons lurking in there somewhere.  Furthermore, the idea that feedback projections are inhibitory is at odds with the basic anatomy, where all long-range connections in the neocortex are excitatory \cite{JohnsonBurkhalter97,ShaoBurkhalter96}, and the excitatory nature of these top-down connections is compatible with the well-supported biased-competition model \cite{DesimoneDuncan95,MillerCohen01}.  Although there are ways of reshuffling connections to make biased-competition and generative models more mathematically consistent \cite{Spratling08}, this approach still retains the requirement of inhibitory top-down connections (biased competition is made to be more like a generative model, where lateral pooled inhibition is replaced with top-down inhibition, and also activations and synapses that can be either positive or negative), which \incite{Spratling08} acknowledges are biologically implausible.


In summary, although our framework shares the overall generative model goal, it achieves this goal in a fundamentally different way from most generative models, which we argue has both computational and biological plausibility advantages.   Furthermore, our model is distinct in being architecturally founded on making true predictions about the future, instead of just re-generating the current sensory inputs.  Despite these differences, it is likely that many of these theorists would recognize our model as fitting well within their broader vision for how neocortex works.

\subsubsection{Deep Auto-encoder Neural Networks}

The Restricted Boltzmann Machine (RBM) framework \cite{Hinton02,HintonSalakhutdinov06,Hinton07a} represented a critical bridge between the Bayesian generative model framework, and the now-dominant resurgence of neural network models.  The RBM was derived from a mathematically well-characterized generative-model framework, but required a final training phase using error backpropagation.  Eventually, it became apparent that the initial RBM training could be skipped entirely, with the development of various important tricks for making deep (i.e., having many hidden layers) models converge effectively \cite{CiresanMeierGambardellaEtAl10,CiresanMeierSchmidhuber12,KrizhevskySutskeverHinton12,BengioCourvilleVincent13,LeCunBengioHinton15}.  One of the most important such tricks is the use of weight sharing among topographically organized groups of units in lower layers, which mathematically is the same as {\em convolution} by a filter defined by this set of shared weights \cite{LeCunBoserDenkerEtAl90,LeCunBengioHinton15}.

Most of the deep neural networks (i.e., {\em deep nets}) are trained to produce localist category labels for bitmap images, and do not include generative-model aspects.  Nevertheless, these models appear to capture some important properties of the ventral {\em What} pathway \cite[e.g.,]{MajajHongSolomonEtAl15}, building on insights from earlier more neuroscience-inspired frameworks \cite{RiesenhuberPoggio99}.  However, they require vast amounts of hand-labeled image data, and are thus not plausible models of the largely self-organizing nature of human visual learning.  Indeed, we argue that these models are somewhat like powerful 3D printers, that instead print brain circuits mimicking those in the human brain.  Their performance is proportional to the sample size of human behavior available (e.g., number of samples of human object categorization applied to a wide range of images), which is analogous to how fine-grained the scan of an object is for a 3D printer --- the finer the scan, the more accurate the reproduction.  Because the mapping function from image to object label present in human brains is very high-dimensional, a very large number of samples is needed to reproduce it accurately.  To continue the analogy, a deep convolutional neural net also constitutes a good raw material to ``render'' in, as it starts out with structural biases etc that match those of the visual system.  And, several tricks that improve performance are also biologically-supported properties such as winner-take-all learning and pressure to develop sparse representations, which are also included in our Leabra framework.  By contrast, our model represents an attempt to reconstruct the complex interactive dynamics that shape the human visual system based on raw visual input, without relying on any direct sampling of the mature system.

There has also been some renewed focus on deep versions of auto-encoder models, which are the neural network equivalent of a generative model \cite{BengioYaoAlainEtAl13,Valpola14,RasmusBerglundHonkalaEtAl15,LeMongaDevinEtAl12}.  Many of these models adopt a denoising training strategy to prevent the model from just learning a degenerate ``copy the input'' strategy \cite{BengioYaoAlainEtAl13}, and include a strongly hierarchical outside-in training strategy in the form of a {\em ladder} network \cite{Valpola14,RasmusBerglundHonkalaEtAl15}.  Very recently, this auto-encoder paradigm has been extended into a true predictive learning framework like that in the present model \cite{LotterKreimanCox16}.  This model is trained in a purely unsupervised manner on movies, predicting the next frame, which is effectively what we are doing.  The model learns to generate realistic-looking images and achieves overall good predictive error scores.  The analysis of the internal learned representations focused on lower-level visual parameters such as camera pan and roll, and there did not appear to be any invariant object representations that self-organized.  The model was also trained to decode faces using subsequent supervisory training, with similar overall results to comparable auto-encoders.

Thus, there are considerable similarities at a broad level between these models and our framework, but overall these models are more closely aligned with traditional Bayesian generative models than our framework.  For example, they adopt a strict hierarchical structure to the layers, with each higher layer attempting to encode the layer below it, instead of the multi-pathway, collaborative-across-levels approach characteristic of our model.  Furthermore, they do not typically include any bidirectional constraint satisfaction processing, so the inference process is strictly feedforward.  Finally, these models are not used in a purely self-organizing manner --- the final step is generally to train on standard human-labeled supervised datasets, and the key measure of interest is the extent to which the auto-encoder pretraining reduces the amount of supervised training required to achieve a given level of performance.

Biologically, there has been a long history of skepticism about the biological plausibility of error-driven backpropagation learning \cite[e.g.,]{Crick89}.  As noted earlier, we have long argued that these issues can be overcome through the use of bidirectional excitatory connectivity and temporal-difference based synaptic plasticity, which closely approximate error backpropagation \cite{OReilly96} (see also \nopcite{Movellan90,XieSeung03,ScellierBengio17}).  Furthermore, we have shown how models using these learning mechanisms can learn like these other deep neural networks, while also exhibiting important bidirectional dynamics \cite{OReillyWyatteHerdEtAl13,WyatteHerdMingusEtAl12,WyatteCurranOReilly12}.  

\subsubsection{Forward Models}

A major, well-established application of predictive learning is for {\em forward models} that predict the outcome of actions \cite{KawatoFurukawaSuzuki87,JordanRumelhart92,MiallWolpert96}.  The LIP predictive remapping from saccades is really a form of forward model (predicting the next sensory state that follows from the motor action of moving the eyes), and our model advances the idea that every area of cortex has a deep-layer forward model associated with it.  Besides driving the self-organization of the entire visual system, one might ask what other potential benefits all these forward models might have?  One popular idea is that they can be used to select actions that achieve desired outcomes, by effectively running them backward \cite{Hommel04,James90,PezzuloCastelfranchi09,Friston10}.  Although this {\em ideomotor} principle is attractive, it is not clear if it is tractable for realistic motor actions \cite{HerbortButz12,JordanRumelhart92}.  We are particularly skeptical of prevalent models that hypothesize long sequences of chained predictions to generate action plans \cite{BurgessOKeefe97,PastalkovaItskovAmarasinghamEtAl08,LismanRedish09}.  Such chains are only as strong as their weakest links, and the working memory demands required to keep such a process going seem excessive, especially for rodents.  Instead, we suggest that one-step predictions can be generated over many different time scales, and particularly in the prefrontal cortex, longer-time-scale predictions of outcomes are used to guide planful action \cite{OReillyHazyMollickEtAl14,OReillyPetrovCohenEtAl14,OReillyHazyHerd15}.  Nevertheless, it is plausible that the same basic predictive learning mechanisms exploited in posterior cortex for fast-time-scale predictive learning could also be important for these longer-time-scale learning processes in frontal areas.  

Due to the simple one-to-one retinotopic nature of saccade motor plans relative to the current visual input, this domain does not capture the more general challenges in motor learning.  Therefore, we plan to explore the motor control implications of pervasive predictive learning in the context of the auditory pathway, including predicting the effects of speech output, to study the process of learning to imitate speech sounds, as has been explored using forward models \cite{GuentherVladusich12}.

One major issue raised in this context is the relationship between the hypothesized forward models learned in the cerebellum \cite{WolpertMiallKawato98,Verduzco-FloresOReilly15,Shadmehr17} relative to those in the neocortex.  Although both systems may be learning predictive models, the cerebellum appears to be specialized for shorter, faster time scales of motor control (e.g., with around 10 msec resolution).  Furthermore, differential effects of cerebellar lesions early vs. later in life suggest that the cerebellum serves to shape learning in the neocortex, which can then take on much of the learned functionality.  The primary cortical output of the cerebellum goes to frontal and some parietal thalamic areas \cite{StrickDumFiez09}, so it may teach cortex by providing a plus-phase training signal, thereby plugging directly into the same learning system described here (similar to the superior colliculus inputs to the second pulvinar map as mentioned above; \nopcite{Shipp03}).  We will investigate this possibility in future work.

\subsubsection{Hawkins' Model}

The importance of predictive learning and temporal context are central to the theory advanced by Jeff Hawkins \cite{HawkinsBlakeslee04}.  This theoretical framework has been implemented in various ways, and mapped onto the neocortex \cite{GeorgeHawkins09}.  In one incarnation, the model is similar to the Bayesian generative models described above, and many of the same issues apply (e.g., this model predicts explicit error coding neurons, among a variety of other response types).  Another more recent incarnation diverges from the Bayesian framework, and adopts various heuristic mechanisms for constructing temporal context representations and performing inference and learning.  We think our model provides a computationally more powerful mechanism for learning how to use temporal context information, and learning in general, based on error-driven learning mechanisms.  At the biological level, the two frameworks appear to make a number of distinctive predictions that could be explicitly tested, although enumerating these is beyond the scope of this paper.

\subsubsection{Granger's Model}

Another model which has a detailed mapping onto the thalamocortical circuitry was developed by Granger and colleagues \cite{RodriguezWhitsonGranger04}.  The central idea behind this model is that there are multiple waves of sensory processing, and each is progressively differentiated from the previous ones, producing a temporally-extended sequence of increasingly elaborated categorical encodings ({\em iterative hierarchical clustering}).  The framework also hypothesizes that temporal sequences are encoded via a chaining-based mechanism.  In contrast with the DeepLeabra framework, there does not appear to be a predictive learning element to this theory, nor does it address the functional significance of the alpha frequency modulation of these circuits.

\subsubsection{Other Frameworks for Cortical Oscillations}

There have been a number of different computational functions ascribed to cortical oscillations and synchrony, which are not reflected in our model.  Perhaps the most influential such idea is that different phases of cortical synchrony can support multiple interleaved {\em bindings} of separate features \cite[e.g.,]{WangBuhmannvonderMalsburg90,GrayEngelKonigEtAl92,EngelKonigKreiterEtAl92,ZemelWilliamsMozer95,HummelBiederman92}.  We have argued against such models in favor of coarse-coded distributed representations that naturally support binding without requiring an elaborate and brittle synchrony-based mechanism that ultimately requires decoding mechanisms that obviate most of the benefit of the binding in the first place \cite{OReillyBusby02,OReillyBusbySoto03,CerOReilly06,OReillyPetrovCohenEtAl14}.  The function of cortical oscillations in the current model serve instead to coordinate and organize the entire distributed network, which is generally widely accepted and uncontroversial.  We have also developed models of the role of the theta rhythm in the hippocampus \cite{KetzMorkondaOReilly13}, and the beta rhythm in the basal ganglia (BG) and prefrontal cortex (PFC) \cite{KetzJensenOReilly15,OReillyPetrovCohenEtAl14,JilkLebiereOReillyEtAl08}.

Briefly, we think that the hippocampal episodic memory system integrates over two alpha cycles in its theta frequency (5 hz, 200 msec) encoding and retrieval cycle, while the BG/PFC system operates at a faster cycle rate (beta = 20 hz, 50 msec) to allow more rapid behavioral responding and updating of working memory representations.  Interestingly, the 50 msec time frame for BG function was independently established in the ACT-R model based on fitting behavioral data \cite{StoccoLebiereAnderson10,AndersonLebiere98,JilkLebiereOReillyEtAl08}.  These functional roles contrast with the influential model of Lisman and colleagues, based on the numerical observation that 8 or so 40 hz gamma cycles can be embedded in one theta cycle, which seemed to correspond to the ``magic number 7'' working memory capacity constraint \cite{IdiartLisman95,LismanJensen13}.  However, outside of specialized phonological processing pathways, the pervasive representational capacity of any given brain area appears to be more like 2-4 \cite{Cowan01}, and may have more to do with use of the two different hemispheres plus the ability to (barely) support at most two different distributed representations within a given area \cite{BuschmanSiegelRoyEtAl11}.

\subsubsection{Hinton's Joint View and Object Model}

One of the major ideas behind our model is that the spatial and object pathways must be jointly active and learning to generate predictions about what will happen next.  A related idea was proposed by \incite{Hinton81}, who advocated solving the joint spatial configuration and object identification problems at the same time, with the goal of producing a canonical object representation that would then be easier to recognize.  However, the ill-posed and very high-dimensional nature of this problem proved intractable.  Our approach avoids these problems by {\em first} developing the spatial prediction pathway independent of object recognition, using abstracted spatial blob representations, which is entirely tractable and easily learned.  Then, we do not require a canonical object representation, but rather rely on well-established principles of hierarchical topographic connectivity to develop invariant object representations in the high levels of the {\em What} pathway \cite{Fukushima80,RiesenhuberPoggio99,OReillyWyatteHerdEtAl13}.

\subsubsection{Mumford's Models}

David Mumford's early theoretical papers on the thalamus and cortex come the closest overall to capturing the central ideas in the current model, including the notion of the pulvinar as a kind of blackboard \cite{Mumford91} and the cortex as a generative model \cite{Mumford92}.  Although we only read these papers after developing our model, and there are many important differences in our approaches, the degree of concordance at the big-picture level is nevertheless remarkable.

\subsection{Broader Implications of our Framework}

Next, we consider a few of the most important broader implications of our framework.

\subsubsection{Nature vs. Nurture in Development}

There are many important developmental implications for a predictive learning approach in general \cite[e.g.,]{ElmanBatesKarmiloff-SmithEtAl96,MunakataMcClellandJohnsonEtAl97}, and, as noted above, for the specific developmental requirements of our what-where-integration model.  First, if you have a learning process that operates at a rate of 10 times per second, then a great deal of learning can accumulate very quickly.  For example, the full sequence of training used in our model would represent just 21 hours of real-time learning at this rate.   Of course, real-world environmental events may not be quite as dense a source of learning opportunities, and babies are certainly not awake very much at the start, but nevertheless it seems likely that a huge amount of predictive learning could be acquired by 4 months, when various studies indicate that babies have a decent understanding of basic physics \cite[e.g.,]{Spelke94,KellmanSpelke83}.  Thus, this knowledge, which has been characterized as innate core knowledge \cite{Spelke94}, may well be better described as learned.  Nevertheless, given the ubiquitous nature of physics, coupled with genetically-coded learning mechanisms and developmental wiring processes, it is likely inevitable that all neurologically-intact babies will develop the same systematic predictive knowledge of this basic physics, so for all practical purposes, it certainly seems to be innate.  Thus, the utility of simplistic nature vs. nurture dichotomies must be entirely rethought in the context of strong interactions between genetically-specified features of the brain and experience-expectant learning mechanisms \cite[e.g.,]{ElmanBatesKarmiloff-SmithEtAl96,GreenoughBlackWallace87}.

In any case, there is now a great opportunity to explore more detailed data on the development of visual expectations about the world, using a more advanced version of our model and environment that contains multi-body interactions of various types (collisions, support, occlusion, etc).  Furthermore, as noted above, the object representations learned by our model likely provide the foundation for subsequent word learning, and there is a large and somewhat contentious literature on this topic, which a more advanced multi-modal version of our model could hopefully contribute to \cite[e.g.,]{StevensGleitmanTrueswellEtAl17,YuSmith12,ColungaSmith05,WaxmanGelman09}.  This area is especially ripe for such models given a recent emphasis on collecting real-world experience samples that provide considerable insight and constraints \cite{YuSmith12,RoyFrankDeCampEtAl15,StevensGleitmanTrueswellEtAl17}.

\subsubsection{Consciousness and Qualia}

There are some potentially important implications of our framework for understanding the nature of consciousness, and what it feels like to be conscious of the visual world (qualia).  The pulvinar plays a central role in our model as a kind of {\em projection screen}, but this naturally raises the question: is ``anyone'' watching this screen?  Indeed, subjectively, there is a widespread seductive feeling that our brains have a kind of theater where the conscious part watches the incoming reports from the senses. \incite{Dennett91} refers to this as the {\em Cartesian Theater}, to deride the implicit dualism present in many theories (i.e., between the conscious part that watches the screen, and the unconscious part that projects representations onto it).  But what if our brains really do have a kind of ``silver screen of the Cartesian Theater'' in the pulvinar (updating at film-appropriate alpha frame rates no less!), which underlies this pervasive subjective feeling of there being a kind of internal movie screen in our minds?

Without adopting any form of materialistic dualism, it is still possible that the pulvinar can play a critical role in organizing and coordinating diffuse brain areas around a common focus on the collaboratively-generated prediction of what will happen next.  In so doing, we could say that this naturally contributes to the unitary nature of conscious experience, and provides a plausible substrate for how many different brain areas can share in a common perceptual-level sensory ``qualia'', which, because it is so strongly anchored by low-level visual areas (V1, V2), would have a distinctly ``visual'' feel to it.  This kind of architecture would seem likely to produce a different emergent subjective experience than one where each area only interacts with its nearest neighbors, and is thus more ``isolated'' (higher-order areas in particular would be more strongly detached from low-level sensory details).  This may also explain some of the mechanisms behind an embodied, sensory-motor foundation to higher-level cognitive function \cite{Barsalou08,Barsalou09,Anderson03a}. 

Critically, we avoid any strong localization of consciousness by virtue of the fact that each brain area is both a contributor to, and receiver of, this pulvinar projection screen, so there is no dualism of the form targeted by the Cartesian Theater notion --- consciousness remains an emergent process characterized by coordination of processing across diffuse brain areas, which is a common notion across many different accounts \cite{Baars83,Baars02,DehaeneNaccache01,CrickKoch03,Tononi04,Lamme06,SethDienesCleeremansEtAl08}.  In particular, the pulvinar may represent a different kind of global workspace than other accounts have postulated \cite{Baars02,DehaeneNaccache01}, but with perhaps similar functional implications.

Finally, it is essential to recognize that consciousness is {\em inescapably dualist} --- it is a property of {\em subjective} experience, which can {\em never} be described in purely {\em objective} terms.  This is not substance dualism, but rather {\em perspective dualism} --- it is literally definitionally impossible to transplant yourself into (another) human brain (you would become the other person, with no trace of yourself left, or some weird hybrid that is neither), so unless you happen to already be a human brain, you'll {\em never} know subjectively what it feels like to be one (and likewise for one individual brain to the next).  This perspective dualism likely accounts for much of what is attributed to the {\em hard problem} of consciousness \cite{Chalmers95}, without requiring any kind of substance dualism, and without preventing the attempt to map objective properties of the brain onto the subjective nature of experience.  For example, it would be really interesting if we could selectively deactivate the pulvinar and subjectively report the effect on the nature of our experience.  But that report would not enable others to actually experience the same thing, in the same way that attempting to convey the feeling of being on LSD or other powerful drugs is ultimately insufficient (no matter how poetic you get), if you haven't tried them yourself (and even then, you only truly know your own experience).  Thus, while it is impossible to prove, the image of all these brain areas gathered around the silver screen of the pulvinar may underlie some important aspects of our subjective experience, and hence the seductive pull of the Cartesian Theater notion.

\subsection{Predictions}

A paper on the importance of predictive learning certainly must include a section on predictions from this framework!  As in predictive learning, enumerating predictions from a theory provides a way of testing internal representations and refining them in light of observed data.  There are so many possible predictions from our framework, and a good deal of the existing data has already been discussed above, so here we highlight a few of the most central tests.

\begin{itemize}
\item Early developmental damage to the pulvinar should massively impair visual learning, but similar damage after developmental learning is complete should mainly affect attention (and also carefully-constructed learning tests that require learning in affected visual areas).
\item Early developmental damage to MT (and probably DP) should paradoxically impair object recognition, by interfering with the partitioning of prediction error, but later in development the stabilized {\em What} pathway representations should be much less affected.  The same applies to area LIP, but that might have even broader direct impairments that make it difficult to interpret.  Given the relative homogeneity and plasticity of neocortex, other areas might be able to partially compensate, so this could be challenging to test effectively.
\item The quantitative differences in response properties in the {\em What * Where} vs. {\em What} pathways as shown in Figure~\ref{fig.actrf} and Table~\ref{tab.actrf_stats} should be testable using sufficiently large samples of neural recordings.  The fuller integration in the {\em What * Where} pathway may emerge in the areas above MT (DP, MST, V6) in the larger scale context of the primate brain compared to our small-scale model.
\item If it were possible to selectively block the 5IB intrinsic bursting neurons, or perhaps disable their bursting behavior in some other way, we would predict that this would have a significant impact on any task requiring temporal integration of information over time.  For example, discriminating different individuals based on their walking motion, or recognizing a musical tune.  More generally, if any person was brave enough to attempt taking a pharmacological agent that selectively interfered with 5IB bursting, we would predict that it would significantly disrupt the basic continuity of consciousness --- everything would feel more fragmented and discontinuous and incoherent.  Indeed, perhaps certain existing psychoactive substances can be understood in part in terms of their modulation of alpha bursting?
\item Neocortical learning should also be significantly impaired with blockage of 5IB intrinsic bursting dynamics, because these contribute to the hypothesized plus phase of learning.  To test this prediction, the widely-used statistical learning paradigm would be ideal, where sequences of tones or visual stimuli are presented, with various forms of statistical regularities \cite[e.g.,]{AslinSaffranNewport98}.
\item Using large-scale lamina-specific neural recording techniques, it should be possible to quantify the information encoded in the layer 6 regular spiking (RS) neurons just after 5IB bursting, compared to the information in the superficial layers just prior.  Because we think that the layer 6 RS neurons convey the temporal context information from the prior alpha cycle, these two should be more strongly correlated in their information content, as compared to for example the information in superficial layers during the subsequent alpha cycle.  Also, these layer 6 neurons should exhibit more rapid representational changes immediately post 5IB bursting compared to later in the cycle.
\item A critical and only indirectly supported \cite{LimMcKeeWoloszynEtAl15,JedlickaBenuskovaAbraham15,ZenkeGerstnerGanguli17} property of our synaptic plasticity mechanism is the rapid updating of the plasticity threshold determining the boundary between LTD and LTP at the alpha time scale (as compared to the slower adaptation assumed in the BCM algorithm) --- this could be tested much more directly using standard {\em in vitro} techniques.  However, there may be important features of the awake {\em in vivo} environment that are essential for how the learning actually works, so that would be the ideal and only definitive test environment.  Potentially modern optogenetic and imaging techniques would be capable of addressing this question.
\item It should take at least two alpha cycles to process information from a new, exploratory fixation in a complex visual scene --- the first alpha cycle will only have weak predictive and attentional deep layer representations associated with it, so a second one is required to generate a reliable prediction and more refined attentional spotlight.  Thus, we predict that the modal fixation time in such cases should be around 200 msec.  We are unsure of what may happen with more complex, novel, or otherwise hard-to-process stimuli: they may require more alpha cycles, or the duration of settling within a given alpha cycle may be stretched out as the constraint satisfaction process converges \cite{WyatteCurranOReilly12}.
\item Instead of computing stable, static representations, the constant predictive pressure in this framework should favor rapidly-updating, dynamic representations that track the environment closely.  For example, working memory representations of spatial locations may be encoded in retinotopic coordinates, and updated with every saccade, instead of using a more allocentric representation that does not require this updating \cite{Wurtz08,CavanaghHuntAfrazEtAl10,FixRougierAlexandre11}.  This dynamic, constantly-updating, environmentally-tied vision of cognition is generally compatible with the embodied cognition approaches \cite{Barsalou08,Barsalou09,Anderson03a,SmithThelen03}.  
\end{itemize}

\subsection{Unresolved Issues and Future Research}

We have mentioned a number of unresolved issues and future directions throughout the paper. Here we highlight a few of the most important.
\begin{itemize}
\item Scaling up: How will the current model scale up to realistic 3D objects, larger spatial scales (allowing a difference between microsaccades and regular saccades), binocular and color vision, etc?  We are confident in the basic principles, but much hard computational work remains to scale up the model to handle more realistic visual inputs, including likely adding additional high-level areas to specialize on encoding the relevant new dimensions in an efficient, systematic, and compact manner (e.g., CIP, next to LIP, appears to be specialized for 3D shape information, and interacts with the IT {\em What} pathway; \nopcite{FreudPlautBehrmann16,DrommePremereurVerhoefEtAl16,TsutsuiJiangYaraEtAl01}).
\item Scaling n: The attentional properties of our framework are only relevant in cluttered scenes with multiple different objects that could be tracked --- these kinds of complex environments also need to be explored for many basic physical phenomena (collisions, support, occlusions etc).   Will the LIP spatial blob representations provide a central organizing ``FINST'' pointer that coordinates attention and prediction across multiple brain areas, for the attentionally selected objects \cite{Pylyshyn89,CavanaghHuntAfrazEtAl10,OReillyPetrovCohenEtAl14}?
\item Scaling out: how does visual predictive learning interact with auditory and/or somatosensory predictive learning?  As noted earlier, including auditory inputs is essential for exploring language learning, and forward-model-like predictive learning in speech, and motor control more broadly.
\item Scaling on: how do predictions and representations of longer time-scale events and episodes build upon the fast alpha-rhythm sensory predictive learning loop?  We noted that the medial temporal lobe can encode two alpha trials in one of its characteristic theta cycles, but how are yet longer time scales encoded?  Robust active maintenance in the prefrontal cortex likely plays a critical role, but how are its representations trained in the context of predictive learning?
\item Biological and mechanical motion: living things and machines move differently than inert physical objects --- if we are to accurately predict the visual world, a strong interaction between {\em What * Where} is necessary for these things.  From the principles of our framework, we would predict that that a specialized higher-level area above the basic {\em What * Where} pathway, with strong input from the {\em What} pathway, would be needed to learn these higher-order cases, and indeed just such an area in the STS, anatomically above the MT, MST pathway, has been identified \cite{PucePerrett03}.  
\item Dynamic alpha: there is considerable evidence that the alpha rhythm can be entrained by external stimuli, which is important for ensuring that the temporal context updates track relevant events in the environment.  The current model just uses a fixed trial timing, so relevant mechanisms to support alpha phase entrainment need to be incorporated into our model.
\item Dynamic activations: As reviewed above, there are many short-time-scale dynamics that may play an important role in shaping the time-evolution of neural representations at the alpha time scale --- these may affect the dynamics of prediction updating in important ways and should be thoroughly explored.
\item To what extent do the lessons from our pulvinar-based model apply to the LGN, in its interconnectivity with the retina and V1?  A fundamental difference is that there are no alpha-bursting plus phase driver inputs to the LGN as far as we know, so it would seem that the V1 / LGN system learns in a purely Hebbian manner without the benefit of predictive error signals, which is consistent with many Hebbian models of V1 learning \cite[e.g.,]{MillerKellerStryker89,BednarMiikkulainen03}.  However, the same prediction-generation pathway from layer 6CT to LGN does exist --- likely this is playing a largely attentional role as it also plays in the attentional aspect of our model.  Nevertheless, all of these issues bear deeper reexamination to see if there might be some other interesting kinds of thalamocortical learning dynamics taking place, which would likely also apply to other modalities.
\end{itemize}

\subsection{Conclusions}

In conclusion, our model clearly builds on ideas that have long been advocated in understanding neocortical function, while also adding some important new elements, that together have produced a coherent, functional, first pass working model demonstrating the sufficiency of the framework to achieve significant forms of learning through the predictive mechanism.  There are many outstanding questions still, so a pessimist may not yet be convinced of the value of this framework, and certainly we have a tremendous amount left to learn.  Finally, it is worth observing that the odds of discovering a model of this complexity through a purely bottom up, empirically-driven approach seem rather small.  Similarly, purely computational or cognitive-level theorists would probably not have arrived at some of the key insights provided by the biology.  Thus, a systems-focused, computational-modeling approach that integrates elements from all these different levels of analysis can play a critical role in advancing our understanding of the complexities of brain function.

\clearpage

\clearpage
\section{Appendix: Computational Model Details}

This appendix provides more information about the {\em What-Where Integration (WWI)} model.  The purpose of this information is to give more detailed insight into the model's function beyond the level provided in the main text, but with a model of this complexity, the only way to really understand it is to explore the model itself.  It is available for download at:

\noindent\verb\http://grey.colorado.edu/CompCogNeuro/\
\verb\index.php/CCN_Repository\.

And the best way to understand this model is to understand the framework in which it is implemented, which is explained in great detail, with many running simulations explaining specific elements of functionality, at \verb\http://ccnbook.colorado.edu\.  

\subsection{Layer Sizes and Structure}

\begin{table*}
  \centering
\begin{tabular}{llrrlll}
\hline
     &      & \multicolumn{2}{c}{{\bf Units}} & \multicolumn{2}{c}{{\bf Groups}} & \\
{\bf Area} & {\bf Name} & {\bf X} & {\bf Y} & {\bf X} & {\bf Y} & {\bf Receiving Projections} \\
\hline
V1 & V1s & 4 & 4 & 8 & 8 &  \\
   & V1p & 4 & 4 & 8 & 8 & V1s V2d V3d V4d TEOd  \\
Eyes & EyePos & 21 & 21 & & &  \\
     & SaccadePlan & 11 & 11 & & &  \\
     & Saccade & 11 & 11 & & &  \\
Obj & ObjVel & 11 & 11 & & & \\
V2 & V2s & 10 & 10 & 8 & 8 & V1s LIPs V3s V4s TEOd V1p \\
   & V2d & 10 & 10 & 8 & 8 & V2s V1p LIPd LIPp V3d V4d V3s TEOs \\
   & V2p & 10 & 10 & 8 & 8 & V2s V3d V4d TEOd \\
LIP & MtPos& 1 & 1 & 8 & 8 & V1s \\
    & LIPs & 4 & 4 & 8 & 8 & MtPos ObjVel SaccadePlan EyePos LIPp \\
    & LIPd & 4 & 4 & 8 & 8 & LIPs LIPp ObjVel Saccade EyePos \\
    & LIPp & 1 & 1 & 8 & 8 & V1s LIPd \\
V3 & V3s & 10 & 10 & 4 & 4 & V2s V4s TEOs MTs LIPs V1p MTp TEOd \\
   & V3d & 10 & 10 & 4 & 4 & V3s V1p MTp LIPd MTd V4d V4s MTs TEOs \\
   & V3p & 10 & 10 & 4 & 4 & V3s V2d MTd TEOd \\
MT & MTs & 10 & 10 & & & V2s V3s TEOs V1p V3p TEOp OFCp \\
   & MTd & 10 & 10 & & & MTs V1p MTp OFCp TEOd \\
   & MTp & 10 & 10 & & & MTs V2d V3d MTd TEOd \\
V4 & V4s & 10 & 10 & 4 & 4 & V2s TEOs V1p OFCp \\
   & V4d & 10 & 10 & 4 & 4 & V4s V1p V4p OFCp TEOd TEOs \\
   & V4p & 10 & 10 & 4 & 4 & V4s V2d V3d V4d TEOd \\
TEO & TEOs & 8 & 8 & 4 & 4 & V4s V1p \\
    & TEOd & 8 & 8 & 4 & 4 & TEOs TEOd V1p V4p TEOp OFCp \\
    & TEOp & 8 & 8 & 4 & 4 & TEOs V3d V4d TEOd \\
\hline
\end{tabular}
\caption{\footnotesize Layer sizes, showing numbers of units in one unit group (or entire layer if Group is missing), and the number of Groups of such units, along X,Y axes.  Each area has three associated layers: {\em s} = superficial layer, {\em d} = deep layer, {\em p} = pulvinar layer (driven by 5IB neurons from associated area).}
\label{tab.layer_sizes}
\end{table*}

Figure~\ref{fig.wwi_model} shows the general configuration of the model, and Table~\ref{tab.layer_sizes} shows the specific sizes of each of the layers, and where they receive inputs from.  The main text contains figures showing the patterns of connectivity, which establish the three pathways (Figures~\ref{fig.model_cons}, \ref{fig.model_cons_pulv}).

All the activation and general learning parameters in the model are at their standard Leabra defaults.

\subsection{Projections}

Detailing each of the specific parameters associated with the different projections shown in Table~\ref{tab.layer_sizes} would take too much space --- those interested in this level of detail should download the model from the link shown above.  There are topographic projections between many of the lower-level retinotopically-mapped layers, consistent with our earlier vision models \cite{OReillyWyatteHerdEtAl13}.  For example the 8x8 unit groups in V2 are reduced down to the 4x4 groups in V3 via a 4x4 unit-group topographic projection, where neighboring units have half-overlapping receptive fields (i.e., the field moves over 2 unit groups in V2 for every 1 unit group in V3), and the full space is uniformly tiled by using a wrap-around effect at the edges.  Similar patterns of connectivity are used in current deep convolutional neural networks.  However, we do not share weights across units as in a true convolutional network.

The projections from ObjVel (object velocity) and SaccadePlan layers to LIPs,d were initialized with a topographic sigmoidal pattern that moved as a function of the position of the unit group, by a factor of .5, while the projections from EyePos were initialized with a gaussian pattern.  These patterns multiplied uniformly distributed random weights in the .25 to .75 range, with the lowest values in the topographic pattern having a multiplier of .6, while the highest had a multiplier of 1 (i.e., a fairly subtle effect).  This produced faster convergence of the LIP {\em Where} pretraining compared to purely random initial weights, consistent with the basis function theory and related empirical observations \cite{ZipserAndersen88,PougetSejnowski97}.

In addition to exploring different patterns of overall connectivity, we also explored differences in the relative strengths of receiving projections, which can be set with a \texttt{wt\_scale.rel} parameter in the simulator.  All feedforward pathways have a default strength of 1.  For the feedback projections, which are typically weaker (consistent with the biology), we explored a discrete range of strengths, typically .5, .2, .1, and .05.  The strongest top-down projections were into V2s from LIP and V3, while most others were .2 or .1.  Likewise projections from the pulvinar were weaker, typically .1.  These differences in strength sometimes had large effects on performance during the initial bootstrapping of the overall model structure, but in the final model they are typically not very consequential for any individual projection.

\subsection{Training Parameters}

As noted in the main text, training typically consisted of 512 alpha trials per epoch (51.2 seconds of real time equivalent), for 1,000 such epochs.  Each trial was generated from the dynamic visual environment as described in the main text.  Because the start of each sequence of 4 trials is unpredictable, we turned off learning for that trial, which improves learning overall.  We have recently developed an automatic such mechanism based on the running-average (and running variance) of the prediction error, where we turn off learning whenever the current prediction error z-normalized by these running average values is below 1.5 standard deviations, which works well, and will be incorporated into future models.  Biologically, this could correspond to a connection between pulvinar and neuromodulatory areas that could regulate the effective learning rate in this way.

The plots of learning trajectories have been smoothed with a gaussian kernel with a half-width of 8 epochs, sigma = 4 epochs, to make the different lines more easily discriminable --- there is a reasonably high level of random noise in performance due to random variation in the environment parameters etc, so this smooths that out and allows the mean level to be visible.

\subsection{Model Algorithms}

The model was implemented using the Leabra framework, which is described in detail in previous publications \cite{OReillyHazyHerd15,OReillyMunakataFrankEtAl12,OReillyMunakata00,OReilly01,OReilly98,OReilly96}, and summarized here.  The main implementation of Leabra is in the {\em emergent} software \cite{AisaMingusOReilly08}, and another detailed explanation of the algorithm, and simple implementations of all the equations in Python and MATLAB, are available from:

\noindent\verb\https://grey.colorado.edu/emergent/\
\verb\index.php/Leabra\

These same equations and standard parameters have been used to simulate over 40 different models in \incite{OReillyMunakataFrankEtAl12} and \incite{OReillyMunakata00}, and a number of other research models.  Thus, the model can be viewed as an instantiation of a systematic modeling framework using standardized mechanisms, instead of constructing new mechanisms for each model.  

\providecommand{\tightlist}{%
  \setlength{\itemsep}{0pt}\setlength{\parskip}{0pt}}

\subsection{Leabra Algorithm
Equations}\label{leabra-algorithm-equations}

The pseudocode for Leabra is given here, showing exactly how the pieces
of the algorithm fit together, using the equations and variables from
the actual code. The implementation contains a number of optimizations
(including vectorization and GPU code), but this provides the core math
in simple form.

See the \texttt{Matlab} directory in the emergent \url{svn} source
directory for a complete implementation of these equations in Matlab,
coded by Sergio Verduzco-Flores --- this can be a lot simpler to read
than the highly optimized C++ source code.

\subsubsection{Timing}\label{timing}

Leabra is organized around the following timing, based on an
internally-generated alpha-frequency (10 Hz, 100 msec periods) cycle of
expectation followed by outcome, supported by neocortical circuitry in
the deep layers and the thalamus, as hypothesized in the
\url{DeepLeabra} extension to standard Leabra:

\begin{itemize}
\tightlist
\item
  A \textbf{Trial} lasts 100 msec (10 Hz, alpha frequency), and
  comprises one sequence of expectation --- outcome learning, organized
  into 4 quarters.

  \begin{itemize}
  \tightlist
  \item
    Biologically, the deep neocortical layers (layers 5, 6) and the
    thalamus have a natural oscillatory rhythm at the alpha frequency.
    Specific dynamics in these layers organize the cycle of expectation
    vs. outcome within the alpha cycle.
  \end{itemize}
\item
  A \textbf{Quarter} lasts 25 msec (40 Hz, gamma frequency) --- the
  first 3 quarters (75 msec) form the expectation / minus phase, and the
  final quarter are the outcome / plus phase.

  \begin{itemize}
  \tightlist
  \item
    Biologically, the superficial neocortical layers (layers 2, 3) have
    a gamma frequency oscillation, supporting the quarter-level
    organization.
  \end{itemize}
\item
  A \textbf{Cycle} represents 1 msec of processing, where each neuron
  updates its membrane potential etc according to the above equations.
\end{itemize}

\subsubsection{Variables}\label{variables}

LeabraUnits are organized into LeabraLayers, which sometimes have unit
groups (which are now typically purely virtual, not actual Unit\_Group
objects). The LeabraUnit has the following key parameters, along with a
number of others that are used for other non-default algorithms and
various optimizations, etc.

\begin{itemize}
\tightlist
\item
  \textbf{act} = activation sent to other units
\item
  \textbf{act\_nd} = non-depressed activation --- prior to application
  of any short-term plasticity
\item
  \textbf{net\_raw} = raw netinput, prior to time-averaging
\item
  \textbf{net} = time-averaged excitatory conductance (net input)
\item
  \textbf{gc\_i} = inhibitory conductance, computed from FFFB inhibition
  function typically
\item
  \textbf{I\_net} = net current, combining excitatory, inhibitory, and
  leak channels
\item
  \textbf{v\_m} = membrane potential
\item
  \textbf{v\_m\_eq} = equilibrium membrane potential --- not reset by
  spikes --- just keeps integrating
\item
  \textbf{adapt} = adaptation current
\item
  \textbf{avg\_ss} = super-short term running average activation
\item
  \textbf{avg\_s} = short-term running average activation, integrates
  over avg\_ss, represents plus phase learning signal
\item
  \textbf{avg\_m} = medium-term running average activation, integrates
  over avg\_s, represents minus phase learning signal
\item
  \textbf{avg\_l} = long-term running average activation, integrates
  over avg\_m, drives long-term floating average for Hebbian learning
\item
  \textbf{avg\_l\_lrn} = how much to use the avg\_l-based Hebbian
  learning for this receiving unit's learning --- in addition to the
  basic error-driven learning --- this can optionally be dynamically
  updated based on the avg\_l factor and average level of error in the
  receiving layer, so that this Hebbian learning constraint can be
  stronger as a unit gets too active and needs to be regulated more
  strongly, and in proportion to average error levels in the layer.
\item
  \textbf{avg\_s\_eff} = effective avg\_s value used in learning ---
  includes a small fraction (.1) of the avg\_m value, for reasons
  explained below.
\end{itemize}

Units are connected via synapses parameterized with the following
variables. These are actually stored in an optimized vector format, but
the LeabraCon object contains the variables as a template.

\begin{itemize}
\tightlist
\item
  \textbf{wt} = net effective synaptic weight between objects ---
  subject to contrast enhancement compared to fwt and swt
\item
  \textbf{dwt} = delta-wt --- change in synaptic weights due to
  learning
\item
  \textbf{dwavg} = time-averaged absolute value of weight change, for
  normalizing weight changes
\item
  \textbf{moment} = momentum integration of weight changes
\item
  \textbf{fwt} = fast weight --- used for advanced fast and slow weight
  learning dynamic --- otherwise equal to swt --- stored as
  non-contrast enhanced value
\item
  \textbf{swt} = slow weight --- standard learning rate weight ---
  stored as non-contrast enhanced value --- optional
\end{itemize}

\subsubsection{Activation Update Cycle (every 1 msec): Net input,
Inhibition,
Activation}\label{activation-update-cycle-every-1-msec-net-input-inhibition-activation}

For every cycle of activation updating, compute the net input,
inhibition, membrane potential, and activation:

\begin{itemize}
\tightlist
\item
  \textbf{Net input} (see LeabraUnitSpec.cpp for code):

  \begin{itemize}
  \tightlist
  \item
    \textbf{\texttt{net\_raw}}\texttt{\ +=\ (sum\ over\ recv\ connections\ of:)\ scale\_eff\ *\ act\ *\ wt}

    \begin{itemize}
    \item
      \textbf{scale\_eff} =
      \verb\https://grey.colorado.edu/emergent/\
      \verb\index.php/Leabra_Netin_Scaling\ factor that includes 1/N to compute an average, plus
      wt\_scale.rel and abs relative and absolute scaling terms.
    \item
      \textbf{act} = sending unit activation
    \item
      \textbf{wt} = receiving connection weight value between sender and
      receiver
    \item
      does this very efficiently by using a sender-based computation,
      that only sends \emph{changes} (deltas) in activation values ---
      typically only a few percent of neurons send on any given cycle.
    \end{itemize}
  \item
    \textbf{\texttt{net}}\texttt{\ +=\ dt.integ\ *\ dt.net\_dt\ *\ (net\_raw\ -\ net)}

    \begin{itemize}
    \tightlist
    \item
      time integration of net input, using net\_dt (1/1.4 default), and
      global integration time constant, dt.integ (1 = 1 msec default)
    \end{itemize}
  \end{itemize}
\end{itemize}

\begin{itemize}
\tightlist
\item
  \textbf{Inhibition} (see LeabraLayerSpec.cpp for code) -- earlier versions of Leabra used an explicit k-Winners-Take-All inhibition function, but the FFFB equations here are much simpler and produce desirable flexibility in overall activation levels:

  \begin{itemize}
  \tightlist
  \item
    \textbf{\texttt{ffi}}\texttt{\ =\ ff\ *\ MAX(netin.avg\ -\ ff0,\ 0)}

    \begin{itemize}
    \tightlist
    \item
      feedforward component of inhibition with ff multiplier (1 by
      default) --- has ff0 offset and can't be negative (that's what
      the MAX(.. ,0) part does).
    \item
      \textbf{netin.avg} is average of net variable across unit group or
      layer, depending on what level this is being computed at (both are
      supported)
    \end{itemize}
  \item
    \textbf{\texttt{fbi}}\texttt{\ +=\ fb\_dt\ *\ (fb\ *\ acts.avg\ -\ fbi)}

    \begin{itemize}
    \tightlist
    \item
      feedback component of inhibition with fb multiplier (1 by default)
      --- requires time integration to dampen oscillations that
      otherwise occur --- fb\_dt = 1/1.4 default
    \end{itemize}
  \item
    \textbf{\texttt{gc\_i}}\texttt{\ =\ gi\ *\ (ffi\ +\ fbi)}

    \begin{itemize}
    \tightlist
    \item
      total inhibitory conductance, with global gi multiplier ---
      default of gi=1.8 typically produces good sparse distributed
      representations in reasonably large layers (25 units or more)
    \end{itemize}
  \end{itemize}
\end{itemize}

\begin{itemize}
\tightlist
\item
  \textbf{Membrane potential} (see LeabraUnitSpec.cpp for code)

  \begin{itemize}
  \tightlist
  \item
    \textbf{\texttt{I\_net}}\texttt{\ =\ net\ *\ (e\_rev.e\ -\ v\_m)\ +\ gc\_l\ *\ (e\_rev.l\ -\ v\_m)\ +\ gc\_i\ *\ (e\_rev.i\ -\ v\_m)\ +\ noise}

    \begin{itemize}
    \tightlist
    \item
      net current = sum of individual ionic channels: e = excitatory, l
      = leak (gc\_l is a constant, 0.1 default), and i = inhibitory
    \item
      e\_rev are reversal potentials: in normalized values derived from
      biophysical values, e\_rev.e = 1, .l = 0.3, i = 0.25
    \item
      noise is typically gaussian if added
    \end{itemize}
  \item
    if ex:
    \textbf{\texttt{I\_net}}\texttt{\ +=\ g\_bar.l\ *\ exp\_slope\ *\ exp((v\_m\ -\ thr)\ /\ exp\_slope)}

    \begin{itemize}
    \tightlist
    \item
      this is the exponential component of AdEx, if in use (typically
      only for discrete spiking), exp\_slope = .02 default
    \end{itemize}
  \item
    \textbf{\texttt{v\_m}}\texttt{\ +=\ dt.integ\ *\ dt.vm\_dt\ *\ (I\_net\ -\ adapt)}

    \begin{itemize}
    \tightlist
    \item
      in , we use a simple midpoint method that evaluates v\_m with a
      half-step time constant, and then uses this half-step v\_m to
      compute full step in above I\_net equation. vm\_dt = 1/3.3
      default.
    \item
      v\_m is always computed as in discrete spiking, even when using
      rate code, with v\_m reset to vm\_r etc --- this provides a more
      natural way to integrate adaptation and short-term plasticity
      mechanisms, which drive off of the discrete spiking.
    \end{itemize}
  \item
    \textbf{\texttt{I\_net\_r}}\texttt{\ =\ net\ *\ (e\_rev.e\ -\ v\_m\_eq)\ +\ gc\_l\ *\ (e\_rev.l\ -\ v\_m\_eq)\ +\ gc\_i\ *\ (e\_rev.i\ -\ v\_m\_eq)\ +\ noise}

    \begin{itemize}
    \tightlist
    \item
      rate-coded version of I\_net, to provide adequate coupling with
      v\_m\_eq.
    \end{itemize}
  \item
    \textbf{\texttt{v\_m\_eq}}\texttt{\ +=\ dt.integ\ *\ dt.vm\_dt\ *\ (I\_net\_r\ -\ adapt)}

    \begin{itemize}
    \tightlist
    \item
      the \emph{equilibrium} version of the membrane potential does
      \emph{not} reset with spikes, and is important for rate code per
      below
    \end{itemize}
  \end{itemize}
\end{itemize}

\begin{itemize}
\tightlist
\item
  \textbf{Activation} (see LeabraUnitSpec.cpp for code)

  \begin{itemize}
  \tightlist
  \item
    \textbf{\texttt{g\_e\_thr}}\texttt{\ =\ (gc\_i\ *\ (e\_rev\_i\ -\ thr)\ +\ gc\_l\ *\ (e\_rev\_l\ -\ thr)\ -\ adapt)\ /\ (thr\ -\ e\_rev.e)}

    \begin{itemize}
    \tightlist
    \item
      the amount of excitatory conductance required to put the neuron
      exactly at the firing threshold, thr = .5 default.
    \end{itemize}
  \item
    \texttt{if(v\_m\ \textgreater{}\ spk\_thr)\ \{\ spike\ =\ 1;\ v\_m\ =\ vm\_r;\ I\_net\ =\ 0.0\ \}\ else\ \{\ spike\ =\ 0\ \}}

    \begin{itemize}
    \tightlist
    \item
      spk\_thr is spiking threshold (1.2 default, different from rate
      code thr), vm\_r = .3 is the reset value of the membrane potential
      after spiking --- we also have an optional refractory period
      after spiking, default = 3 cycles, where the vm equations are
      simply not computed, and vm remains at vm\_r.
    \item
      if using spiking mode, then \textbf{act} = spike, otherwise, rate
      code function is below
    \end{itemize}
  \item
    \texttt{if(v\_m\_eq\ \textless{}=\ thr)\ \{\ }\textbf{\texttt{new\_act}}\texttt{\ =\ NXX1(v\_m\_eq\ -\ thr)\ \}\ else\ \{\ }\textbf{\texttt{new\_act}}\texttt{\ =\ NXX1(net\ -\ g\_e\_thr)\ \}}

    \begin{itemize}
    \tightlist
    \item
      it is important that the time to first ``spike'' be governed by
      v\_m integration dynamics, but after that point, it is essential
      that activation drive directly from the excitatory conductance
      (g\_e or net) relative to the g\_e\_thr threshold --- activation
      rates are linear in this term, but not even a well-defined
      function of v\_m\_eq --- earlier versions of Leabra only used the
      v\_m\_eq-based term, and this led to some very strange behavior.
    \item
      NXX1 = noisy-x-over-x+1 function, which is implemented using a
      lookup table due to the convolving of the XX1 function with a
      gaussian noise kernel
    \item
      \texttt{XX1(x)\ =\ gain\ *\ x\ /\ (gain\ *\ x\ +\ 1)}
    \item
      gain = 100 default
    \end{itemize}
  \item
    \textbf{\texttt{act\_nd}}\texttt{\ +=\ dt.integ\ *\ dt.vm\_dt\ *\ (new\_act\ -\ act\_nd)}

    \begin{itemize}
    \tightlist
    \item
      non-depressed rate code activation is time-integrated using same
      vm\_dt time constant as used in v\_m, from the new activation
      value
    \end{itemize}
  \item
    \textbf{\texttt{act}}\texttt{\ =\ act\_nd\ *\ syn\_tr\ (or\ just\ act\_nd)}

    \begin{itemize}
    \tightlist
    \item
      if short-term plasticity is in effect, then syn\_tr variable
      reflects the synaptic transmission efficacy, and this product
      provides the net signal sent to the receiving neurons. otherwise
      syn\_tr = 1.
    \end{itemize}
  \item
    \textbf{\texttt{adapt}}\texttt{\ +=\ dt.integ\ *\ (adapt.dt\ *\ (vm\_gain\ *\ (v\_m\ -\ e\_rev.l)\ -\ adapt)\ +\ spike\ *\ spike\_gain)}

    \begin{itemize}
    \tightlist
    \item
      adaptation current --- causes rate of activation / spiking to
      decrease over time, adapt.dt = 1/144, vm\_gain = 0.04, spike\_gain
      = .00805 defaults
    \end{itemize}
  \end{itemize}
\end{itemize}

\subsubsection{Learning}\label{learning}

Learning is based on running-averages of activation variables, described
first:

\begin{itemize}
\tightlist
\item
  \textbf{Running averages} computed continuously every cycle, and note
  the compounding form (see LeabraUnitSpec.cpp for code)

  \begin{itemize}
  \tightlist
  \item
    \textbf{\texttt{avg\_ss}}\texttt{\ +=\ dt.integ\ *\ ss\_dt\ *\ (act\_nd\ -\ avg\_ss)}

    \begin{itemize}
    \tightlist
    \item
      super-short time scale running average, ss\_dt = 1/2 default ---
      this was introduced to smooth out discrete spiking signal, but is
      also useful for rate code
    \end{itemize}
  \item
    \textbf{\texttt{avg\_s}}\texttt{\ +=\ dt.integ\ *\ act\_avg.s\_dt\ *\ (avg\_ss\ -\ avg\_s)}

    \begin{itemize}
    \tightlist
    \item
      short time scale running average, s\_dt = 1/2 default --- this
      represents the ``plus phase'' or actual outcome signal in
      comparison to avg\_m
    \end{itemize}
  \item
    \textbf{\texttt{avg\_m}}\texttt{\ +=\ dt.integ\ *\ act\_avg.m\_dt\ *\ (avg\_s\ -\ avg\_m)}

    \begin{itemize}
    \tightlist
    \item
      medium time-scale running average, m\_dt = 1/10 average --- this
      represents the ``minus phase'' or expectation signal in comparison
      to avg\_s
    \end{itemize}
  \item
    \textbf{\texttt{avg\_l}}\texttt{\ +=\ avg\_l.dt\ *\ (avg\_l.gain\ *\ avg\_m\ -\ avg\_l);\ avg\_l\ =\ MAX(avg\_l,\ min)}

    \begin{itemize}
    \tightlist
    \item
      long-term running average --- this is computed just once per
      learning trial, \emph{not every cycle} like the ones above ---
      gain = 2.5 (or 1.5 in some cases works better), min = .2, dt = .1
      by default
    \item
      same basic exponential running average as above equations
    \end{itemize}
  \item
    \textbf{\texttt{avg\_s\_eff}}\texttt{\ =\ m\_in\_s\ *\ avg\_m\ +\ (1\ -\ m\_in\_s)\ *\ avg\_s}

    \begin{itemize}
    \tightlist
    \item
      mix in some of the medium-term factor into the short-term factor
      --- this is important for ensuring that when neuron turns off in
      the plus phase (short term), that enough trace of earlier
      minus-phase activation remains to drive it into the LTD weight
      decrease region --- m\_in\_s = .1 default.
    \item
      this is now done at the unit level --- previously was done at the
      connection level which is much less efficient!
    \end{itemize}
  \end{itemize}
\end{itemize}

\begin{itemize}
  \tightlist
  \item \emph{Optional, on by default:} dynamic modulation of amount of
    Hebbian learning, based on avg\_l value and level of err in a given
    layer --- these factors make a small (few percent) but reliable
    difference in overall performance across various challenging tasks
    --- they can readily be omitted in favor of a fixed avg\_l\_lrn
    factor of around 0.0004 (with 0 for target layers --- it doesn't
    make sense to have any Hebbian learning at output layers):

    \begin{itemize}
    \tightlist
    \item
      \textbf{\texttt{avg\_l\_lrn}}\texttt{\ =\ avg\_l.lrn\_min\ +\ (avg\_l\ -\ avg\_l.min)\ *\ ((avg\_l.lrn\_max\ -\ avg\_l.lrn\_min)\ /\ avg\_l.gain\ -\ avg\_l.min))}

      \begin{itemize}
      \tightlist
      \item
        learning strength factor for how much to learn based on avg\_l
        floating threshold --- this is dynamically modulated by
        strength of avg\_l itself, and this turns out to be critical
        --- the amount of this learning increases as units are more
        consistently active all the time (i.e., ``hog'' units).
        avg\_l.lrn\_min = 0.0001, avg\_l.lrn\_max = 0.5. Note that this
        depends on having a clear max to avg\_l, which is an advantage
        of the exponential running-average form above.
      \end{itemize}
    \item
      \textbf{\texttt{avg\_l\_lrn}}\texttt{\ *=\ MAX(1\ -\ cos\_diff\_avg,\ 0.01)}

      \begin{itemize}
      \tightlist
      \item
        also modulate by time-averaged cosine (normalized dot product)
        between minus and plus phase activation states in given
        receiving layer (cos\_diff\_avg), (time constant 100) --- if
        error signals are small in a given layer, then Hebbian learning
        should also be relatively weak so that it doesn't overpower it
        --- and conversely, layers with higher levels of error signals
        can handle (and benefit from) more Hebbian learning. The
        MAX(0.01) factor ensures that there is a minimum level of .01
        Hebbian (multiplying the previously-computed factor above). The
        .01 * .05 factors give an upper-level value of .0005 to use for
        a fixed constant avg\_l\_lrn value --- just slightly less than
        this (.0004) seems to work best if not using these adaptive
        factors.
      \end{itemize}
    \end{itemize}
\end{itemize}

\begin{itemize}
\tightlist
\item
  \textbf{Learning equation} (see LeabraConSpec.h for code) --- most of
  these are intermediate variables used in computing final dwt value

  \begin{itemize}
  \tightlist
  \item
    \textbf{\texttt{srs}}\texttt{\ =\ ru-\textgreater{}avg\_s\_eff\ *\ su-\textgreater{}avg\_s\_eff}

    \begin{itemize}
    \tightlist
    \item
      short-term sender-receiver co-product --- this is the
      intracellular calcium from NMDA and other channels
    \end{itemize}
  \item
    \textbf{\texttt{srm}}\texttt{\ =\ ru-\textgreater{}avg\_m\ *\ su-\textgreater{}avg\_m}

    \begin{itemize}
    \tightlist
    \item
      medium-term sender-receiver co-product --- this drives dynamic
      threshold for error-driven learning
    \end{itemize}
  \item
    \textbf{\texttt{dwt}}\texttt{\ +=\ lrate\ *\ {[}\ m\_lrn\ *\ XCAL(srs,\ srm)\ +\ ru-\textgreater{}avg\_l\_lrn\ *\ XCAL(srs,\ ru-\textgreater{}avg\_l){]}}

    \begin{itemize}
    \tightlist
    \item
      weight change is sum of two factors: error-driven based on
      medium-term threshold (srm), and BCM Hebbian based on long-term
      threshold of the recv unit (ru-\textgreater{}avg\_l)
    \item
      in earlier versions, the two factors were combined into a single
      threshold value, using normalized weighting factors --- this was
      more elegant, but by separating the two apart, we allow the
      hebbian component to use the full range of the XCAL function (as
      compared to the relatively small avg\_l\_lrn factor applied
      \emph{inside} the threshold computation). By multiplying by
      avg\_l\_lrn outside the XCAL equation, we get the desired contrast
      enhancement property of the XCAL function, where values close to
      the threshold are pushed either higher (above threshold) or lower
      (below threshold) most strongly, and values further away are less
      strongly impacted.
    \item
      m\_lrn is a constant and is typically 1.0 when error-driven
      learning is employed (but can be set to 0 to have a completely
      Hebbian model).
    \item
      XCAL is the ``check mark'' linearized BCM-style learning function
      (see figure) that was derived from the Urakubo Et Al (2008) STDP
      model, as described in more detail in the CCN textbook:
      \url{http://ccnbook.colorado.edu}
    \item
      \texttt{XCAL(x,\ th)\ =\ (x\ \textless{}\ d\_thr)\ ?\ 0\ :\ (x\ \textgreater{}\ th\ *\ d\_rev)\ ?\ (x\ -\ th)\ :\ (-x\ *\ ((1-d\_rev)/d\_rev))}
    \item
      d\_thr = 0.0001, d\_rev = 0.1 defaults
    \item
      x ? y : z terminology is C syntax for: if x is true, then y, else
      z
    \end{itemize}
  \end{itemize}
\end{itemize}

\begin{itemize}
\tightlist
\item
  \textbf{Momentum} --- as of version 8.2.0, momentum is turned on by
  default, and has significant benefits for preventing hog units by
  driving more rapid specialization and convergence on promising error
  gradients.

  \begin{itemize}
  \tightlist
  \item
    \textbf{\texttt{dwavg}}\texttt{\ =\ MAX(dwavg\_dt\_c\ *\ dwavg,\ ABS(dwt))}

    \begin{itemize}
    \tightlist
    \item
      increment the running-average weight change magnitude (dwavg),
      using abs (L1 norm) instead of squaring (L2 norm), and with a
      small amount of decay: dwavg\_dt\_c = 1 - .001 --- software uses
      dwavg\_tau = 1000 as a time-constant of this decay:
      \texttt{dwavg\_dt\_c\ =\ 1\ -\ 1/dwavg\_tau}.
    \end{itemize}
  \item
    \textbf{\texttt{moment}}\texttt{\ =\ m\_dt\_c\ *\ moment\ +\ dwt}

    \begin{itemize}
    \tightlist
    \item
      increment momentum from new weight change ---
      \texttt{m\_dt\_c\ =\ 1\ -\ 1/m\_tau} where m\_tau = 20 trial time
      constant for momentum integration by default, which works best
      (i.e., m\_dt\_c = .95 --- .9 (m\_tau = 10) is a
      traditionally-used momentum value that also works fine but .95
      (m\_tau = 20) works better for most cases.
    \end{itemize}
  \item
    \texttt{if(dwavg\ !=\ 0)\ dwt\ =\ moment\ /\ MAX(dwavg,\ norm\_min);\ else\ dwt\ =\ moment}

    \begin{itemize}
    \tightlist
    \item
      set the weight change used by following weight update equation to
      use momentum, normalized by dwavg if available (nonzero) --- this
      normalization is used in RMSProp, ADAM, and other related
      algorithms.
    \end{itemize}
  \end{itemize}
\end{itemize}

\begin{itemize}
\tightlist
\item
  \textbf{Weight update equation} (see LeabraConSpec.h for code) (see
  below for alternative version using differential fast vs. slow
  weights, not used by default)

  \begin{itemize}
  \tightlist
  \item
    The \textbf{fwt} value here is the linear, non-contrast enhanced
    version of the weight value, while \textbf{wt} is the sigmoidal
    contrast-enhanced version, which is used for sending netinput to
    other neurons. One can compute fwt from wt and vice-versa, but
    numerical errors can accumulate in going back-and forth more than
    necessary, and it is generally faster to just store these two weight
    values (and they are needed for the slow vs. fast weights version
    show below).
  \item
    \texttt{dwt\ *=\ (dwt\ \textgreater{}\ 0)\ ?\ (1-fwt)\ :\ fwt}

    \begin{itemize}
    \tightlist
    \item
      soft weight bounding --- weight increases exponentially
      decelerate toward upper bound of 1, and decreases toward lower
      bound of 0. based on linear, non-contrast enhanced fwt weights.
    \end{itemize}
  \item
    \textbf{\texttt{fwt}}\texttt{\ +=\ dwt}

    \begin{itemize}
    \tightlist
    \item
      increment the linear weights with the bounded dwt term
    \end{itemize}
  \item
    \textbf{\texttt{wt}}\texttt{\ =\ SIG(fwt)}

    \begin{itemize}
    \tightlist
    \item
      new weight value is sigmoidal contrast enhanced version of fast
      weight
    \item
      \texttt{SIG(w)\ =\ 1\ /\ (1\ +\ (off\ *\ (1-w)/w)\^{}gain)}
    \end{itemize}
  \item
    \textbf{\texttt{dwt}}\texttt{\ =\ 0}

    \begin{itemize}
    \tightlist
    \item
      reset weight changes now that they have been applied.
    \end{itemize}
  \end{itemize}
\end{itemize}

\begin{itemize}
\tightlist
\item
   \emph{Optional, on by default:} \textbf{Weight Balance} --- this option attempts to
  maintain more balanced weights across units, to prevent some units
  from hogging the representational space, by changing the rates of
  weight increase and decrease in the soft weight bounding function, as
  a function of the average receiving weights:

  \begin{itemize}
  \tightlist
  \item
    \texttt{dwt\ *=\ (dwt\ \textgreater{}\ 0)\ ?\ wb\_inc\ *\ (1-fwt)\ :\ wb\_dec\ *\ fwt}

    \begin{itemize}
    \tightlist
    \item
      wb\_inc = weight increase modulator, and wb\_dec = weight decrease
      modulator (when these are both 1, this is same as standard, and
      this is the default value of these factors)
    \end{itemize}
  \item
    \texttt{wt\_avg\ =\ }

    \begin{itemize}
    \tightlist
    \item
      average of all the receiving weights --- computed \emph{per
      projection} (corresponding to a dendritic branch perhaps)
    \end{itemize}
  \item
    \texttt{if\ (wt\_avg\ \textgreater{}\ hi\_thr)\ then\ wbi\ =\ gain\ *\ (wt\_avg\ -\ hi\_thr);\ wb\_inc\ =\ 1\ -\ wbi;\ wb\_dec\ =\ 1\ +\ wbi}

    \begin{itemize}
    \tightlist
    \item
      If the average weights are higher than a high threshold (hi\_thr =
      .4 default) then the increase factor wb\_inc is reduced, and the
      decrease factor wb\_dec is increased, by a factor wbi that is
      determined by how far above the threshold the average is. Thus,
      the higher the weights get, the less quickly they can increase,
      and the more quickly they decrease, pushing them back into
      balance.
    \end{itemize}
  \item
    \texttt{if\ (wt\_avg\ \textless{}\ lo\_thr)\ then\ wbd\ =\ gain\ *\ (wt\_avg\ -\ lo\_thr);\ wb\_inc\ =\ 1\ -\ wbd;\ wb\_dec\ =\ 1\ +\ wbd}

    \begin{itemize}
    \tightlist
    \item
      This is the symmetric version for case when weight averages are
      below a low threshold (lo\_thr = .2), and the weight balance
      factors go in the opposite direction (wbd is negative), causing
      weight increases to be favored over decreases.
    \end{itemize}
  \item
    The hi\_thr and lo\_thr parameters are specified in terms of a
    target weight average value \texttt{trg\ =\ .3} with a threshold
    \texttt{thr=.1} around that target value, with these defaults
    producing the default .4 and .2 hi and lo thresholds respectively.
  \item
    A key feature of this mechanism is that it does not change the sign
    of any weight changes, including not causing weights to change that
    are otherwise not changing due to the learning rule. This is not
    true of an alternative mechanism that has been used in various
    models, which normalizes the total weight value by subtracting the
    average. Overall this weight balance mechanism is important for
    larger networks on harder tasks, where the hogging problem can be a
    significant problem.
  \end{itemize}
\end{itemize}


\subsubsection{Deep Context}

At the end of every plus phase, a new deep-layer context net input is computed from the dot product of the context weights times the sending activations, just as in the standard net input:
\begin{equation}
 \eta = \langle x_i \wij \rangle = \oneo{n} \sum_i x_i \wij
 \label{eq.net_in_ti}
\end{equation}
This net input is then added in with the standard net input at each cycle of processing.

Learning of the context weights occurs as normal, but using the sending activation states from the {\em prior} time step's activation.

\subsubsection{Computational and Biological Details of SRN-like Functionality}

Predictive auto-encoder learning has been explored in various frameworks, but the most relevant to our model comes from the application of the SRN to a range of predictive learning domains \cite{Elman90,Elman91,Jordan89,ElmanBatesKarmiloff-SmithEtAl96}.  One of the most powerful features of the SRN is that it enables error-driven learning, instead of arbitrary parameter settings, to determine how prior information is integrated with new information.  Thus, SRNs can learn to hold onto some important information for a relatively long interval, while rapidly updating other information that is only relevant for a shorter duration \cite[e.g.,]{CleeremansServan-SchreiberMcClelland89,Cleeremans93}.  This same flexibility is present in our DeepLeabra model.  Furthermore, because this temporal context information is hypothesized to be present in the deep layers throughout the entire neocortex (in every microcolumn of tissue), the DeepLeabra model provides a more pervasive and interconnected form of temporal integration compared to the SRN, which typically just has a single temporal context layer associated with the internal ``hidden'' layer of processing units.

\begin{figure}
  \centering\includegraphics[width=3in]{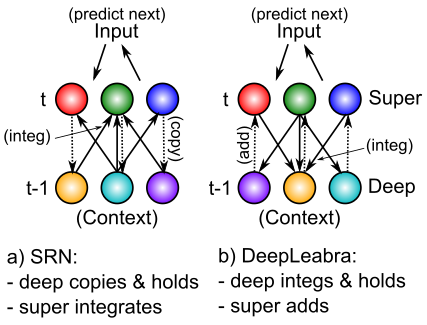}
  \caption{\footnotesize How the DeepLeabra temporal context computation compares to the SRN mathematically. {\bf a)} In a standard SRN, the context (deep layer biologically) is a copy of the hidden activations from the prior time step, and these are held constant while the hidden layer (superficial) units integrate the context through learned synaptic weights.  {\bf b)} In DeepLeabra, the deep layer performs the weighted integration of the soon-to-be context information from the superficial layer, and then holds this integrated value, and feeds it back as an additive net-input like signal to the superficial layer.  The context net input is pre-computed, instead of having to compute this same value over and over again.  This is more efficient, and more compatible with the diffuse interconnections among the deep layer neurons.  Layer 6 projections to the thalamus and back recirculate this pre-computed net input value into the superficial layers (via layer 4), and back into itself to support maintenance of the held value.}
  \label{fig.srn_vs_ti}
\end{figure}

An extensive computational analysis of what makes the SRN work as well as it does, and explorations of a range of possible alternative frameworks, has led us to an important general principle: {\em subsequent outcomes determine what is relevant from the past}.  At some level, this may seem obvious, but it has significant implications for predictive learning mechanisms based on temporal context.  It means that the information encoded in a temporal context representation cannot be learned at the time when that information is presently active.  Instead, the relevant contextual information is learned on the basis of what happens next.  This explains the peculiar power of the otherwise strange property of the SRN: the temporal context information is preserved as a {\em direct copy} of the state of the hidden layer units on the previous time step (Figure~\ref{fig.srn_vs_ti}), and then learned synaptic weights integrate that copied context information into the next hidden state (which is then copied to the context again, and so on).  This enables the error-driven learning taking place in the {\em current} time step to determine how context information from the {\em previous} time step is integrated.  And the simple direct copy operation eschews any attempt to shape this temporal context itself, instead relying on the learning pressure that shapes the hidden layer representations to also shape the context representations.  In other words, this copy operation is essential, because there is no other viable source of learning signals to shape the nature of the context representation itself (because these learning signals require future outcomes, which are by definition only available later).

The direct copy operation of the SRN is however seemingly problematic from a biological perspective: how could neurons copy activations from another set of neurons at some discrete point in time, and then hold onto those copied values for a duration of 100 msec, which is a reasonably long period of time in neural terms (e.g., a rapidly firing cortical neuron fires at around 100 Hz, meaning that it will fire 10 times within that context frame).  However, there is an important transformation of the SRN context computation, which is more biologically plausible, and compatible with the structure of the deep network (Figure~\ref{fig.srn_vs_ti}). Specifically, instead of copying an entire set of activation states, the context activations (generated by the phasic 5IB burst) are immediately sent through the adaptive synaptic weights that integrate this information, which we think occurs in the 6CC (corticortical) and other lateral integrative connections from 5IB neurons into the rest of the deep network \cite{Thomson10,ThomsonLamy07,SchubertKotterStaiger07}.  The result is a {\em pre-computed net input} from the context onto a given hidden unit (in the original SRN terminology), not the raw context information itself.  Computationally, and metabolically, this is a much more efficient mechanism, because the context is, by definition, unchanging over the 100 msec alpha cycle, and thus it makes more sense to pre-compute the synaptic integration, rather than repeatedly re-computing this same synaptic integration over and over again (in the original feedforward backpropagation-based SRN model, this issue did not arise because a single step of activation updating took place for each context update --- whereas in our bidirectional model many activation update steps must take place per context update).

There are a couple of remaining challenges for this transformation of the SRN.  First, the pre-computed net input from the context must somehow persist over the subsequent 100 msec period of the alpha cycle.  We hypothesize that this can occur via NMDA and mGluR channels that can easily produce sustained excitatory currents over this time frame.  Furthermore, the reciprocal excitatory connectivity from 6CT to TRC and back to 6CT could help to sustain the initial temporal context signal.  Second, these contextual integration synapses require a different form of learning algorithm that uses the sending activation from the prior 100 msec, which is well within the time constants in the relevant calcium and second messenger pathways involved in synaptic plasticity \cite{UrakuboHondaFroemkeEtAl08,BearMalenka94}.

Finally, we note that we had proposed a different, more limited version of this overall DeepLeabra framework previously, which we referred to as {\em LeabraTI} (temporal integration) \cite{KachergisWyatteOReillyEtAl14}.  The LeabraTI model hypothesized that higher areas attempt to reconstruct the activation states over the superficial layers of the areas below them, which raised many problems having to do with creating a plausible (and computationally effective) difference between the minus and plus phase states of these areas.  Thus, from the perspective of our current framework, the configuration of the TRC neurons within the overall network seems suspiciously ideal for their use as a projection-screen-like substrate for predictive auto-encoder learning.  Furthermore, using a single layer driven bidirectionally for the visible layer neurons as we do with the TRC neurons is much more efficient and natural than the two separate layers (input and output) that are required in the typical feedforward SRN framework.



\end{document}